%% file: main.tex
\documentclass[a4paper]{article}

\usepackage[english]{babel}
\usepackage[utf8]{inputenc}
\usepackage[T1,T2A]{fontenc}
\usepackage[a4paper,top=3cm,bottom=2cm,left=3cm,right=3cm,marginparwidth=2cm]{geometry}

\usepackage{lineno}

\usepackage{amsfonts,amsmath}
\usepackage{amssymb}
\usepackage{siunitx}
\usepackage{slashed}
\usepackage{jinstpub}
\usepackage{graphicx}
\usepackage{url}
\usepackage[colorinlistoftodos]{todonotes}

\usepackage{xspace}
\def\bit{\begin{itemize}}
\def\eit{\end{itemize}}
\def\beqr{\begin{eqnarray}}
\def\bfig{\begin{figure}}
\def\efig{\end{figure}}

\def\ttbar{$t\bar{t}~$}

\def\GeV{GeV}
\def\MeV{MeV}

\def\mumu{\ensuremath{\mu^+\mu^-}}
\def\ee{\ensuremath{e^+e^-}}

\DeclareUnicodeCharacter{2212}{-}

\newcommand{\iab}{{\,{\rm ab}^{-1}}}

\newcommand\snowmass{\begin{center}\rule[-0.2in]{\hsize}{0.01in}\\\rule{\hsize}{0.01in}\\
\vskip 0.1in Cross-Frontier Report Submitted to the US Community Study\\ 
on the Future of Particle Physics (Snowmass 2021)\\ 
\rule{\hsize}{0.01in}\\\rule[+0.2in]{\hsize}{0.01in} \end{center}}

\newcommand*{\SRotp}{SR\ensuremath{^{\gamma}_{1t}}\xspace}
\newcommand*{\SRttp}{SR\ensuremath{^{\gamma}_{2t}}\xspace}
\newcommand*{\GEANT}{\textsc{Geant}\xspace}
\newcommand\memoriam{\begin{center}\vskip 0.3in
{\small\it{In memory of Meenakshi Narain, \\ 
a friend and a colleague, who passed away on January 1, 2023. \\She fought tirelessly so that the next generation of particle physicists could continue the quest to discover \\the underlying laws of the universe. Her vision and support ensured the creation of the forum and\\ this report. She will be deeply missed.\\}}\end{center}}
\graphicspath{{Figures/}}

\title{\textbf{Muon Collider Forum Report}\memoriam}

\input{authors}

\abstract{
A multi-TeV muon collider offers a spectacular opportunity in the direct exploration of the energy frontier. Offering a combination of unprecedented energy collisions in a comparatively clean leptonic environment, a high energy muon collider has the unique potential to provide both precision measurements and the highest energy reach in one machine that cannot be paralleled by any currently available technology. The topic generated a lot of excitement in Snowmass meetings and continues to attract a large number of supporters, including many from the early career community. In light of this very strong interest within the US particle physics community, Snowmass Energy, Theory and Accelerator Frontiers created a cross-frontier Muon Collider Forum in November of 2020. The Forum has been meeting on a monthly basis and organized several topical workshops dedicated to physics, accelerator technology, and detector R\&D. Findings of the Forum are summarized in this report.
}

\begin{document}
\snowmass
\maketitle

%\renewcommand*{\thefootnote}{\fnsymbol{footnote}}
%\footnote{Corresponding author, sergo@fnal.gov}
%\tableofcontents
%\begin{linenumbers}
\newpage
\section{Executive Summary}
\input{exec-summary}
\newpage
\section{Introduction}
\label{sec:introduction}
\input{introduction}

%%% PHYSICS PART

\section{Physics Case}
    \subsection{General Introduction}
    \label{sec:physIntro}
    \input{physics-intro} 

    \subsection{Physics cases at different energies}
    \label{sec:physdiffE}
    %Physics cases at different energies, Main ~10 TeV, sub-TeV with Maximum Lumi
    \input{physics-diffE}

    \subsection{Higgs Boson}
    \label{sec:physhiggs} %(Patrick, Matt, Zhen) 
    \input{physics-Higgs}

    \subsection{Dark Matter}\label{sec:physDM} % (Zhen, JiJi)
    \label{sec:physDM}

\input{physics-DM} 
    %Zhen can you coordinate what goes in here with what goes in the full sim/fast sim comparison. 
     %Sure. I will do that $_{ZL}$.
        
    \subsection{Naturalness}
    \label{sec:physnaturalness}%(Matt, LianTao, JiJi) JiJi's blurb below
    \input{physics-Natur}

    \subsection{Complementary Probes}\label{sec:physcomplementarity}
    \input{physics-compintro}
    \subsubsection{Flavor and CP violation}\label{sec:physflavor}
    \input{physics-Flav} 
   
    \subsubsection{g-2 Anomaly}\label{sec:physgminus2}%(Yoni,JiJi)
    \input{physics-gminus2}

    \subsubsection{Heavy Neutral Lepton}\label{sec:HNL}%(Zhen, Kunfeng)
    \input{physics-HNL}

    \subsection{Future Theory Development for Simulation}\label{sec:physfuture}%(Fabio)
    \input{physics-theorysimulations}

    %what should leave here versus the path forwards section?

%%% ACCELERATOR PART

\section{Accelerator}
    
\input{general} % Mark, Diktys, please edit general.tex

\input{feasibility} % intro section to "miracles are technical challenges" if any (All?)
    \input{proton_linac} % Dave, Vladimir, please edit proton_linac.tex
    \input{accumulation} % Dave, Vladimir, please edit accumulation.tex
    \input{targets} % Katsuya, Sasha, please edit targets.tex
    \input{ionization_cooling} % Katsuya, Diktys, Sasha, please edit cooling.tex
    \input{rf} % Derun, Yagmur, Scott, Katsuya, Sasha, please edit rf.tex
    \input{final_cooling} % Scott, Katsuya, Sasha, please edit final.tex
    \input{acceleration} % Scott, Dave, Sasha, please edit acceleration.tex
    \input{collider_ring} % Nikolai, Eliana, Dave, Sasha, please edit collider_ring.tex
    \input{neutrino_flux} % Nikolai, please edit neutrino_flux.tex

\input{imcc} % Mark, Diktys, please edit imcc.tex

\input{rd_priorities} % Derun, Diktys, All

\input{fermilab_site} % David, please edit fermilab_site.tex

%%% DETECTOR PART  

\section{Detectors} 
 
    \input{det-intro} %SergoJ
 
    \subsection{Environment} %Simone
    \label{sec:det:environment}
    \input{det-environment} 
 
    \subsection{Current Configuration} % Kevin B.
    \label{sec:det:current-configuration}
    \input{det-currentconf}

    \subsection{Feasibility Statement}% Isobel O./Tova H.
    \label{sec:det:det-new}
    \input{det-new}
    
    \subsection{Simulated Performance} 
    \label{sec:det:performance}
    \input{det-perf}

    \subsection{Fast- to Full- Simulation comparisons}
        \subsubsection{$H\rightarrow b\bar{b}$ Cross Section}
        \label{sec:Hbbcomparative}
        \input{fullsim-Hbb} 
        \subsubsection{Dark Matter with Disappearing Track} 
        \label{sec:DTcomparative}
        \input{fullsim-DT}

    \subsection{R\&D Priorities for Muon Collider Detectors} 
        \label{sec:detimprov}
        \input{det-improv}

%%% SYNERGIES PART    
    
\section{Synergies}\label{sec:synergy}
    \subsection{Neutrino Frontier}%(Pedro, Andre)
    \input{neutrinos}

    \subsection{Intensity Frontier}%(Sam)
    %{\color{blue}[ {\bf SH:} better name for this subsection? It's really the ``Intensity Frontier'', which was renamed.]}
    \input{beamdump}
    \subsection{Muon-Ion Collider}%Darin, Wei
    \input{muIC}

    \subsection{Applications outside HEP} 
    
    Various industry applications of high intensity muon beams have been mentioned in literature.     For example, muon spin rotation/relaxation/resonance ($\mu$SR) is a collection of methods that use the muon’s spin to examine structural and dynamical processes in bulk materials at the atomic scale. 
    With $\mu$SR beams of polarized muons are shot into a material. The muons’ spins precess around the local magnetic fields in the material. One can then examine how the internal magnetic fields of different materials have affected the muons’ spins by observing the directions in which the positrons produced in decays of the muons are emitted.
    
    Due to a strong penetrating power of muons, cosmic-ray muons have been widely used for imaging of large scale objects. The muon scattering angle in such an imaging process depends not only on the amount of material but also on the muon energy. A large disadvantage of the cosmic ray based muon imaging is that the flux is low and the muons have a wide spread in energy. On the other hand, artificially produced and accelerated muon beams can have high flux and be monochromatic, enabling higher resolution imaging in less time.
    
    These applications could clearly benefit from better quality muon beam. Because a muon beam is generated as a tertiary beam and initially has a large phase-space volume, cooling techniques are necessary to reduce the phase space volume, thus increasing the quality of the muon beam. Further advancements are also necessary to achieve muon acceleration for uses across numerous industry applications.

\section{Path Forward}
\input{pathforward}

%\section{Summary}
%\end{linenumbers}

%\printbibliography
\bibliographystyle{JHEP}
\bibliography{main.bib}

\end{document}

%% file: authors.tex
\newcounter{instituteref}

\author[1]{K.M.~Black}
\author[2]{S.~Jindariani}
\author[3]{D.~Li}
\author[4,5]{F.~Maltoni} 
\author[6]{P.~Meade}
\author[2]{D.~Stratakis} 
\author[7]{D.~Acosta}
\author[3]{R.~Agarwal}
\author[8]{K.~Agashe}
\author[26]{C.~Aim\`{e}}
% ...

\affiliation[1]{University of Wisconsin-Madison, Madison, WI, United States}
\affiliation[2]{Fermi National Accelerator Laboratory, Batavia, IL, United States}
\affiliation[3]{Lawrence Berkeley National Laboratory, Berkeley, CA, United States}
\affiliation[4]{Center for Cosmology, Particle Physics and Phenomenology, Universit\'e catholique de Louvain, B-1348 Louvain-la-Neuve, Belgium}
\affiliation[5]{Dipartimento di Fisica e Astronomia, Universit\`a di Bologna, Bologna, Italy}
\affiliation[6]{C. N. Yang Institute for Theoretical Physics, Stony Brook University, Stony Brook, NY, United States}
\affiliation[7]{Rice University, Houston, Texas,  United States}
\affiliation[8]{University of Maryland, College Park, MD, United States}
% ...
\affiliation[26]{Dipartimento di Fisica, Universit\`a di Pavia, and INFN, Sezione di Pavia, Pavia, Italy}
% ...

\emailAdd{sergo@fnal.gov}

%% file: exec-summary.tex
\begin{center} 
    ``One ring to rule them all, one ring to find them,..'' J.R. R. Tolkien
\end{center}

A multi-TeV muon collider offers a spectacular opportunity in the direct exploration 
%of the TeV scale. 
of the energy frontier. 
\textbf{Offering a combination of unprecedented %partonic
energy collisions in a 
%constrained and 
comparatively clean leptonic environment, a high energy muon collider 
%offers a unique expedition to
has the unique potential to provide both precision measurements and the highest energy reach in one machine that cannot be paralleled by any currently available technology.}

%This unique machine offers a direct window to both precision measurements and the highest energies in 
%one machine a single setup that cannot be paralleled by any currently available %technology.} 

The LHC has fundamentally altered the high energy particle physics landscape in the last ten years.  The discovery of the Higgs boson has laid out a clear path to study numerous fundamental questions about our universe. However, also crucial is the glaring absence of definitive signs of new physics 
%from the LHC 
up to $\sim$ the TeV scale.  We now have a 
%qualitative 
definite picture from data which did not exist a decade ago, where there is a Higgs like particle at 125 GeV and then  a (possibly large) gap to the scale of physics beyond the Standard Model (SM). While this reinforces the need to study the Higgs boson properties in greater detail, it also sharpens the need to significantly extend the reach of the energy frontier to its utmost. Motivated by theoretical studies, experimental results, and the desire to explore the energy scale  well beyond the LHC, a natural goal for the next Energy Frontier collider is the 10+ TeV scale. 
%for partonic collisions.  
Such a scale can be directly probed at lepton colliders running at the corresponding energy, while for hadron colliders one would need much higher energies, in the range of 100 TeV. However, the only indisputable target set by the LHC thus far is the Higgs boson, which resides at the weak scale.  Maximizing the physics potential at the energy frontier for precision Higgs studies can profit enormously from a cleaner environment than that offered in hadron collisions. \textbf{Therefore, given the LHC results, a 10+ TeV lepton collider going beyond the classic precision versus energy dichotomy is an ideal machine.}  Unfortunately with conventional technology electron based colliders cannot reach this scale.  Consequently, to probe this scale with leptons we must investigate completely new options, namely muon colliders or wakefield acceleration of electrons.

A high energy muon collider represents %as the next project after the HL-LHC 
 a paradigm change for the community. Its  attractiveness is at least twofold. First, it offers a unique range of physics opportunities opened by an accelerator design  where luminosity goes in par with the energy, providing the ideal combination of precision and direct reach needed for the exploration of the 10 TeV scale. Second, it is timely, not only from the physics needs, but also from the technological developments. The accelerator challenges can {\em now} be overcome using the technological advances achieved over the past decade. Significant progress has been made in the development of high power targets and of high-field HTS magnets, in the demonstration of operation of normal conducting RF cavities in magnetic fields, and of the self-consistent lattice designs of the various subsystems. 
\textbf{No fundamental show-stoppers have been identified. Nevertheless, engineering challenges exist in many aspects of the design and targeted R\&D is necessary in order to make further engineering and design progress.} Consideration of issues related to sustainability makes a high-energy muon collider particularly attractive due to its relatively small footprint and excellent power efficiency. 

There is an established plan for supporting muon collider related R\&D activities in Europe and it is 
imperative for Snowmass/P5 to reestablish R\&D efforts in the US and to enable participation of US physicists in the International Muon Collider Collaboration (IMCC). Successful execution of the IMCC program relies heavily on US participation, in particular in the areas of unique technology expertise (e.g. ionization cooling, targets, high-field magnets, etc) acquired during the Muon Accelerator Program in the U.S. Furthermore, renewed US contributions in muon collider efforts provides for an ideal opportunity to make designs for a potential US siting. The existing accelerator facilities at Fermilab (e.g. PIP-II) and the proposed booster upgrade or extension of the PIP-II linac could be developed into proton drivers for a Muon Collider. Furthermore, synergies with the neutrino and intensity frontiers programs in the US, as well as overlaps with nuclear science and industry applications provide additional benefits. This makes Fermilab a particularly attractive siting option. A set of Muon Collider design options, with potential siting at FNAL, would allow for deliberations with the IMCC and the international committees to facilitate a global consensus on the selection of siting. 

%{\color{blue} Comment here about a possible fit at FNAL?}
%While the physics case exists, and the accelerators can be constructed, one is presented with unique challenges for extracting the physics from muon colliders given that they are unstable particles. 

Despite the compelling physics case and the feasibility of the accelerators, a series of challenges arise from the unstable nature of muon beams that have a wide range of implications. 
Firstly, the decay of muons in the collider ring produces an intense flux of neutrinos that exit from the ground at a significant distance from the collider. Secondary particle production induced by these high-energy neutrinos has been of concern in the past. Novel ideas to reduce the density of this flux have been developed, including placing the beam line components on movers and tilting the ring periodically in small steps such that the muon beam does not point to a specific location for an extended period of time. It is expected that these will reduce the flux to a negligible level. It should be noted though, that high-energy neutrinos while posing an experimental challenge provide also an opportunity of 
%synergistic physics
synergies with other physics projects, as outlined in this report. Secondly, in the context of experiments/detectors, %the challenge comes from the overwhelming beam induced background originating from muon decays. 
muon decays induce a possibly overwhelming background,  i.e., a cloud of low momentum particles that end up in the detector volume that can impact physics performance and has to be curbed. 
Feasibility of suppressing this background to the levels necessary for precision physics was often questioned in the past. However, major advancements in collider detector technologies have been achieved in the last decade, driven in large by needs of the High Luminosity LHC experiments. These advancements allow one to incorporate precision timing and particle-flow into the tracking and calorimetry systems, which in turn translates into a dramatic reduction of the beam induced background (BIB). Full simulation studies demonstrate that necessary performance can be achieved with a reasonable extrapolation of the known technology. Further improvements are certainly conceivable and can be achieved by further optimization studies, including deployment of Machine Learning techniques. 

The most fruitful path forward towards the development of a conceptual design of a Muon Collider would be the engagement of the U.S. community in the IMCC. The U.S. Muon Collider community is well positioned to provide crucial contributions to physics studies, further advance the accelerator technology and detector instrumentation, and explore options for domestic siting of a muon collider. An Integrated National Collider R\&D Initiative being discussed in Snowmass can provide a much needed platform for R\&D funding for such accelerator and detector development.

%It is evident that the Muon Collider has the potential to illuminate various aspects of electroweak symmetry breaking, dark matter, and naturalness.

From the exploratory studies performed so far, it is already evident that the muon collider has the potential to illuminate various open questions in fundamental interactions, from the origin of electroweak symmetry breaking to the nature of dark matter, to naturalness. It is also capable of covering a broad swathe of well-motivated physics beyond the Standard Model and probing explanations for potential signals in experiments across the many frontiers of particle physics. The accelerator and detector technologies have advanced in the last decade, making the collider feasible on the timescale of approximately 20 years. \textbf{Engaging in the development of a muon collider will reinvigorate the U.S. HEP community and provide benefit to the field across multiple frontiers.  %"Much remains before us. But the muons are calling, and we must go."~\cite{AlAli:2021let}.
}

\begin{center}
    “We choose to go to the moon. We choose to go to the moon in this decade and do the other things (accomplishments and aspirations), not because they are easy, but because they are hard, because that goal will serve to organize and measure the best of our energies and skills, because that challenge is one that we are willing to accept, one we are unwilling to postpone, and one which we intend to win.” John F. Kennedy
\end{center}

%% file: introduction.tex
\subsection{Muon Collider Forum Report}
There has been a recent explosion of interest in muon colliders, as evident from a ten-fold increase in the number of related publications submitted to arXiv in the last few years.  The topic generated enormous interest during the US particle physics Snowmass planning process that evolved into a cross-frontier Muon Collider Forum. The purpose of the report is to summarize findings of the Forum which represents a state of the art summary of muon collider efforts spanning the physics motivation, detector requirements, and accelerator status. In writing of this report, emphasis was made on a significant shift in physics motivation and technology readiness since the previous muon collider efforts. While the previous US push focused mainly on the 125~GeV Higgs factory concept, more recently the primary interest of theoretical and experimental communities has shifted to a higher energy machine, with $\sim 10$~TeV collision energy as the target. This is due to the fact that some of the $e^+e^-$ Higgs factories are technologically more mature than $\mu^+\mu^-$, while the physics potential of a high-energy muon collider is truly unmatched. Staging at 125~GeV is still motivated and provides a complementary to $e^+e^-$ program with model-independent measurements of the Higgs boson mass and width, and a highly precise muon Yukawa coupling. Additional staging options are also being studied, including (a) operation at $\sim 1$ TeV with maximum luminosity to enable physics above \ttbar threshold and to probe certain beyond Standard Model (SM) scenarios; and (b) running at 3 TeV inspired by the Compact Linear Collider (CLIC) physics case~\cite{Aicheler:2012bya,Linssen:2012hp}.

In the report we also demonstrate that over the past decade we have seen the beginnings of several transformative new developments in accelerator and detector technologies. These developments address many of past concerns about muon collider feasibility. We present a technically limited R\&D timeline and argue that investments into both accelerator and detector development are necessary to make muon colliders a reality on the timescale of approximately two decades. We believe that U.S. HEP community possesses critical expertise and is uniquely positioned to make leading contributions to the global muon collider efforts. Activities within the Snowmass Muon Collider Forum already identified key areas of interest and expertise, assuming that P5 will support a revival of the Muon Collider R\&D program.  

\begin{figure}[th]
\begin{center}
\includegraphics[width=0.98\textwidth]{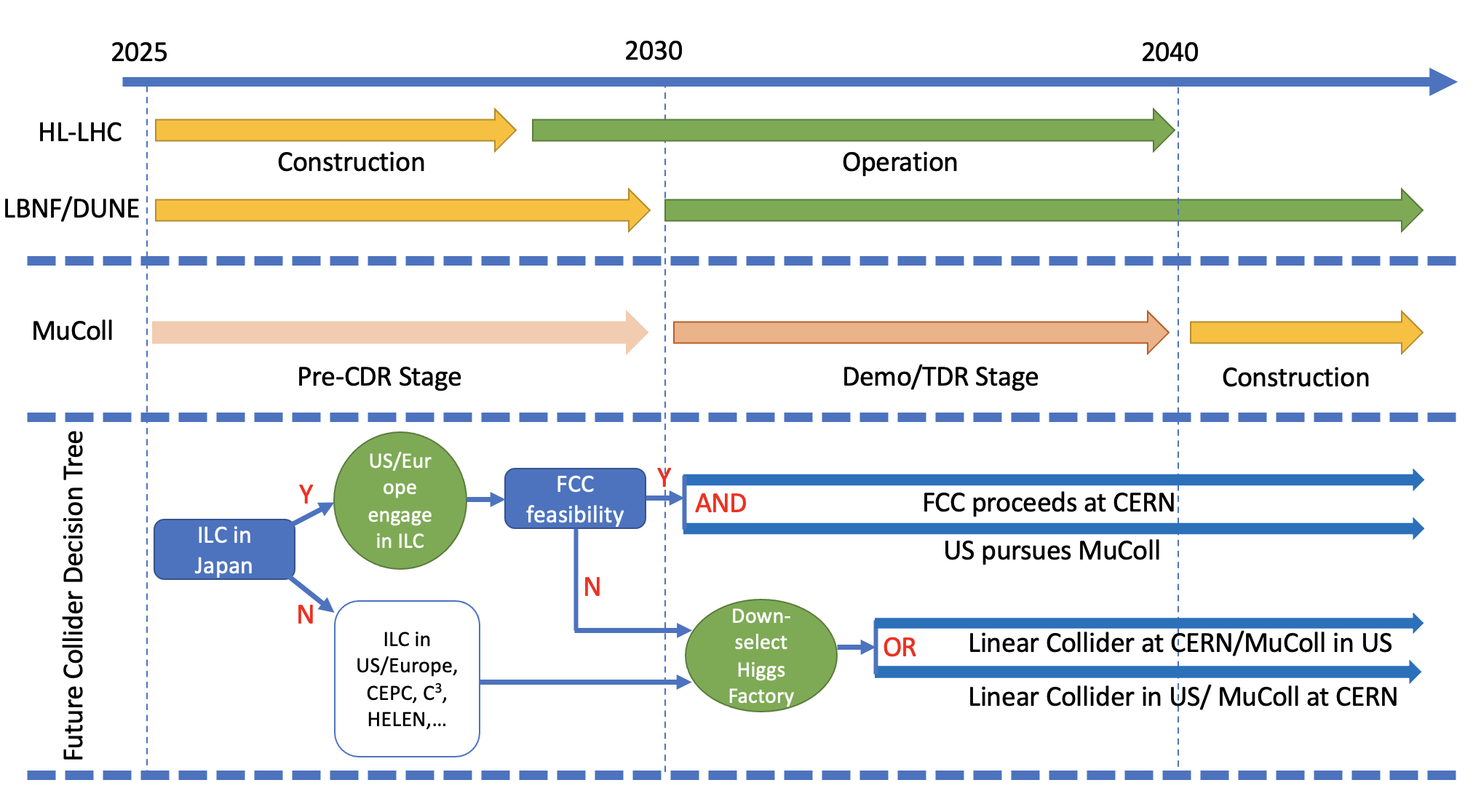}
\caption{A sketch of what the Muon Collider timeline could look like, superimposed with approximate HL-LHC and LBNF/DUNE schedules. Future collider decision tree adopted from Ref.~\cite{Pushpa:EFtalk} is also shown. The decision tree is "optimistic" in the sense that the timeline is driven by physics goals and technology readiness rather than financial considerations. We also assume that globally more than one future collider can be pursued at the same time.}%
\label{fig:timeline}
\end{center}
\end{figure}

The Muon Collider R\&D program would fit well within the timeline of U.S. HEP funding. In order to see this, it is useful to look at the landscape of currently ongoing large-scale facilities. A possible Muon Collider timeline is sketched in Fig.~\ref{fig:timeline}. For approximately the next decade, most of U.S. and Europe construction funds will be devoted to LBNF/DUNE facilities~\cite{DUNE:2016hlj} and HL-LHC upgrades~\cite{Apollinari:2017lan}. Successful completion of these mega-projects and execution of their physics program are essential for the future of particle physics. Without a significant increase in the HEP budget, it is difficult to envision a large scale investment into future colliders before completion of DUNE and HL-LHC upgrades. However, a modest amount of funding for the Muon Collider R\&D would enable further technological progress and a Reference Design Report around the time of the next Snowmass process. Following the completion of DUNE in the early 2030s, more funds can become available for a larger investment towards the construction of muon collider demonstration facilities with the goal of producing a Technical Design Report and declaring the project ready for construction in early 2040s. It should be noted that decisions about International Linear Collider (ILC)~\cite{Behnke:2013xla} in Japan and feasibility of Future Circular Collider (FCC)~\cite{FCC:2018byv} tunnel are expected within the next five years. While these decisions can alter the collider landscape, in all scenarios a high-energy muon collider will remain a highly attractive future collider option with unique physics capabilities.

\subsection{Why Collide Muons?}
Despite the incredible success of the SM at predicting various particle physics phenomena, many questions remain open. Discovery of the Higgs boson in 2012 helped to shed light on the origin of mass but does not explain why  electroweak symmetry breaking occcurs and what sets its scale. Other unanswered questions have to do with the origin of Dark Matter (DM), the origin of flavor, and the nature of the neutrino sector. We also do not know if there is a fundamental reason for the gauge symmetry and what kind of unification of the known forces may exist at the higher energy scales. Conventionally, answers to these questions are pursued by probing small distances with either precision (indirectly) or energy (directly). The Muon Collider has the potential to provide both, leveraging full energy of the accelerator with a relatively clean environment.

A facility colliding high-energy muon beams has a number of advantages when compared to its electron-positron and proton-proton counterparts ~\cite{Chao2014pea}. First, since the muon is a lepton, all of the beam energy is available in the collision. Second, since the muon is roughly 200 times heavier than the electron and thus emits around $10^9$ times less synchrotron radiation at the same energy, it is possible to produce multi-TeV collisions in a reasonably compact circular collider. Finally, a high-energy muon collider is the most efficient machine in terms of power per luminosity~\cite{Long:2020wfp}, a very important consideration in light of the global push for a more energy efficient and sustainable future.

In principle, muon colliders can reach very high energies in excess of 100 TeV. In order for this to happen, the size of the accelerator ring will have to be sufficiently large (e.g. 100 km ring would enable a 40-60 TeV collider). Further considerations such as cost, power consumption, and construction time may also impose practical limitations for what energy and luminosity are achievable, e.g. for energies much greater than 10 TeV synchrotron radiation must be taken into account similar to electron colliders at much lower energy. However, there are no fundamental physics reasons that would prevent it from going well beyond what is achievable by any other currently proposed technology. 

While the above arguments are highly appealing, there are several challenges with muons. First, muons are obtained from decay of pions made by higher energy protons impinging on a target. The proton source must have a high intensity, and very efficient capture of pions is required. Second, muons have very large emittance and must be cooled quickly before their decay. Given their short time, ionization cooling~\cite{Neuffer:1983jr} is the only viable option. Moreover, conventional synchrotron accelerators are too slow and recirculating accelerators and/or pulsed synchrotrons must be considered. Because they decay while stored in the collider, muons irradiate the ring and detector with decay electrons. Shielding is essential and backgrounds need to be strongly suppressed.

\subsection{Muon Collider History}
The concept of a muon collider is not new.  Muon storage rings were mentioned in the literature in 1965~\cite{Tinlot:1965ab} and concepts for a muon collider and for the required muon cooling were developed in the 1970s and 1980s. A muon collider collaboration was formed in the U.S. in the 1990s which delivered a design study in 1999~\cite{Ankenbrandt:1999cta}.  In 2000 the Neutrino Factory and Muon Collider Collaboration (NFMCC) was formed~\cite{Zisman:2000dn} which set out to perform a multi-year R\&D program aimed at validating the critical design concepts for the Neutrino Factory (NF) and the Muon Collider (MC). The Muon Accelerator Program (MAP)~\cite{Palmer:2013/07/02bta} was a follow-on (approved in 2011) program to the NFMCC and was tasked to assess the feasibility of the technologies required for the construction  of the NF and the MC. At the conclusion of MAP the program had produced a number of significant milestones summarized in Section~\ref{gen:accel}.

Although MAP was terminated in 2016, work continued on documenting the program's results and has provided a ``jumping-off" point for the recently formed International Muon Collider Collaboration (section~\ref{sec:IMCC}).

\subsection{International Muon Collider Collaboration}
\label{sec:IMCC}
The 2019 update of the European Strategy for Particle Physics (ESPPU) identified muon colliders as a highly promising path to reaching very high center-of-mass (COM) energies in leptonic collisions. In response to these findings, the European Laboratory Directors Group (LDG) formed a muon beam panel and charged it with delivering input to the European Accelerator R\&D Roadmap covering the development and evaluation of a muon collider option. In parallel, CERN initiated formation of a new International Muon Collider Collaboration (IMCC) to assess feasibility of building a high energy muon collider, identify critical challenges, and develop an R\&D program aimed to address them. The effort includes development of the machine-detector interface (MDI), detector concepts, and an evaluation of the physics potential. 

\begin{figure}[h]
\begin{center}
\includegraphics[width=0.98\textwidth]{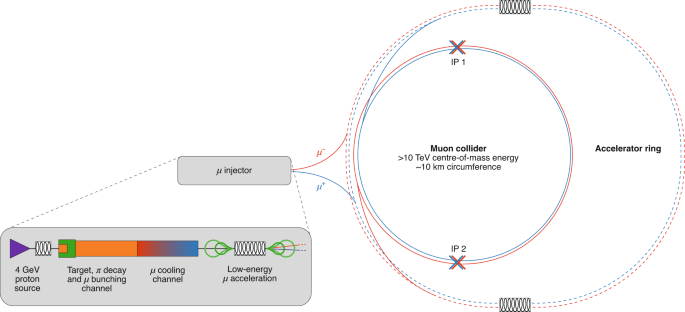}
\caption{Schematic layout of 10 TeV-class muon collider complex being studied within the International Muon Collider Collaboration.  From https://muoncollider.web.cern.ch/}%
\label{fig:IMCC}
\end{center}
\end{figure}

The collaboration is hosted by CERN. The near-term goal is to establish whether an investment into a full Conceptual Design Report and a demonstration program are scientifically justified for the next European Strategy for Particle Physics Update. Depending on the outcome of this study and the decisions made at the next ESPPU, the design can be further optimised and a demonstration program can be executed in the following years. The latter contains one or more test facilities as well as the development and testing of individual components and potentially dedicated beam tests. The resulting conceptual design will demonstrate the possibility to technically commit to the collider. In this case a technical design phase will follow to prepare the approval and ultimate implementation of the collider.

The design strategy taken by IMCC relies heavily on the concepts developed by the MAP collaboration. In the baseline design, muons are produced in decays on pions produced by colliding a multi-megawatt proton beam onto a target. The muons are then cooled to the emittances necessary to achieve target luminosities, rapidly accelerated to the desired energies in order to minimize the number of muon decays, and injected into a collider ring with two interaction points. IMCC envisions a staged approach with the first stage collider operating at the COM energy of 3 TeV and the second stage at 10+ TeV (Fig.~\ref{fig:IMCC}). Staging allows for demonstration of performance at the lower energy and also facilitates stretching out the construction time, while executing a vibrant physics program. The front end and most of the cooling chain in the accelerator complex are common to all stages. An alternative Low EMittance Muon Accelerator (LEMMA) approach~\cite{Alesini:2019tlf}, which uses positrons to produce muon pairs at threshold, was also considered but had difficulties with achieving a high muon beam current and hence the necessary luminosity.

Despite strong interest and expertise, U.S. participation in IMCC has been limited to the work done in the context of Snowmass. The European muon beam panel included two representatives (including the co-chair) from the U.S., and a large number of scientists helped to organize the IMCC working group activities. U.S. scientists made key contributions to most areas of the IMCC design development and planning, including magnets, RF cavities, muon production and cooling, muon acceleration, beam dynamics, machine-detector interface, and the high-energy complex. Besides the accelerator design, the Energy and Theory Frontier communities in the U.S. provided strong contributions in the areas of physics studies and detector design.

%% file: physics-intro.tex
 The most fundamental physics case for the energy frontier is not collider specific.  The investigation of the unknown is an ubiquitous motivation in any area of science.  Therefore exploring the shortest distances, the highest energies, and the earliest times in our universe will always be the prime motivation for the energy frontier.  
 Any collider that can push the furthest in these directions will always be compelling regardless of the particles collided.  However, what differentiates a muon collider at high energies is {\em how} it reaches the energy frontier and the unique range of physics that comes along with it. Colliders are often grouped into two distinct categories of precision (electron based) and discovery (proton based) machines. Muon colliders, however, do not really fit this naive classification and have to be considered as a fundamentally different option for our field. Muon colliders being circular {\em and}  compact  provide a unique combination of energy, precision, {\em and}  high luminosity. Thus they are a distinctly attractive option.
  
  At its core what enables the remarkable physics potential of a high-energy muon collider is that it accelerates fundamental rather than composite particles.  This has two key advantages which are normally competing in usual electron and proton based colliders, i.e., i)  equivalent high energy collisions  reached in a more compact setting and ii) a cleaner (non QCD dominated) environment to undertake physics studies in. 
  
  The direct reach of a muon collider at energy $E_\mathrm{COM}=\sqrt{s}$ for a heavy particle of mass $M$, can be easily estimated considering the $\mu^+\mu^-$ $s$-channel annihilation into a pair of heavy particles, whose kinematical threshold is at $M=\sqrt{s}/2$. Heavier states, such as $Z'$, $W'$, or heavy Higgses~\cite{Han:2022edd} could  be produced singly in association with a soft/collinear vector boson ($\gamma,Z,W^\pm$), extending the mass reach beyond $M=\sqrt{s}/2$ and possible almost to $\sqrt{s}$.
  
  A useful way to estimate the reach of a muon collider is to compare it to a $pp$ machine by means of an effective parton luminosity.  At a hadron collider high-$Q^2$ events are produced through the collision of elementary constituents of the protons, the partons. By employing standard collinear factorization, the inclusive cross section for producing a particle final state $F$ can be written as
  \begin{equation}
      \sigma(pp\rightarrow F+X)=\int_{\tau_0}^1 d\tau \sum_{ij}\frac{d\mathcal{L}_{ij}}{d\tau}\hat{\sigma}(ij\rightarrow F),
  \end{equation}
  where $\hat{\sigma}$ represents the partonic cross section of partons (quarks or gluons) $i,j$ to produce the final state $F$, and $\mathcal{L}_{ij}$ represents the parton luminosity constructed from the parton distribution functions(PDFs) as a function of $\tau=\hat{s}/s$.  Given that a muon collider is colliding fundamental particles, in a $2\rightarrow 2$ process it can in principle produce states of mass $M$ up to $\sqrt{s}/2$, while at a proton collider there is an extreme suppression from the falloff of PDFs with $\sqrt{s}$.  To estimate when a proton collider and muon collider would have the same inclusive cross section 
  \begin{equation}
     \sigma_\mathrm{p} = \sigma_\mathrm{\mu}
 \end{equation}
  to produce a pair of particles of mass $M$ in a $2\rightarrow 2$ process, one needs the partonic cross section.  Production processes  could be different for a proton collider compared to a muon collider, yet we can characterize it by the ratio
 \begin{equation}
     \beta\equiv \frac{\hat{\sigma}_\mathrm{p}}{\hat\sigma_\mathrm{\mu}},
 \end{equation}
where the annihilation cross section $\sigma_{\mu}$ is calculated close to its maximum, {\it i.e.},  slightly above threshold $\sqrt{s}=M/2$.  For annihilation processes involving EW processes $\beta\sim 1$, whereas for producing colored states one could imagine a scaling due to the coupling of order $(\alpha_\mathrm{S}/\alpha)^2$.   

\begin{figure}
\centering
\includegraphics[width=0.9\linewidth]{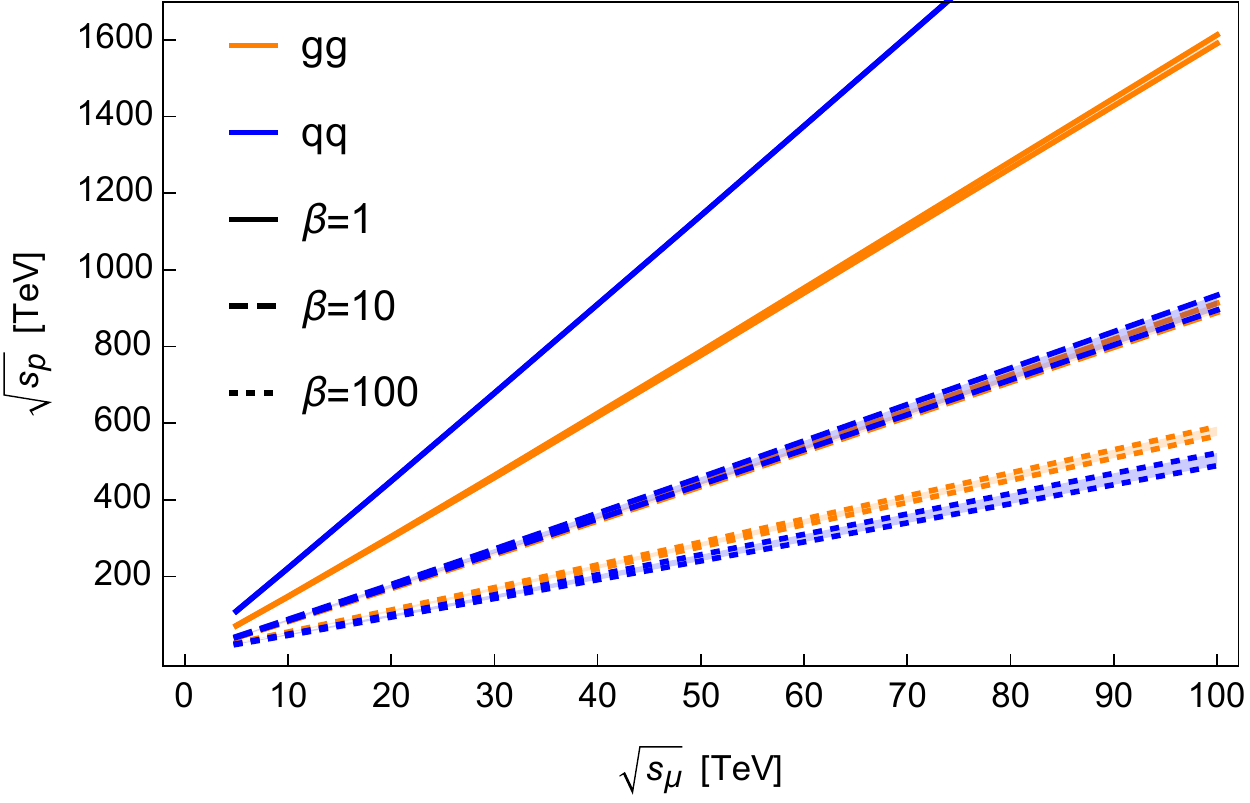}
\caption{ 
The COM energy for a proton collider $\sqrt{s}_\mathrm{p}$ and a muon collider $\sqrt{s}_\mathrm{\mu}$ such that the $2\rightarrow 2$ cross sections are the same based on different assumptions for the partonic cross sections characterized by $\beta$.  Separate curves for gluon and quark annihilation channels are shown, with the bands given by two choices of PDFs, i.e. NNPDF3.0LO and CT18NNLO~\cite{AlAli:2021let}.
}
\label{fig:energycompare}
\end{figure}

In Figure~\ref{fig:energycompare}, we show the resulting COM energies which yield equivalent cross sections. For example, a muon collider with $\sqrt{s_\mathrm{\mu}}\sim 10$ TeV is equivalent to a pp collider of $\sqrt{s_\mathrm{p}}\sim 70$ TeV. This is only illustrative since it depends on the details of the partonic cross section. For example, in some cases the equivalent muon collider scale could be  under 5 TeV~\cite{AlAli:2021let}. However, as shown in numerous studies~\cite{AlAli:2021let}, the details depend on the specific physics cases, some of which we will be reviewed in this section. 

Based on what we have learned from the LHC thus far since the last Snowmass/P5 process, {\em the ability to reach higher energies in a compact collider is crucial.}   
As of today, ATLAS and CMS have discovered a Higgs-like state at 125 GeV and nothing else. This determines two priorities for future colliders. First, it is crucial to make as large of step possible in energy, e.g., to the $\gtrsim 10$ TeV scale. From what we know now, there may be a significant energy gap between the TeV scale that the LHC/HL-LHC is probing to the scale of new physics. The largest step possible in energy is motivated not just for the sake of exploration, but also by specific targets related to the origin of electroweak symmetry breaking (EWSB), Naturalness, DM, complementarity with other frontiers, and even existing experimental anomalies as we will discuss.  
Second, it is mandatory to achieve the most precise determination of the Higgs boson properties and its interactions.  The Higgs boson is the only unambigious discovery we have made at the LHC so far.  While this is often thought as the domain of ``Higgs Factories", a high energy muon collider also offers unique abilities that can take us beyond the first generation of Higgs factories. This is because not only a muon collider provides a compact energy frontier machine, but also because at high energies emission of $W$ and $Z$ bosons from the initial state muon is enhanced and vector boson fusion becomes dominant. This is the reason why a muon collider is often also dubbed a ``vector boson collider" as well~\cite{Buttazzo:2018qqp,Costantini:2020stv,Han:2020uid,AlAli:2021let,Han:2021kes}. The vector boson fusion channel provides the dominant production mechanism not only for single but also for multi-Higgs final states. Cross sections are significant, increase with the COM energy and do not suffer from the copious QCD backgrounds that come along with an energy frontier proton collider, e.g., the FCC-hh.  The dual nature of a muon collider, i.e.,  "high energy with high precision", allows for an unparalleled number of Higgs bosons that can be produced in the cleanest environment.  This very fact has further ramifications. The Higgs precision measurement program is normally thought as staged in two main phases: the first phase at a  Higgs factory, where deviations of Higgs properties predicted by the SM are detected, followed by a second phase at a different high energy machine to {\em directly} search for the cause of the deviations, e.g., at CERN the FCC-ee followed by the FCC-hh.  This is because, when viewed through the lens of an Effective Field Theory (EFT), the expected deviations in the Higgs couplings  enter as $\mathcal{O}(v^2/M^2)$ where $M$ is the new physics scale, which typically resides well above the energy probed at Higgs factories. On the other hand, the muon collider opens up  a completely novel possibility for Higgs physics.  A high energy muon collider will provide precision and energy at the same time, achieving the highest precision possible for many Higgs observables {\em and} directly test the physics which causes it. Thus a high energy muon collider compared to a standard multi-staged approach based on different colliders in the same facility, such as LEP and then the LHC, can be thought as a "two-colliders-in-one" or "one-stage" option. Putting this together with the opportunity of increasing the COM energy as the accelerator technology will advance, clearly makes the muon collider option a very attractive one. 
%%% Potnetially put an effective PDF plot of muons vs protons to emphasize the difference? 

The novel nature of a high energy muon collider is clear from these simple arguments.  It allows the highest energy reach with the most compact design in space (actual size of the facility) and in time (high-luminosity can be obtained in a handful of years of running), providing a new type of collider that bridges the usual precision versus energy dichotomy.  
This unique features call for a systematic assessment of the beyond SM (BSM) opportunities that a specific muon collider at 10 TeV could open up, to establish the reach in specific scenarios with physics studies that go beyond the simple parton luminosity scaling argument presented above.  
The 2021 Snowmass process has catalysed an explosion of interest in the high energy theoretical community, leading to a wealth of phenomenological studies, whose current number and breadth make it already impossible to cover all results in detail.  
Several references exist that try to summarize a great deal of the physics case~\cite{AlAli:2021let,Aime:2022flm,Costantini:2020stv}. 
For this forum report we emphasize a few specific areas that will make easy for the reader to grasp the impressive physics reach of a 10 TeV muon collider.  
In section~\ref{sec:physhiggs}, we review the current status of Higgs precision physics with muon colliders.  
In particular, we outline some of the novel probes only available at high energy, and give examples of the aforementioned complementarity between searching for deviations through precision and searching for their causes directly with the same collider.  
In section~\ref{sec:physDM}, we explore one of the other pressing mysteries of particle physics, i.e., the nature and origin of dark matter. By briefly reviewing the current status of the field, we show how muon colliders are able to probe minimal dark matter models featuring Weakly Interacting Massive Particles (WIMPs) that are out of reach of other present and future experiments except for the FCC-hh. In section~\ref{sec:physnaturalness}, we then discuss one of the questions that inexorably comes along with the LHC discovery of the Higgs, i.e.,  naturalness.  
From what we have inferred so far from the data collected at the LHC, and in particular, the values of the mass of the Higgs at 125 GeV and the lack of deviations from the SM predictions or of signs of new states, naturalness has to be understood in ways that were not prominent during the last Snowmass process.  Studies performed so far have indicated that the muon collider has a significant reach in constraining scenarios  motivated by the naturalness guiding principle, to an extent similar to the FCC-hh in some cases and clearly surpassing it in others. 
In Section~\ref{sec:physcomplementarity}, we demonstrate how a muon collider is naturally suited as a complementary probe for many other types of particle physics experiments that are already planned or will be undertaken in the coming future.  As example of this complementarity we highlight how a muon collider is uniquely suited to probe existing anomalies in particle physics such as the Fermilab $g-2$ experiment. Our point here is {\em not} motivating a muon collider by leveraging on the current (yet possibly to be solved) anomalies (such as $g-2$, $B$-anomalies, $W$ mass), but to provide evidence that a muon collider {\em will play a role} in addressing a wide range of possible hints of new physics emerging at scales probed at current and planned experiments.  In Section~\ref{sec:physfuture}, we outline some of the additional lines of research that should be undertaken to further improve the projections for muon colliders.  Finally,  although it is manifest that the full physics potential of a muon collider is unlocked by realizing it at the highest energies possible $\gtrsim 10$ TeV, in Section~\ref{sec:physdiffE} we also consider the physics reach and targets that could be attained by staging it at lower energies.  In particular, we discuss some complementary observables that could be measured at lower energies, which could be used to leverage the physics results of the high energy muon collider operation.  We conclude by stressing in Section~\ref{sec:synergy} further synergistic aspects, beyond the specific physics case, such as the impact that a muon collider could have on the neutrino physics program more so than other colliders and vice versa due to the shared infrastructure.

%% file: physics-diffE.tex
\begin{figure}[h]
    \centering
    \includegraphics[width = 0.9\textwidth]{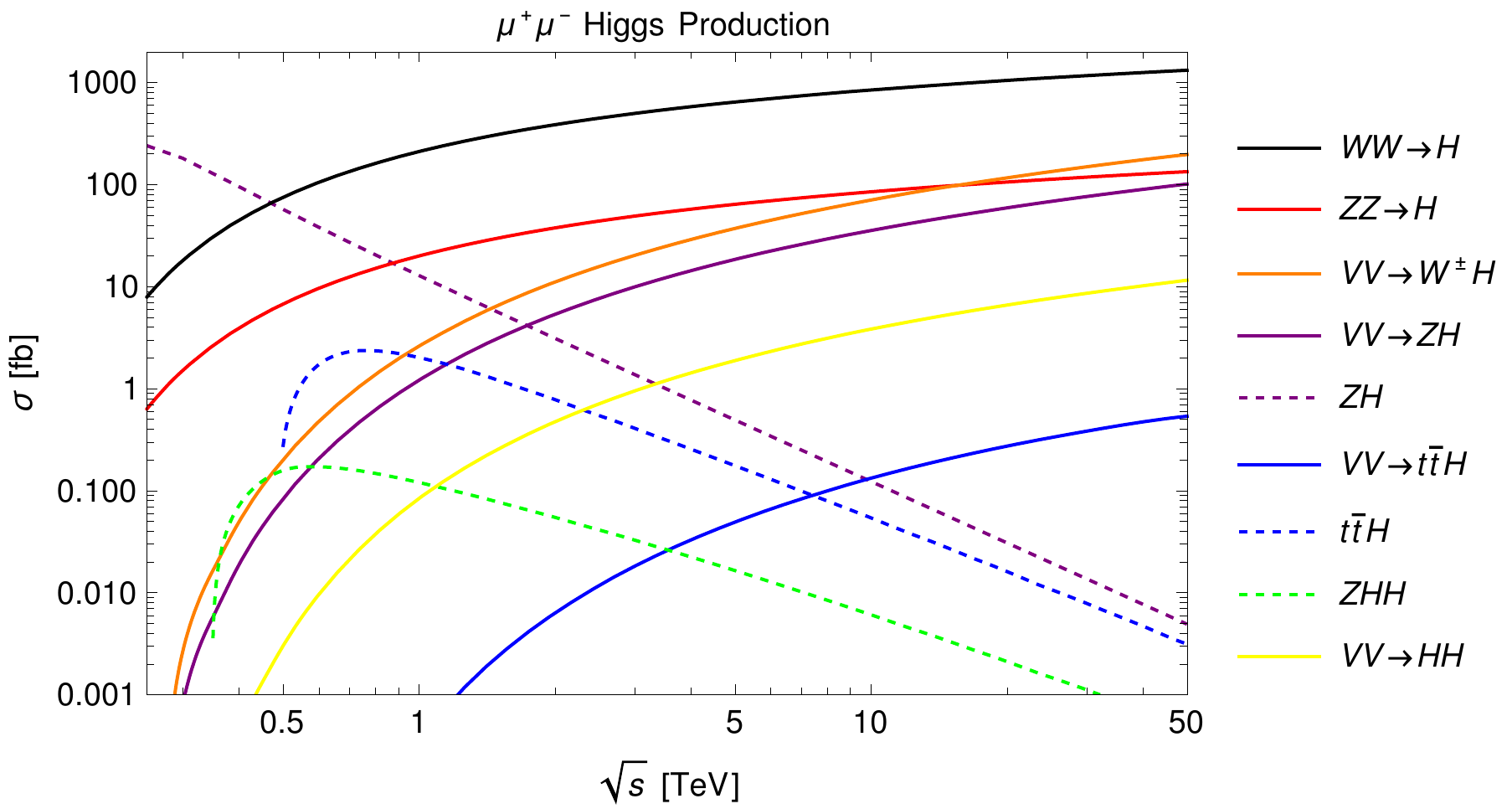}
    \caption{Figure reproduced from~\cite{Forslund:2022xjq} showing various Higgs process as a function of COM energy.  The dashed curves correspond to s-channel annihilation processes, while the solid curves all are from Vector Boson Fusion.}
    \label{fig:singleH}
    \end{figure}

The ultimate physics reach of a muon collider depends on the highest energy that it can achieve. In this report we have focused primarily on the 10 TeV muon collider, because it is realizable with current technology and does not require any ``miracles" to occur. However, there is not a hard upper limit at 10 TeV, in fact some physics studies have been performed up to 100 TeV~\cite{AlAli:2021let}. %{\bf Insert reference to paragraph about upper limit from AF(to be written)}}  
However, in the context of a 10 TeV muon collider, it can be useful to comment about the staging possibilities in reaching this energy.  From the accelerator point of view, staging is a logical and prudent approach. For example, the IMCC investigated the option of a 3 TeV stage before achieving the 10 TeV energy ~\cite{MuonCollider:2022xlm}.  It is therefore important to  consider whether or not a complementary physics program exists with the earlier stage, and whether a multi-staged muon collider could actually be instrumental to the physics program. With the largest possible BSM direct reach in mind, a staged approach is obviously not particularly interesting. However, for accurate SM measurements, or for any process that benefits from lower energies, such as production of new light and weakly interacting states in annihilation, a staged plan could be beneficial. For example, as shown in Fig.~\ref{fig:singleH}, for single Higgs cross sections as a function of $\sqrt{s}$ there are qualitatively different aspects of physics that can be tested depending on the choice of center-of-mass energy.  

In particular, we see that for Higgs physics, 3 TeV, is not a particularly attractive staging option in the absolute sense, as the $s$-channel processes have all fallen significantly from their maximum, and the VBF processes would only have larger cross sections at 10 TeV (including di-Higgs).  Therefore in principle any Higgs measurements done at 3 TeV would only be superseded by 10 TeV measurements without offering new production channels that would be complementary to 10 TeV.  This can be contrasted to the complementarity present potentially for $e^+e^-$ Higgs factories at low energy, where the processes are dominated by associated production rather than VBF.  Although as a first stage 3 TeV would offer the highest precision compared to the HL-LHC and $e^+e^-$ Higgs factories in for example the di-Higgs channel until superseded by higher energy.   Given this general chain of logic, it is therefore useful to consider various staging options based on the physics outcomes that could be achieved. 

The canonical example of a low energy stage of a muon collider is a machine running at 125 GeV COM designed to produce the Higgs through an $s$-channel resonance.  The idea of producing a Higgs through the $s$-channel muon-antimuon annihilation goes back many decades, see for example~\cite{Barger:1996jm}.  However, since the discovery of the Higgs in 2012, more detailed investigations have become available, pointing to the  unique opportunity of directly measuring the Higgs width  by a lineshape mapping process. Such a collider would also provide complementary Higgs coupling measurements, including a measurement of the muon Yukawa coupling at the subpercent level. The scenario where one or more $e^+e^-$ Higgs factories are constructed will not render the 125\,GeV muon collider an uninteresting staging option. On the contrary, as shown in Ref.~\cite{deBlas:2022aow}, there is a strong synergy between a 240\,GeV $e^+e^-$ and a 125\,GeV muon collider due to their different production channels. A combination of the two provides significantly better results on the coupling determination than individual ones.  This will be discussed further in Section~\ref{sec:physhiggs}, where also the complementarity with  measurements at 10 TeV is explicitly discussed.  A possible disadvantage to this particular staging option is that it requires beam conditions that are more challenging than those needed at a higher energy muon collider.

There are of course other staging possibilities than 125 GeV or 3 TeV, although these have been worked out in the most detail thus far.  A balance has to be struck between what makes a high energy muon collider so attractive in terms of luminosity and power consumption and the fact that lower energy stages would necessarily come with lower luminosity. At this moment,  further studies are needed to optimize a staging plan to achieve the largest physics reach of a combined program within a muon collider and in conjunction with other colliders.  Nevertheless, there are some obvious interesting options, some of which have received at least preliminary attention.  For example, it was recently proposed that a muon collider running at the $2m_t$ with foreseeable luminosities could provide a sufficiently precise top mass measurement to answer the question of whether our universe is stable or metastable~\cite{Franceschini:2022veh}. Additional possibilities for measuring the W mass and top mass more precisely are given in Ref.~\cite{Barger:1997yk}. At 10 TeV, a high energy muon collider also does not provide a strong measurement of the top Yukawa coupling compared the the LHC in the standard $t\bar{t}h$ analysis as can be understood from Figure~\ref{fig:singleH}.  Although the ultimate sensitivity of a high energy muon collider still needs investigated, as new methods for measuring couplings can open up with energy, for example the $W^+W^-\rightarrow t\bar{t}$ process is sensitive to deviations in the top Yukawa~\cite{Maltoni:2019aot,Costantini:2020stv,AlAli:2021let}. Nevertheless, similar to the ILC, a muon collider stage in the 500 GeV to 1 TeV range, could provide complementary information to a high energy muon collider.  
Finally, if there is new physics at low mass with muon specific couplings there also could be benefits to a sub TeV staging option.  An example of this is models built to account for the current muon g-2 anomaly and discussed further in Section~\ref{sec:physcomplementarity}.

Very preliminary investigations show that leveraging on runs at different energies is very promising and certainly deserves more studies. Staging options could exist that optimize several complementary open physics questions to the high energy muon collider.  Ultimately, a detailed plan might depend on the physics landscape at the end of the HL-LHC and also on other accelerator projects foreseen. Nevertheless, the very possibility for a muon collider to operate at different energies from the sub-TeV to the multi-TeV range could significantly enhance its physics reach. Such studies will therefore be carried out in parallel leading up to the next Snowmass and ESG studies.

%% file: physics-Higgs.tex
%{\it I wrote the Higgs precision part, in connection to EF04, made a new plot. I think this topic somewhat covered here, although only minimally. Feel free to edit/expand/correct $_{Zhen}$}
%    \begin{itemize}
%    \item kappas, EFT (to tie into EF04)
%	\item multi-Higgs from energy
%    \item Alternative probes available with energy
%    \item 10 TeV results
%	\item 125 GeV results and combo
%	\end{itemize}

%	Point of precision is looking for BSM that causes deviations
%	Muon collider unique in that precision deviations and causes can be examined with same machine - preferred weakly interacting examples? singlet, 2hdm, higgsino, stops?
	
%		Should we break out the following and just reuse the results to talk about the implications for these, or separate subsection?
%	\begin{itemize}
%	    \item Higgs potential
%	    \item Higgs portals
%	    \item Higgs stability
%	\end{itemize}
	
\begin{figure}[h]
\centering
\includegraphics[width=0.7\linewidth]{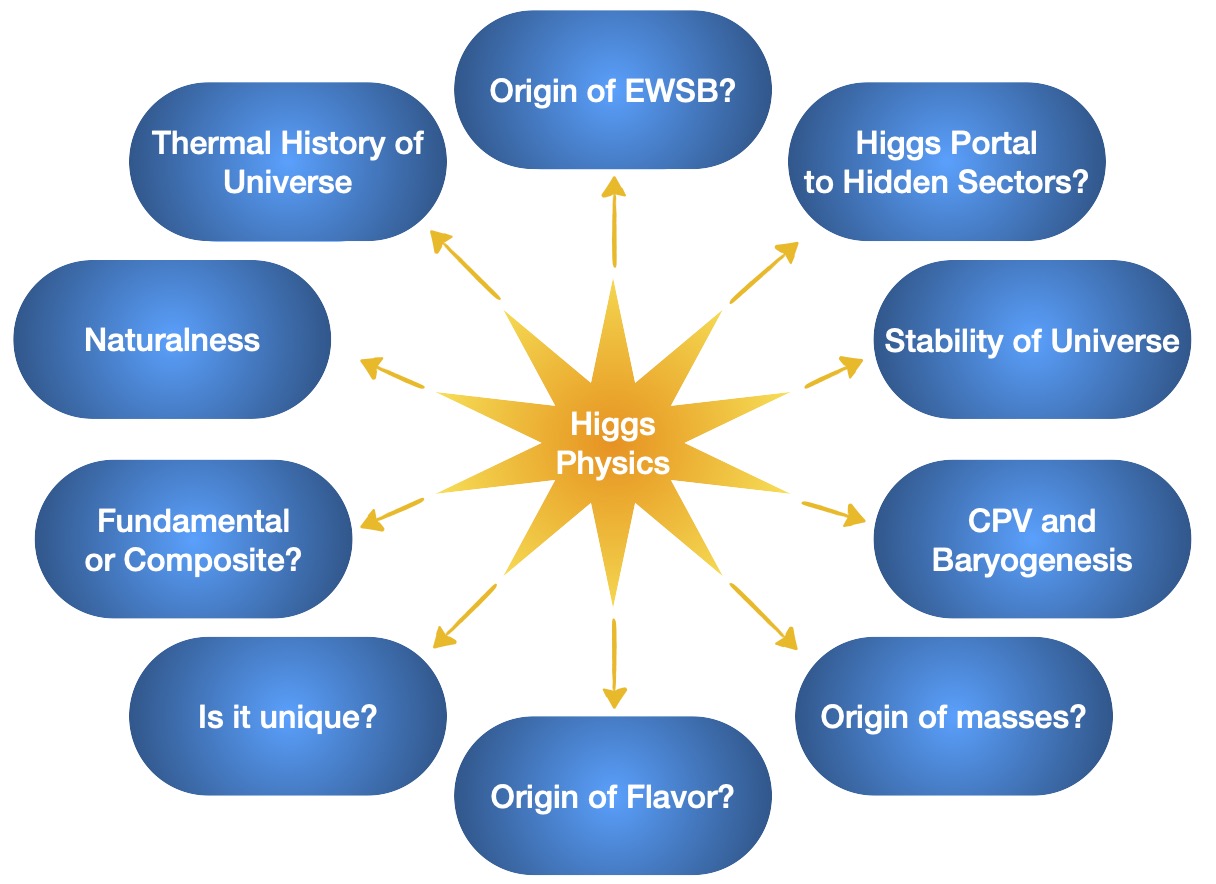}
\caption{Figure from Energy Frontier Higgs topical report illustrating the centrality of the Higgs and the connections to numerous fundamental questions.
}
\label{fig:higgscentral}
\end{figure}

The Higgs boson, the only new fundamental particle discovered so far at the LHC, is the central figure of the Standard Model. As shown in  Figure~\ref{fig:higgscentral}, taken from the corresponding Energy Frontier topical report for the Higgs,  it connects to many of the deepest questions about our universe. Accurately measuring its properties to assess its nature and its role in phenomena that currently escape our understanding, is one of the top priorities of the high-energy community. 

Measurements of the Higgs properties are also a powerful probe of new physics, and they play a central role in the physics programs of all foreseen future colliders.  Muon colliders, in particular, provide a unique setting to probe Higgs properties for two main reasons.  First, multi-TeV energies allow for the production of a larger sample of Higgs bosons than what is attainable at Higgs factories, with a similarly clean environment to study them. In addition, they allow multi-Higgs production and therefore an unmatched probe of the Higgs potential. Second, high energy muon colliders offer the unique ability to simultaneously access Higgs properties with very high precision/accuracy, and in case of deviations, directly probe their origin, as we discuss below.

To demonstrate the first point, we consider the precision on the Higgs couplings that can be achieved at muon colliders. Drawing on the Higgs exclusive channel inputs of Refs.~\cite{Forslund:2022xjq, deBlas:2022aow}, one can perform a global fit analysis. There are two main approaches that are followed for doing the global fits. The first is by assuming the same type of  couplings as in the SM, but associating to each of them a rescaling factor $\kappa_i$. This approach has been dubbed ``kappa framework" and enjoys the simplicity of a direct translation between different channels and the Higgs property precision. A second approach employs the Standard Model Effective Field Theory (SMEFT), which provides a consistent deformation of the SM which allows to perform accurate predictions and combine information across different scales and experiments as long as new physics exists only at a parameterically larger scale than probed. For consistency with the electroweak precision fit group at Snowmass, we use a modified SMEFT framework, where the Higgs width can be considered as an additional free parameter, yet not only Higgs measurements, but also electroweak precision observables and possibly other low-energy measurements are included to achieve a consistent projection of the overall precision. \footnote{We thank EF04 electroweak fitting group for various communications in developing the results.}  

\begin{figure}
\centering
\includegraphics[width=0.9\linewidth]{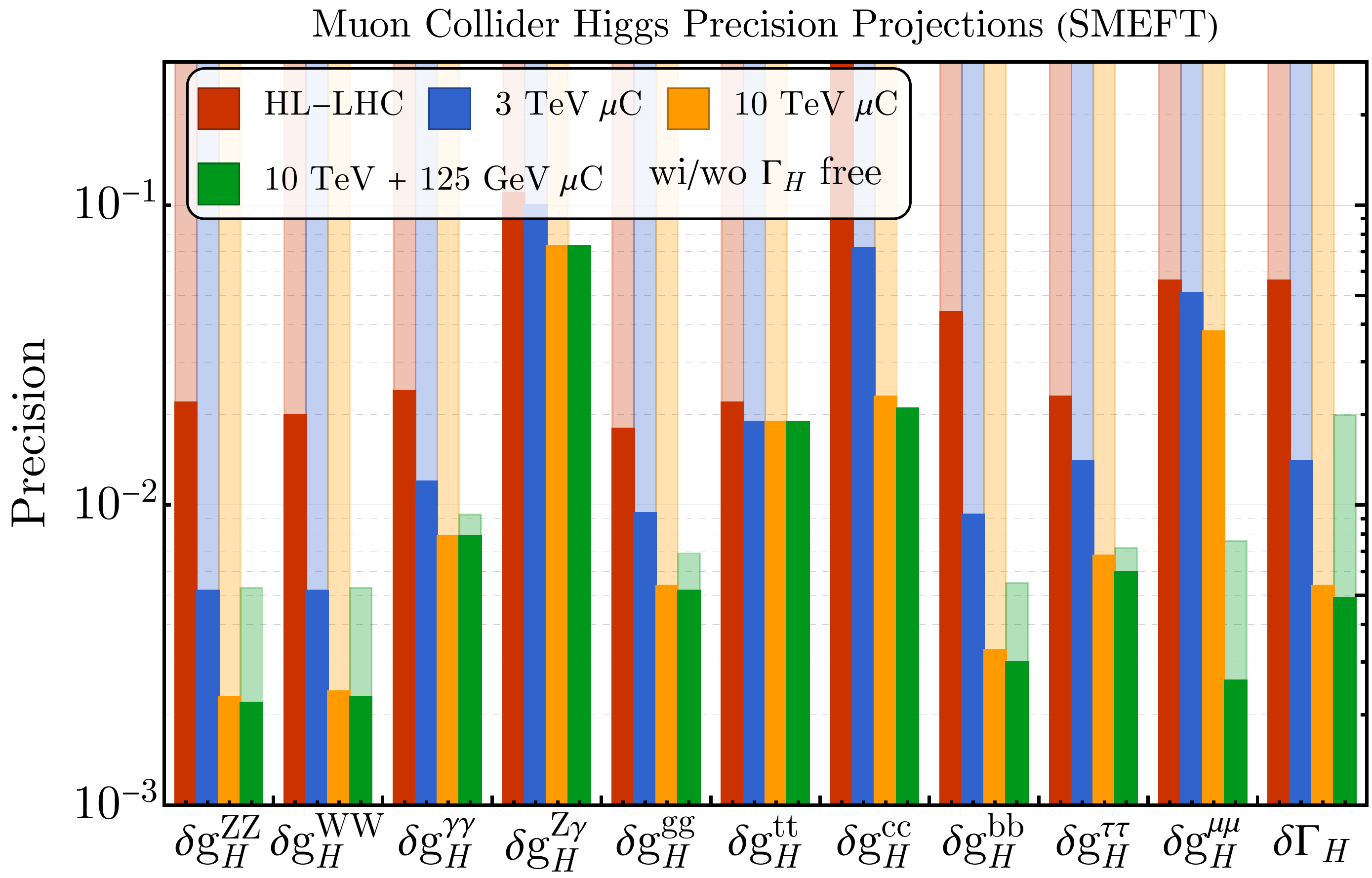}
\caption{
The one-sigma precision reach on the effective Higgs couplings 
from a global fit of the Higgs and electroweak measurements in the SMEFT framework. The first set of (red) columns represents the HL-LHC S2 scenario with electroweak measurements at LEP and SLD. The second (blue) and third (yellow) sets of columns represent the 3 TeV muon collider and 10 TeV muon collider projections, respectively. The fourth (green) sets of columns represent the 10 TeV muon collider combined with a 125 GeV muon collider Higgs factory.
The measurements are combined with the HL-LHC S2 and LEP/SLD measurements for all the muon collider scenarios. The semi-opaque bars represent the results with the Higgs width being a free parameter, e.g., allowing for exotic decays that are hard to constrain through direct searches. The solid bars are for the results without exotic Higgs decays. %{\color{blue} question: are these decays that are indistinguishable from SM hadronic ones or/and invisible?} {\color{blue} Answer: Yes. The sentence before the last is clarifying on this.}
}
\label{fig:HiggsEFT}
\end{figure}

We show the SMEFT projection results in \autoref{fig:HiggsEFT}. Here we only report the Higgs couplings part in the Higgs basis, marginalizing on  other parameters. The corresponding precision for the electroweak sector and trilinear gauge couplings can be found in the Snowmass report~\cite{EF04Report}. In this plot, all muon collider projections are combined with the HL-LHC. The muon collider scenarios considered include a 3 TeV muon collider with 1~$\iab$ of luminosity, a 10 TeV muon collider with 10~$\iab$ and also its combination with a 125~GeV resonant muon collider Higgs factory with 0.02~$\iab$ integrated luminosity. The semi-opaque and opaque bars represent the results with and without the Higgs width $\Gamma_\mathrm{H}$ left as a free parameter. As one can anticipate, considering $\Gamma_\mathrm{H}$ as a calculable parameter in the SMEFT allows to attain a better precision. On the other hand, considering it a free parameter, introduces a "flat" direction in the fit, that needs very specific measurements (such as the direct $\Gamma_\mathrm{H}$ measurement at the resonance peak $\sqrt{s_\mathrm{\mu}}=m_\mathrm{H}$ to be resolved).  At high energies this can also be investigated by using indirect methods such as the ``offshell" methods employed at LHC, and should have roughly the same precision as the direct lineshape measurement but with added theory assumptions.   We would like to emphasize that these different frameworks and/or basis choices can be also associate to different UV hypotheses and are therefore useful also develop an idea of different new physics effects. It is important to keep in mind that there is no best approach that can be single out and on which one solely rely on in establishing the Higgs physics potential of various machines. 
In a rather general way, one can see that the precision on the Higgs couplings increases by an order of magnitude or more at a muon collider compared to the HL-LHC. In simple words, a muon collider could probe with an unprecedented resolution the inner structure of the Higgs boson. Furthermore, it is also evident that a 10 TeV muon collider can generally achieve much better precision compared to the 3 TeV muon collider. This gain in Higgs precision comes from the inherently higher luminosity at higher energies and the logarithmically enhanced weak boson fusion rates for  Higgs boson production.   

The precision Higgs program at the muon collider naturally goes beyond the study  Higgs couplings to lighter SM states. One example is the  Higgs top Yukawa which can be probed in very non-trivial way through the VBF $t\bar t$ process~\cite{Costantini:2020stv,AlAli:2021let}. In fact, a wide variety of differential Higgs measurements could be explored at a high energy muon collider. For example, as shown in Ref.~\cite{MuonCollider:2022xlm}, bounds on composite Higgs models go well beyond a 100 TeV proton collider by exploiting the differential measurements available.  Furthermore a high energy muon collider allows for unprecedented measurements of multi-Higgs production and thereby the Higgs trilinear and even quartic couplings.  These could be determined at better than  10\%~\cite{Han:2020pif,Buttazzo:2020uzc} and $\mathcal{O}(1)$~\cite{Chiesa:2020awd} precision, respectively. One can further connect various Higgs-related processes to new physics hints from low energy precision measurements~\cite{AlAli:2021let,Capdevilla:2021kcf}, as well as directly probing hidden sector physics through Higgs exotic decays~\cite{Liu:2016zki}. 

The second unique aspect of a muon collider is its ability to probe the {\em causes} of possible Higgs property deviations. The discovery of a deviation in the measured Higgs couplings would, at a Higgs factory, generally point to new physics outside the direct discovery reach of that collider. By contrast, a muon collider offers a unique opportunity: a single collider could {\em both} carry out precision measurements illustrating indirect effects of new physics on Higgs properties and directly discover the particles responsible. This is a powerful argument in favor of a high-energy muon collider. Other precision colliders generally aim to make a case for the next energy-frontier collider. The muon collider would already {\em be} such an energy-frontier discovery machine. Here we will give some examples that illustrate this capacity.
	
As a first example, let us consider modifications to the Higgs gluon coupling due the existence of heavy colored partners of the quarks in supersymmetry. The fractional deviation in the Higgs coupling to gluons due to a loop of stops is estimated by~\cite{Ellis:1975ap,Shifman:1979eb}
	\begin{equation}
	    \delta \kappa_g = \frac{1}{4} \left(\frac{m_t^2}{m_{{\tilde t}_1}^2} + \frac{m_t^2}{m_{{\tilde t}_2}^2} - \frac{m_t^2 X_t^2}{m_{{\tilde t}_1}^2 m_{{\tilde t}_2}^2}\right),
	\end{equation}
	where $X_{t}$ is a trilinear mixing term. Using the precision for a 10 TeV muon collider quoted in~\cite{Forslund:2022xjq} as a benchmark, and taking the two stops to be degenerate with $X_t = 0$, a muon collider Higgs precision measurement of $|\kappa_g| \leq \kappa^\mathrm{max}_g$ would translate into a constraint of
	\begin{equation}
	    m_{\tilde t} \gtrsim 1.5\,\mathrm{TeV}\,\sqrt{\frac{0.67\%}{\delta \kappa^\mathrm{max}_g}}.
	\end{equation}
	On the other hand, the direct discovery reach for each stop squark at a high-energy muon collider extends up to very nearly $m_{\tilde t} = \sqrt{s}/2$, or about 5 TeV for a 10 TeV center-of-mass muon collider~\cite{AlAli:2021let}. Thus, the collider will discover the same physics responsible for the measured Higgs coupling deviation. On the other hand, if the stops are sufficiently light, the measurement of $\delta \kappa_g$ could play a role in elucidating the detailed structure of stop mixing by helping to pin down $X_t$.
	
	In composite models, the Higgs couplings to $W$ and $Z$ bosons receive corrections of order $v^2/f^2$, a result that follows from universal model-independent considerations when the Higgs is a pseudo-Nambu-Goldstone boson~\cite{Liu:2018qtb}. In the minimal composite Higgs model~\cite{Agashe:2004rs}, for instance, one finds $\kappa_{W,Z} = \sqrt{1 - v^2/f^2}$. In this case, in the absence of a positive signal, we would obtain a bound on the scale $f$:
	\begin{equation}
	   f \gtrsim 4.8\,\mathrm{TeV}\,\sqrt{\frac{0.13\%}{\delta \kappa^\mathrm{max}_W}}.
	\end{equation}
	The decay constant $f$ does not directly determine the mass scale of all composite states; generic composite mesons lying at the naive dimensional analysis scale $m_\rho \sim 4\pi f/\sqrt{N}$ could remain out of reach of a 10 TeV muon collider if $f \sim 5\,\mathrm{TeV}$. However, composite Higgs models also contain other particles, like top partners, lying around the scale $f$ itself, and these particles play a major role in the naturalness of the theory (see, e.g.,~\cite{Panico:2015jxa}). Thus, much of the parameter space where a precision deviation in $\kappa_W$ is observable could also lead to a direct discovery of new particles associated with compositeness. Because this scenario involves tree-level states 
	%{\color{blue}\bf question: ?} 
	with strong coupling , it is one of the cases where precision is expected to be farthest ahead of direct reach, so this is an encouraging conclusion. 
	
	Similar remarks hold for the Higgs oblique operator $\partial_\mu(H^\dagger H)\partial^\mu(H^\dagger H)$, which can affect the precision measurement of the di-Higgs production rate at a high energy muon collider~\cite{Buttazzo:2020uzc}. One possible origin for such an operator is a singlet scalar mixing with the Higgs, as in the Twin Higgs scenario~\cite{Chacko:2005pe}. A 10 TeV muon collider could probe $f \sim 10\,\mathrm{TeV}$ in direct searches for such a scalar $\phi \to hh \to (b{\bar b})(b{\bar b})$~\cite{AlAli:2021let}. In this case, the precision constraint and the direct search are extremely similar, with the former being a non-resonant search for the di-Higgs process and the latter a resonant search.

% %
% \begin{figure}[t]
% \centering
% \includegraphics[width=\linewidth]{Figures/kappabarplotmuC3_20ifb.pdf}
% \caption{
% The Higgs coupling precision from a global fit of Higgs measurements in the $\kappa$-framework. The four columns represent the HL-LHC S2 scenario, a circular $e^+e^-$ collider at $240\,$GeV, a muon collider at 125\,GeV with a total integrated luminosity of $20\ifb$, and the combination of the $e^+e^-$ and the muon collider, respectively. The measurements are combined with the HL-LHC S2 for all the lepton collider scenarios. The column shows results with $\Gamma_H$ treated as a free parameter; the horizontal marks show the ones assuming that the Higgs has no exotic decay. 
% }
% \label{fig:kapparesult}
% \end{figure}

%% file: physics-DM.tex
    WIPMs are natural cold DM candidates~\cite{Jungman:1995df,Arcadi:2017kky,Roszkowski:2017nbc}. 
    A representative case among the WIMP scenarios is the DM particle being the lightest member of an electroweak (EW) multiplet. 
    The electroweak mass splitting among the members of the same multiplet is small compared to the overall mass scale.  
    The high mass scale and near degeneracy render the DM searches at colliders extremely challenging.
    The model-independent mono-$X$ signals ($X=g,\gamma, W/Z, h...$) are not expected to reach a mass beyond two to three hundreds GeV at the high luminosity upgrade of the LHC (HL-LHC)~\cite{Low:2014cba,Han:2018wus}. On the other hand, the disappearing track-based searches can extend the coverage up to 900 GeV for a triplet ~\cite{CidVidal:2018eel}. This signature class relies on the mass gap between the members of the EW multiplet, which can introduce additional dependencies, in particular for the case of the Higgsino and the scalar multiplets.
    
    A high energy muon collider %\cite{muCcoll:2020,InternationalMuonCollider:2022qki} 
    can make powerful statements about the electroweak WIMP DM for a fermionic DM particle in connection with its thermal relic abundance. 
    We adopt the benchmark choices of the collider energies and the corresponding integrated luminosities:
    \begin{equation}
    %\sqrt{s} = 3,\ 6,\ 10,\ 14, \  30 \ {\rm and}\ 100\ {\rm TeV},\quad {\rm with} \quad  {\mathcal L} = 1,\ 4,\ 10,\ 20,\ 90, \ {\rm and}\ 1000\ {\rm ab}^{-1} ,
    \sqrt{s} = 3,\ 6,\ 10,\ 14, \ {\rm and}\ 30\ {\rm TeV},\quad {\rm with} \quad  {\mathcal L} = 1,\ 4,\ 10,\ 20, \ {\rm and}\ 90\ {\rm ab}^{-1} ,
    \label{eq:para}
    \end{equation}
    respectively. 
    In the universal and inclusive signals, the particles in an EW multiplet are produced in association with at least one energetic SM particle. The soft particles or disappearing tracks are treated as invisible. This signal class is more inclusive and model-independent since is not sensitive to the mass splittings within the EW multiplet. The most obvious channel is the pair production of the EW multiplet associated with a photon, which dominates the sensitivity to higher-dimensional EW multiplets.
 Additionally, vector boson fusion (VBF) channels unique to a high-energy muon collider \cite{Costantini:2020stv} can also contribute. 
    In particular, the mono-muon channel shows the most promise. 
    After considering the inclusive signatures, we also perform a phenomenological estimate of the size of the disappearing track signal.
    Our findings show that the high-energy muon collider could substantially improve the constraints on thermal DM, serving as one of the main physics drivers for a high-energy muon collider.
        
    \begin{figure}[tb]
    \centering
    \includegraphics[width = 0.8\textwidth]{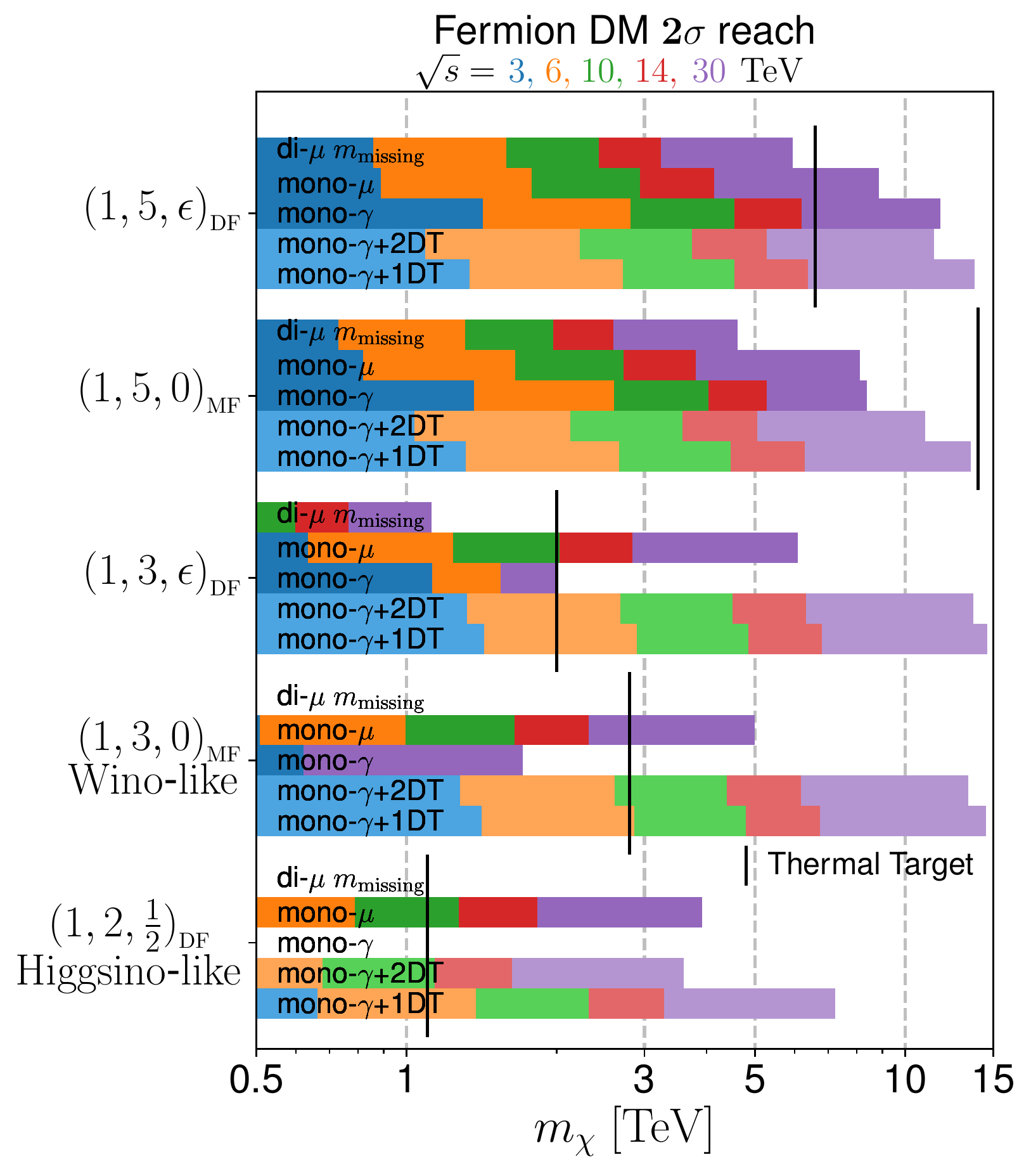}
    \caption{$2\sigma$ exclusion of fermion DM masses with horizontal bars for individual search channels and muon collider energies by the different colors. 
    The vertical bars indicate the thermal mass targets~\cite{Han:2020uak,Han:2022ubw}.}
    \label{fig:summaryF}
    \end{figure}
    
    \begin{figure}[tb]
    \centering
    \includegraphics[width = 0.8\textwidth]{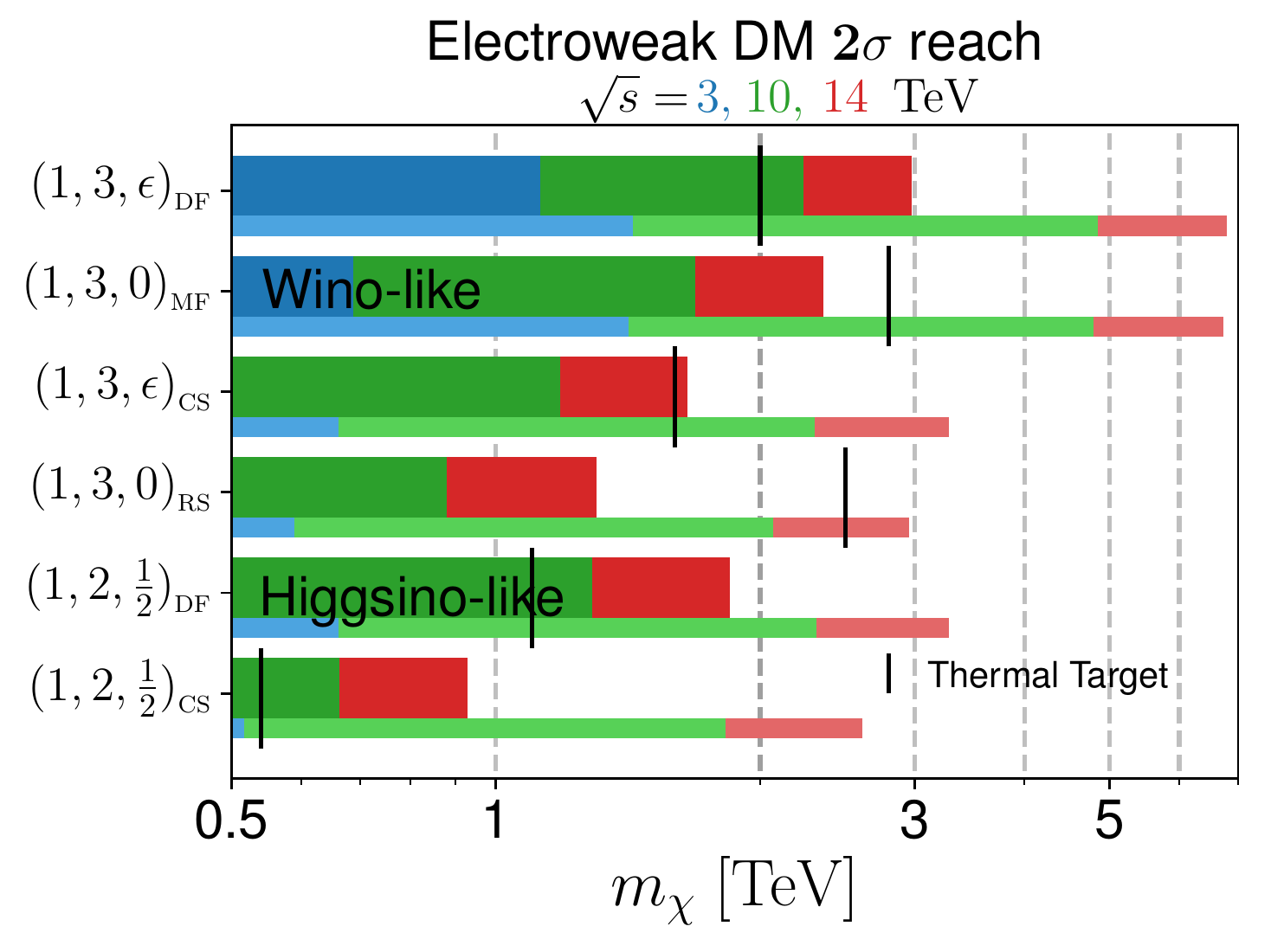}
    \caption{$2\sigma$ exclusion of DM masses with horizontal (thick) bars for combined channels and various muon collider running scenarios for $\sqrt s = 3, 10$ and 14 TeV~\cite{Han:2022ubw}. 
    The thin bars show the reach of the mono-photon plus one disappearing track search. 
    The vertical bars indicate the thermal mass targets.}
    \label{fig:summaryAll2}
    \end{figure}

        We present the sensitivity results for the WIMP DM of Dirac fermion (DF), Majorana fermion (MF) in  Fig.~\ref{fig:summaryF}, obtained from Refs.~\cite{Han:2020uak,Han:2022ubw}.  For related results, and a further discussion of the theoretical uncertainties see Refs.~\cite{Bottaro:2021snn,Bottaro:2022one}. The reaches for $2\sigma$ exclusion are shown for individual search channels, and various muon collider running scenarios are indicated by the color codes.
        The mono-muon channel, a unique signal for muon collider, shows much potential and it is especially promising for lower-dimensional EW multiplets, i.e. with $n\leq 3$. The traditional mono-photon channel at lepton colliders is suitable for higher-dimensional EW multiplets due to the coupling enhancement for high EW $n$-plets and the high multiplicity of the final state. In principle, one can consider radiation of other EW gauge bosons such as $W$ and $Z$ to improve the sensitivity~\cite{Bottaro:2021snn,InternationalMuonCollider:2022qki}. The disappearing track signature will play an indispensable role in searching for EW multiplets. The mono-photon channel with one disappearing track will have the most significant signal rate and can extend the reach significantly for all odd-dimensional cases. Requiring disappearing-track pairs will reduce the reach. However, providing a cleaner signal could turn out instrumental if the single disappearing track signature does now allow for enough background suppression.
        
    The $2\sigma$ reaches for fermionic and scalar DM are summarized in a zoom-in  version with fewer energies of $\sqrt s=3, 10$ and 14 TeV in Fig.~\ref{fig:summaryAll2} (for other work on DM scalars at muon colliders see e.g. Refs.~\cite{Bottaro:2021snn,Bottaro:2022one,Kalinowski:2022fot}).
    The thick (darker) bars represent the reach in DM mass (horizontal axis) by combining different inclusive missing-mass signals. The thin (fainter) bars are the estimates of the mono-photon plus one disappearing track search. 
    We have also included the target masses (vertical bars in black) for which the DM thermal relic abundance is saturated by the EW multiplets DM  under consideration. When combining the inclusive (missing mass) channels, the overall reach is less than the kinematical limit $m_\chi \sim \sqrt{s}/2$, especially for EW multiplets with $n \leq 3$ due to the low signal-to-background ratio. It is possible to cover (with $2 \sigma$) the thermal targets of the doublet and Dirac fermion triplet with a 10 TeV muon collider. A 14 TeV muon collider can cover the complex scalar triplet. For the real scalar and Majorana fermion triplet, a 30 TeV option would suffice. 
    The thermal targets of complex scalar and Dirac fermion (real scalar and Majorana fermion) 5-plet would be covered by 30 (100) 
    TeV muon colliders. 
    %The 100 TeV option will also cover the thermal target for the complex scalar and Dirac fermion 7-plet. The real scalar and Majorana fermion 7-plet can be probed up to $30-40$ TeV in mass at a 100 TeV muon collider, with their thermal target still out of reach. 
    We note that in order to cover the thermal targets, the necessary center-of-mass energy and luminosity in many cases can be much lower than the benchmark values we showed in \autoref{eq:para}. 
    At the same time, the disappearing track signal has excellent potential and could be the leading probe for 5-plet or lower EW multiplet. Based on our study, it could bring the reach very close to the kinematical threshold $m_\chi \sim \sqrt{s}/2$. We note that a 6 TeV muon collider with a disappearing track search can cover the thermal target of the doublet case, motivating further detailed studies in this direction. 
    A 3 TeV muon collider has sufficient energy to access the pure-Higgsino DM through the disappearing track channel kinematically. However, with the current detector layout design~\cite{ILC:2007vrf} and the short lifetime, the signal efficiency would still be too low~\cite{Han:2020uak}. 
    The maximal signal efficiency can be estimated as follows. At $E_{\rm CM} = 3 $ TeV, the Higgsino would be produced relatively close to the threshold. With a lifetime of 0.02~ns, it would have a lab frame lifetime smaller than 0.56~cm, with a smaller {\it transverse} displacement. The single disappearing track reconstruction would have an efficiency at most $2.5\times 10^{-4}$ without taking into account any experimental acceptance. The Higgsino production rate without the requirement of the existence of a 25~GeV $p_T$ photon is 10~fb. After requiring such a photon associated with the single track, the cross-section is 1~fb. Higgsinos will be produced with a pseudorapidity distribution, yielding an even smaller number of signal events in the acceptance region. All of these point towards less than one signal event. At the same time, the beam background would yield around 20 events. For a comparative analysis between our study and the full detector simulation study, see discussions in Sec.~\ref{sec:DTcomparative}.
    
        % {(\it Jiji, is this the part you want to complete?--Zhen)} 

A powerful future constraint on DM via indirect detection, i.e., searching for annihilation products, will come from the Cherenkov Telescope Array (CTA) gamma ray telescope~\cite{CTA:2020qlo}. CTA observations of the Galactic Center will have sensitivity in the range of thermal relic higgsino cross sections. However, substantial astrophysical uncertainties remain, and if the DM distribution in the galaxy is cored on length scales of multiple kiloparsecs, higgsino DM could evade even CTA~\cite{Rinchiuso:2020skh}. Similarly, searches for cosmic ray antiprotons from DM annihilation have the potential to constrain higgsino DM, but currently suffer from considerable modeling uncertainties related to cosmic ray propagation in the galaxy. 
The other class of DM detection, direct detection, tends to be less sensitive to the EW multiplet DM since their cross sections are loop suppressed and are subject to accidental cancellations between different diagrams~\cite{Hill:2013hoa, Hisano:2015rsa}. In addition, direct detection also depends crucially on the local DM energy density and velocity distributions, which could vary the sensitivity a lot. 
In short, the high-energy muon collider could provide the leading and most solid probe of heavy EW dark matter, such as higgsino DM, independent of the astrophysical uncertainties. 

%% file: physics-Natur.tex
The naturalness puzzle of the electroweak scale, also known as the hierarchy problem, has been a major driver of searches for new physics above the weak scale accessible at colliders (for a recent review see the Snowmass contribution~\cite{Craig:2022uua}). It is deeply rooted in our quest for an explanation of electroweak symmetry breaking, or equivalently, a mechanism to generate the familiar Higgs potential, which is put in by hand in the Standard Model.  This problem has only come more to the forefront since the discovery of the Higgs boson rather than some other mechanism like Technicolor. Fine tuning measures the sensitivity of the Higgs potential to the UV physics in new physics models with a dynamically generated weak scale. While the computation of fine tuning might be taken with a grain of salt due to the lack of a sharp definition, the origin of electroweak symmetry breaking is a clear fundamental question of nature, which calls for continuous efforts at the energy frontier.     
    
There are two categories of scenarios addressing the naturalness puzzle with different strategies: solutions of the ``big'' hierarchy problem, stabilizing the Higgs potential from scales around or not far above the weak scale all the way to the Planck scale; and solutions of the ``little'' hierarchy problem, extending the weak scale to some intermediate high energy scale, e.g., the highest energy scale probed experimentally. Currently LHC data has not provided a confirmed statistically significant hint of new physics related to the Higgs boson. This suggests two different possibilities. The first one is that there exists a gap between the weak scale and the new physics scale that solves the (big) hierarchy problem. In other words, the weak scale emerges and is stabilized up to the scale of quantum gravity, but with a large residual fine-tuning. The second possibility contains new particles filling in the gap between the weak scale and the TeV scale which is being probed at the LHC, as suggested in solutions to the little hierarchy problem. But these new particles have exotic collider signatures and are more elusive search targets. A high-energy muon collider demonstrates great potential in covering both possibilities, with either precision measurements (e.g., Higgs couplings) or direct searches. We will sketch some possibilities here. More details could be found in Ref.~\cite{AlAli:2021let}.  
    
So far there are only two well-studied solutions to the big hierarchy problem: supersymmetry and compositeness. Let us consider supersymmetry first. In supersymmetry, three types of key superparticles, higgsinos, stops and gluinos, contribute to the Higgs potential at tree level, one loop and two loops respectively and determine the level of electroweak fine tuning. The specific reach at a muon collider requires detailed simulations. But a simple rule of thumb is that for distinctive final states (i.e., final states that allow efficient cuts to eliminate backgrounds while keeping a high signal efficiency), pair-produced new particles could be discovered up to the production threshold with mass scale $\tilde{m} \lesssim \sqrt{s}/2$, at a high energy muon collider. For higgsinos, in the worst case scenario, the mass splitting between states in the multiplet is due to the Standard Model radiative correction. In this case, the final state consists only soft radiation from transitions within the multiplet, and a more dedicated search is needed, as described in the previous section on dark matter searches. Once on-shell decays within the multiplet are allowed given larger mass splitting, the rule of thumb above applies. In summary, a 10 TeV muon collider with 10 ab$^{-1}$ data could probe a higgsino up to 1 TeV in the worst-case small splitting scenario, and 5 TeV in the large splitting limit, which corresponds to percent level and per mille level fine tuning respectively. 
    
\begin{figure}[h]
\centering
\includegraphics[width=0.7\linewidth]{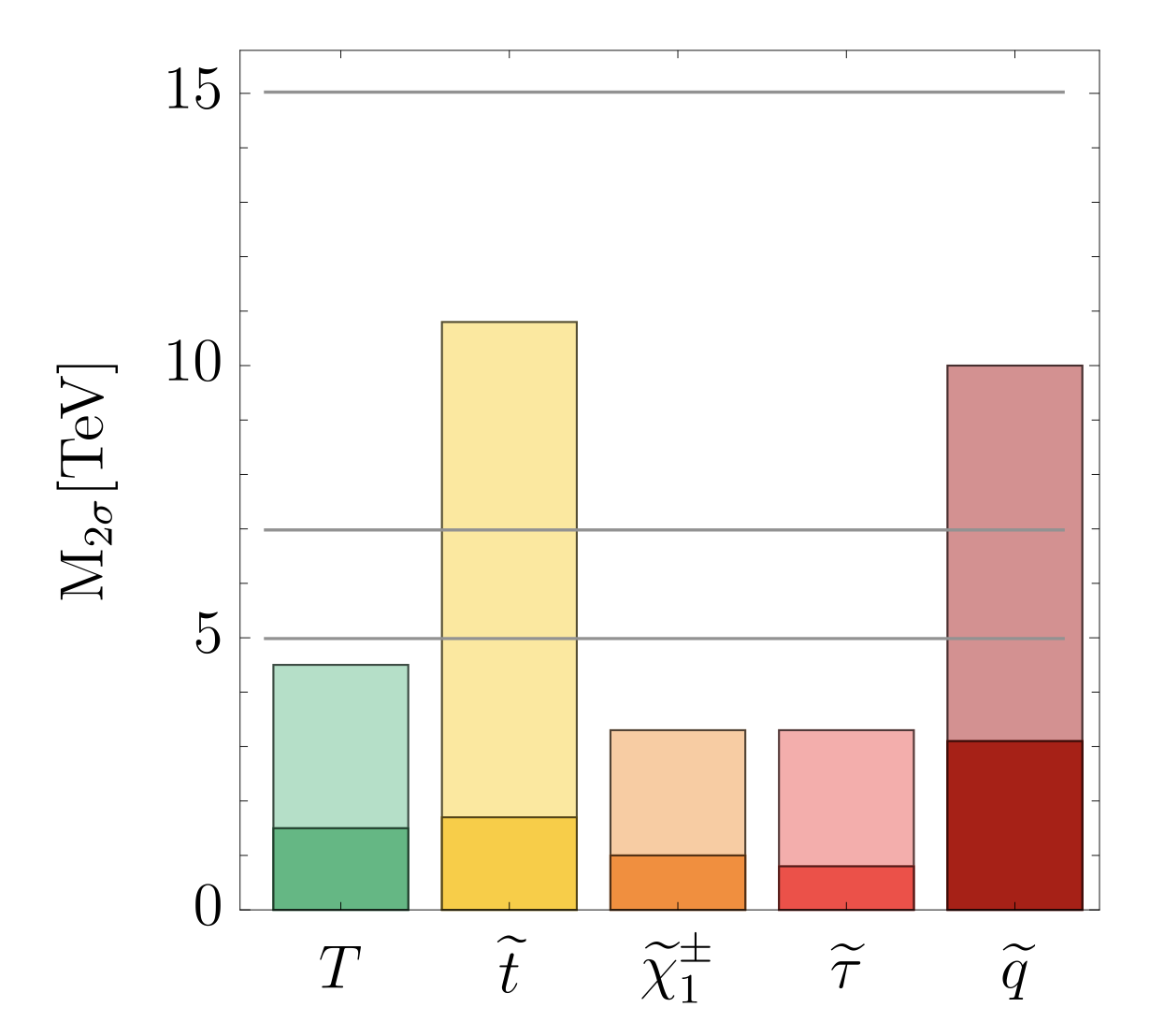}
\caption{Figure from~\cite{Aime:2022flm} showing the discovery reach of a top partner and several supersymetric particles for a 10, 14, and 30 TeV muon collider shown as horizontal lines.  The lightly shaded and darker bars correspond to the 95\% C.L mass reach of the HL-LHC and FCC-hh as determined for the European Strategy Update briefing book. 
}
\label{fig:directreach}
\end{figure}

Next we turn to stops, $\tilde{t}$, the superpartners of the top quark. Away from the limit in which the lightest stop is almost degenerate in mass with one of its decay products, the final states of stop pair production are usually distinctive, e.g., $\tilde{t}\tilde{t}^\dagger \to t \bar{t} +$ missing energy. It is estimated that the mass reach scales as $m_{\tilde{t}} \sim 0.9 \times \sqrt{s}/2$, consistent with the rule of thumb~\cite{AlAli:2021let}. At the HL-LHC, the expected 2$\sigma$ exclusion reach for stops is about 1.7 TeV. To compete with the LHC, a muon collider with $\sqrt{s} \gtrsim 4$ TeV is needed. For the benchmark 10 TeV center of mass energy and 10 ab$^{-1}$ data, stops with masses up to 4.5 TeV could be reached, associated with a $\sim 0.4$\% level fine-tuning. More importantly, the stop mass is crucial to the supersymmetric prediction of the Higgs mass. It is well known that to obtain a 125 GeV Higgs in the minimal supersymmetric Standard Model with no significant left- and right-handed stop mixing, stops have to be in the mass range of (5 -- 10) TeV at large $\tan \beta \gg 1$ (for a review, see Ref.~\cite{Draper:2016pys}). While in principle they could be lighter with large mixing, it is important to understand what is needed to cover the parameter space implied by the Higgs mass.  To test this prediction thoroughly, we need the muon collider to operate at $\gtrsim 20$ TeV. For small $\tan \beta \lesssim 5$, stop masses could be significantly larger, $\sim (100 - 10^4)$ TeV, to accommodate the Higgs mass, in well-motivated models. Though it is a tall order to probe these scales even at future colliders, the models also predict significantly lighter electroweak supersymmetric states, such as electroweakinos.  In particular, it is difficult to construct models that only have the stops as the lightest particles.  Taking into account the Higgs mass, in predictive frameworks for SUSY masses that include the transmission of SUSY breaking effects, typically sleptons and electroweakinos are often in the $\sim$ 1 TeV to few TeV range. These provide accessible motivated targets for a high-energy muon colliders in the $\mathcal{O}(10)$ TeV range.  An example of the reach as compared to the HL-LHC and FCC-hh is shown in Fig.~\ref{fig:directreach}, as reproduced from Ref.~\cite{Aime:2022flm}.

Lastly we want to comment briefly on the gluinos. Since the gluino does not carry electroweak quantum numbers, its production channel such as $\mu^+\mu^- \to \tilde{g} \tilde{g} q \bar{q}$ is of high order with a suppressed cross section. Though the gluino search will lag behind the stop and higgsino searches, a decent reach is still achievable: a 10 TeV muon collider with 10 ab$^{-1}$ of data could match the gluino reach at HL-LHC. 
    
In summary, a 10 TeV muon collider with 10 ab$^{-1}$ data could be sensitive to per cent to per mille level electroweak fine tuning in supersymmetric models, probing important states closely tied with the Higgs potential such as higgsinos and stops. 
    
In the other avenue to solve the big hierarchy problem, the compositeness scenario, fermionic top partners should exist at a scale $m_T \sim y_t f$ with $y_t$ the top Yukawa coupling and $f$ the Goldstone symmetry breaking scale, parametrically below the compositeness scale $m_* \sim g_* f$ with $g_*$ the strong coupling, at which a plethora of bound states appear~\cite{Panico:2015jxa}. The rule of thumb will apply to the direct reach of the fermionic top partners as well. Similar to supersymmetry, a 10 TeV muon collider could detect or exclude top partners up to a few TeV, testing the scenario with per cent to per mille fine tuning. More importantly, the composite scenarios lead to sizable deviations in indirect electroweak observables, such as the Higgs couplings. As discussed in~\cite{Buttazzo:2020uzc}, a high-energy muon collider enjoys an advantage over all other future colliders in testing the Higgs composite scales through constraining dimension-six operators such as 
\begin{equation} 
{\cal O}_H = \frac{1}{2} \left(\partial_\mu |H|^2\right)^2 \, , \quad
{\cal O}_W = \frac{i g}{2} \left(H^\dagger \sigma^a \overleftrightarrow{D}^\mu H\right) D^\nu W^a_{\mu\nu} \, ,
\end{equation}
 due to a combination of high energy and clean backgrounds. In particular, a 10 TeV muon collider could probe the compositeness scale through the indirect precision measurements as high as $m_* \sim$ 45 TeV with 10 ab$^{-1}$ data~\cite{Buttazzo:2020uzc}.  
 
\begin{figure}[h]
\centering
\includegraphics[width=0.5\linewidth]{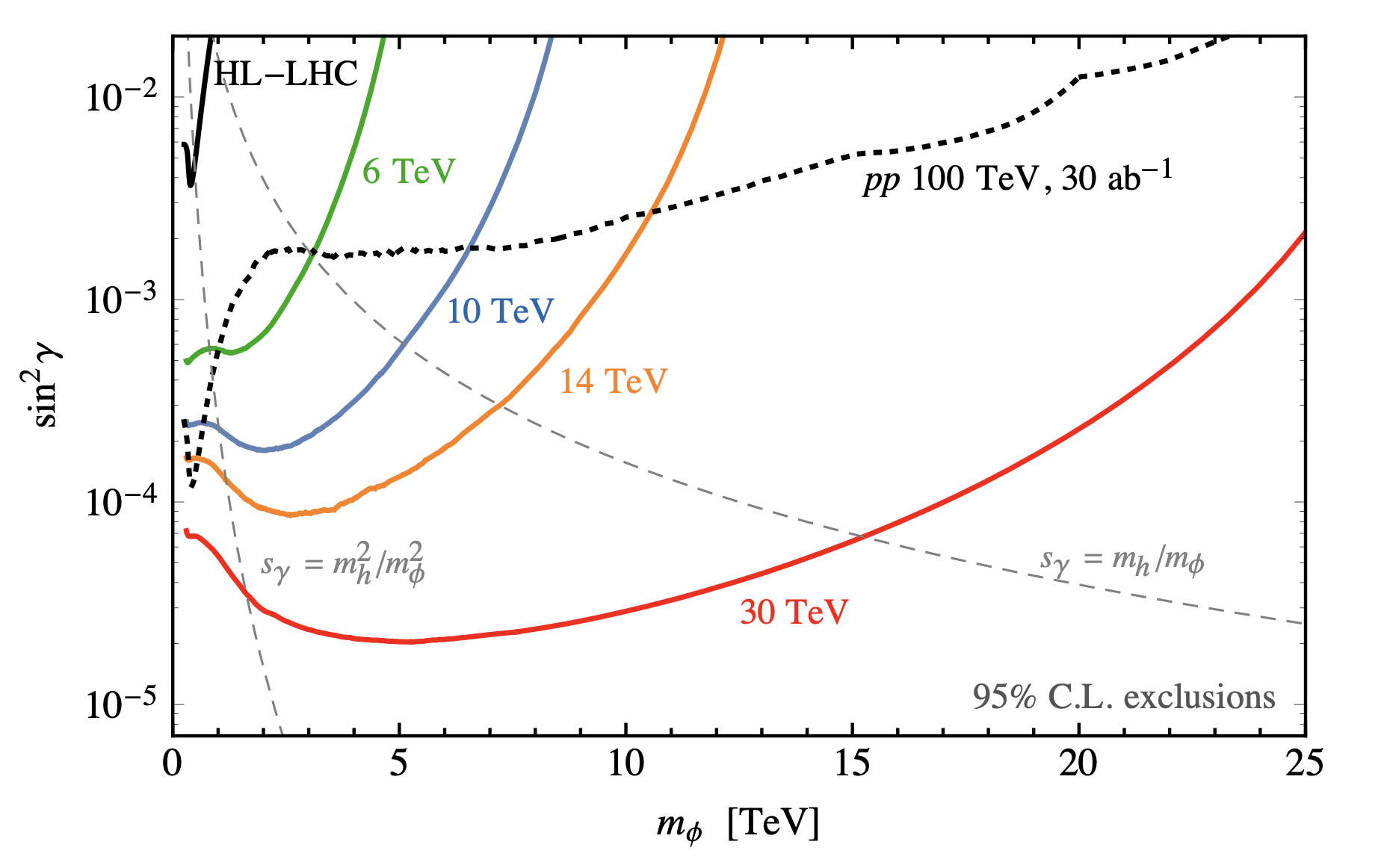}\includegraphics[width=0.5\linewidth]{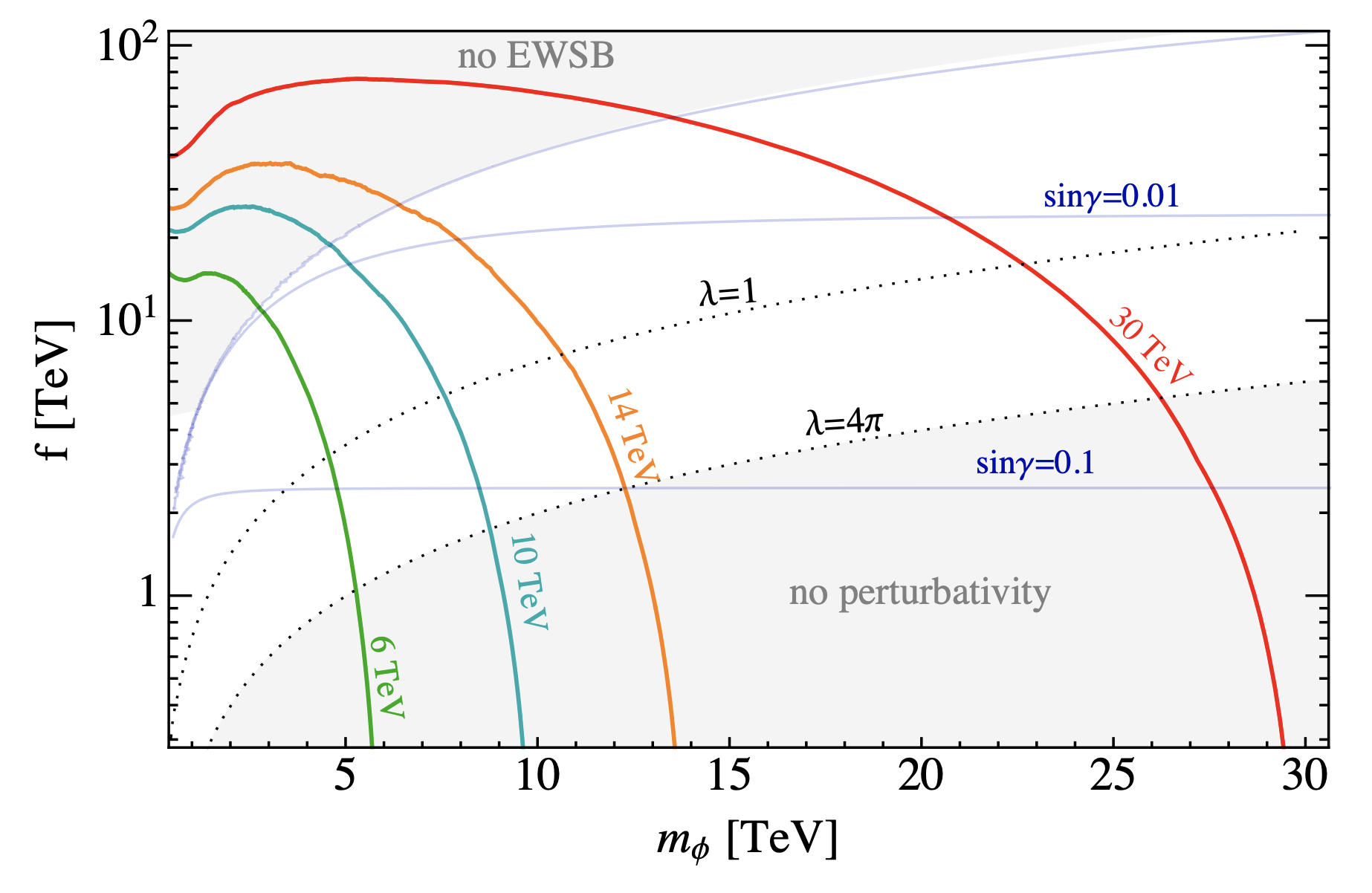}
\caption{Figures from~\cite{AlAli:2021let}, the left hand plot shows the exclusion limits for a massive scalar singlet $\phi$ that mixes with SM Higgs with mixing angle $\gamma$.  The various colored curves correspond to different muon collider COM energies, while the expected limits at HL-LHC (solid) and a FCC-hh (dashed) are shown as black lines for comparison. The thin dashed lines indicate two possible scalings of the mixing angle in realistic models with fixed coupling.  The left hand plot represents a more generic statement than that of naturalness, and is a more general illustration of the powerful Higgs portal reach of a muon collider.  The right hand plot usese the same limits as the LH plot re-interpreted in terms of the reach on the sigma-model scale f in the context of a Twin Higgs model. 
}
\label{fig:neutralnat}
\end{figure}

Solutions to the little hierarchy problem could evade the LHC searches if the partner states are neutral under the Standard Model and related to the visible sector only via a discrete symmetry, as realized in neutral naturalness models, e.g., twin Higgs~\cite{Chacko:2005pe}. A high-energy muon collider, however, is well suited to tackle the subtle signals in such scenarios through either precision measurements or direct searches. More concretely, the muon collider could measure the Higgs coupling deviations due to the scalar mixings, probe exotic Higgs decays with displaced vertices, and search for the Standard Model singlet partners, as well as the radial modes associated with spontaneous breaking of the associated discrete symmetry. In terms of direct searches, searching for the radial modes turns out to be a more promising strategy compared to the other partners. It has been shown that in a twin-Higgs setup, direct searches for the extra scalar at a 10 TeV muon collider could probe most of the parameter space corresponding to a \% level deviations in Higgs couplings, and even explore regions with smaller deviations which will be difficult to be observed in precision measurements~\cite{AlAli:2021let}. In Fig.~\ref{fig:neutralnat}, adapted from Ref.~\cite{AlAli:2021let}, we explictly show the reach for both a neutral naturalness interpretation and more generally the Higgs portal as compared to the HL-LHC and FCC-hh projections.

%% file: physics-compintro.tex
Some of the strongest constraints on new physics come from searches for flavor or CP violation, or other accidental or approximate symmetries of the SM. These include, for example, searches for electron and neutron electric dipole moments (EDMs), lepton flavor-violating (LFV) muon decays, and measurements of Kaon (or other meson) mixing. The sensitivity of many of these searches, in particular for EDMs and LFV decays or transitions, is set to improve by several orders of magnitude in the coming decades~\cite{Alarcon:2022ero, Baldini:2018uhj, Mu2e-II:2022blh, CGroup:2022tli}. Furthermore, new data from Cosmology, such as measurements of a stochastic gravitational wave background~\cite{Caprini:2018mtu, Christensen:2018iqi} or primordial non-Gaussianities in Large Scale Structure and high-redshift 21-cm maps~\cite{Meerburg:2019qqi}, may also provide hints of new physics~\cite{Kamionkowski:1993fg, Huber:2008hg}.

Given the broad scope of these indirect searches for new physics, there is ample reason for optimism that one or more of these experiments might detect a signal of BSM physics with the increased sensitivity available. In fact, as we will discuss below, several potential signals of new physics already exist in the data. However, these indirect probes can only give a signal that new physics exists; they cannot say much about it.  Furthermore, as has been seen on many occasions, indirect signals of BSM physics that at first were thought to be incontrovertible have become swamped by new backgrounds or systematics and remain ambiguous at best.   For this, pushing the energy frontier along with these probes is essential, so that a discovery in a complementary, low-energy experiment can be quickly followed up by in depth studies at the energy scale of the new physics. Fortunately, for motivated extensions of the Standard Model that could lead to such signatures, the relevant scale is in the $\mathcal{O}(1 - 10)$ TeV range, which is amenable to collider exploration.

A high-energy muon collider provides an ideal laboratory for complementing these low energy probes. Colliding muons at high energies can not only extend the reach to new electroweak states in the multi-TeV range, but also study processes that are closely related to those measured in low-energy probes at a different energy scale. Muon colliders are therefore well-suited for following up a signal in a complementary low-energy experiment with unique capabilities for discovering the new states and characterizing the structure of BSM physics.  To demonstrate this complementary nature of high energy muon colliders, we focus on two different types of examples.  The first falls under the class of future precision probes that may point to new physics in Section~\ref{sec:physflavor}.  There are of course numerous other exciting avenues in experiments that reside out of the energy frontier but have complementary ways to study them at muon colliders~\cite{AlAli:2021let}.  The second example we focus on is the fact that there are existing anomalies that may turn out to be a true sign of new physics.  For example the numerous flavor anomalies, the recent CDF $W$-mass measurement, and the Fermilab g-2 measurement.  Remarkably all the current hints in data are well suited to a high energy muon collider.  As an example to demonstrate this we focus on the recent muon g-2 measurement, which clearly has connections to muon colliders in Section~\ref{sec:physgminus2}.

%% file: physics-Flav.tex
To demonstrate the complementary nature of high energy muon colliders, we look at the example of contributions to the electron EDM from loops of new, electroweak-charged particles. If these particles couple to the Higgs, they will produce a 2-loop EDM via Barr-Zee diagrams~\cite{Barr:1990vd} (see Fig.~21 of Ref.~\cite{AlAli:2021let}), whose size is of order
    %%%
    \begin{equation}
    d_e \sim \sin(\delta_{\textrm{CP}}) \frac{e\,m_e}{M^2} \Big(\frac{\alpha}{4\pi}\Big)^2 
    \sim 10^{-32}\, e\,\textrm{cm}\, \sin(\delta_{\textrm{CP}}) \times \bigg(\frac{20\,\textrm{TeV}}{M}\bigg)^2 ,
    \end{equation}
    %%%
    where $\delta_{\textrm{CP}}$ is a $\textrm{CP}$-violating phase and $M$ is the mass scale of the new particles. The current bound from the ACME collaboration is $|d_e| \leq 1.1 \times 10^{-29}\,e\,\textrm{cm}$ at $90\%$ confidence~\cite{ACME:2018yjb}. It is clear then, that a discovery of an EDM within the next two orders of magnitude in sensitivity would point to the $\textrm{few}\,\textrm{TeV}$ scale for new physics. The particles responsible for the EDM could thus be pair-produced at a multi-TeV muon collider, and moreover, the large rates of Higgs and Higgs-pair production at these energies would also allow for precise measurements of their couplings to the Higgs. This allows the possibility of not only discovering the new particles responsible for an EDM, but measuring their interactions that contribute in loops directly. Given the limited information about the structure of new physics available in a single low-energy measurement, these complementary searches are essential for confirming our understanding of any discovery of new physics.

\begin{figure}[t]
    \centering
    \includegraphics[width=0.45\linewidth]{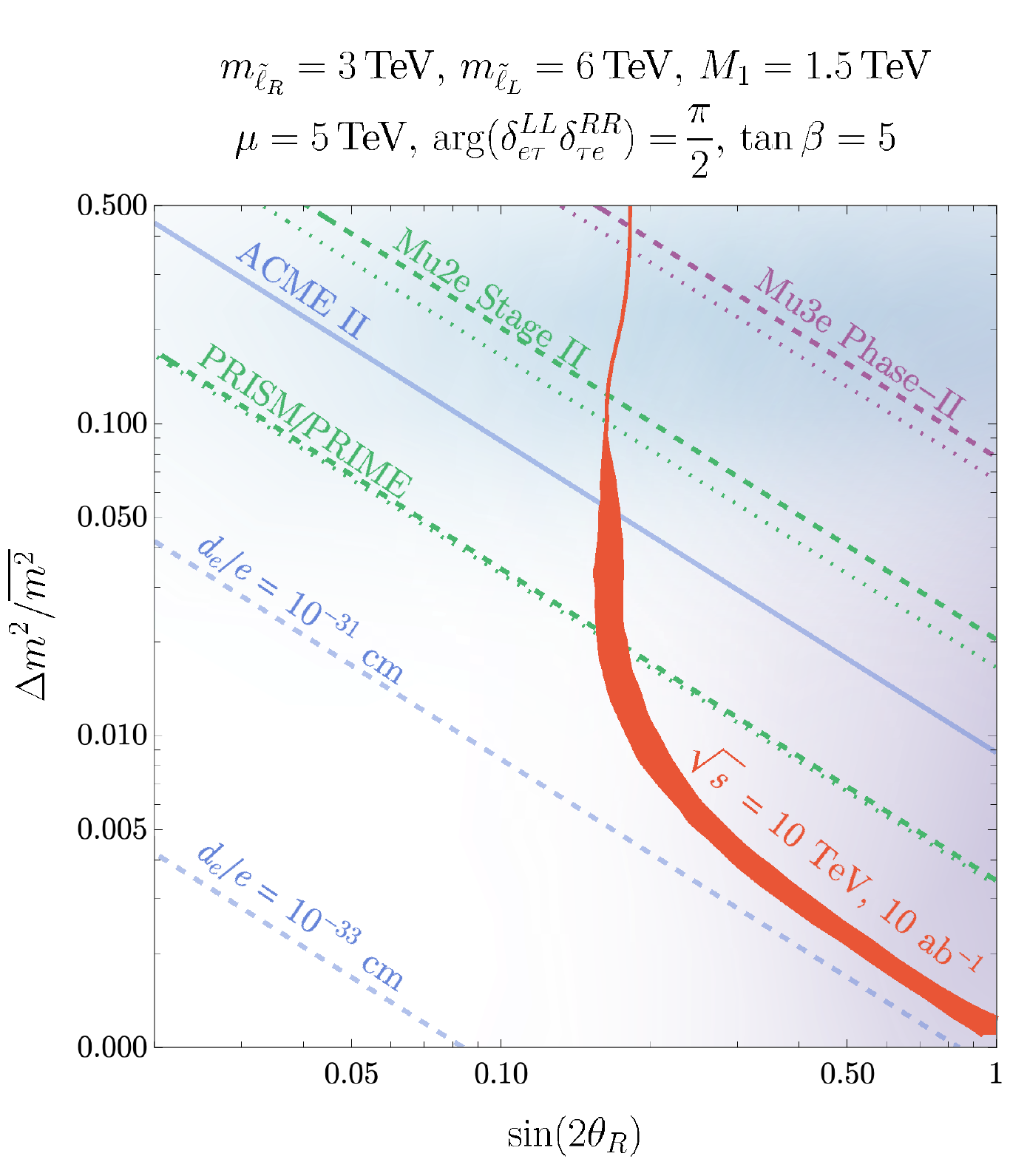}
    \quad
    \includegraphics[width=0.45\linewidth]{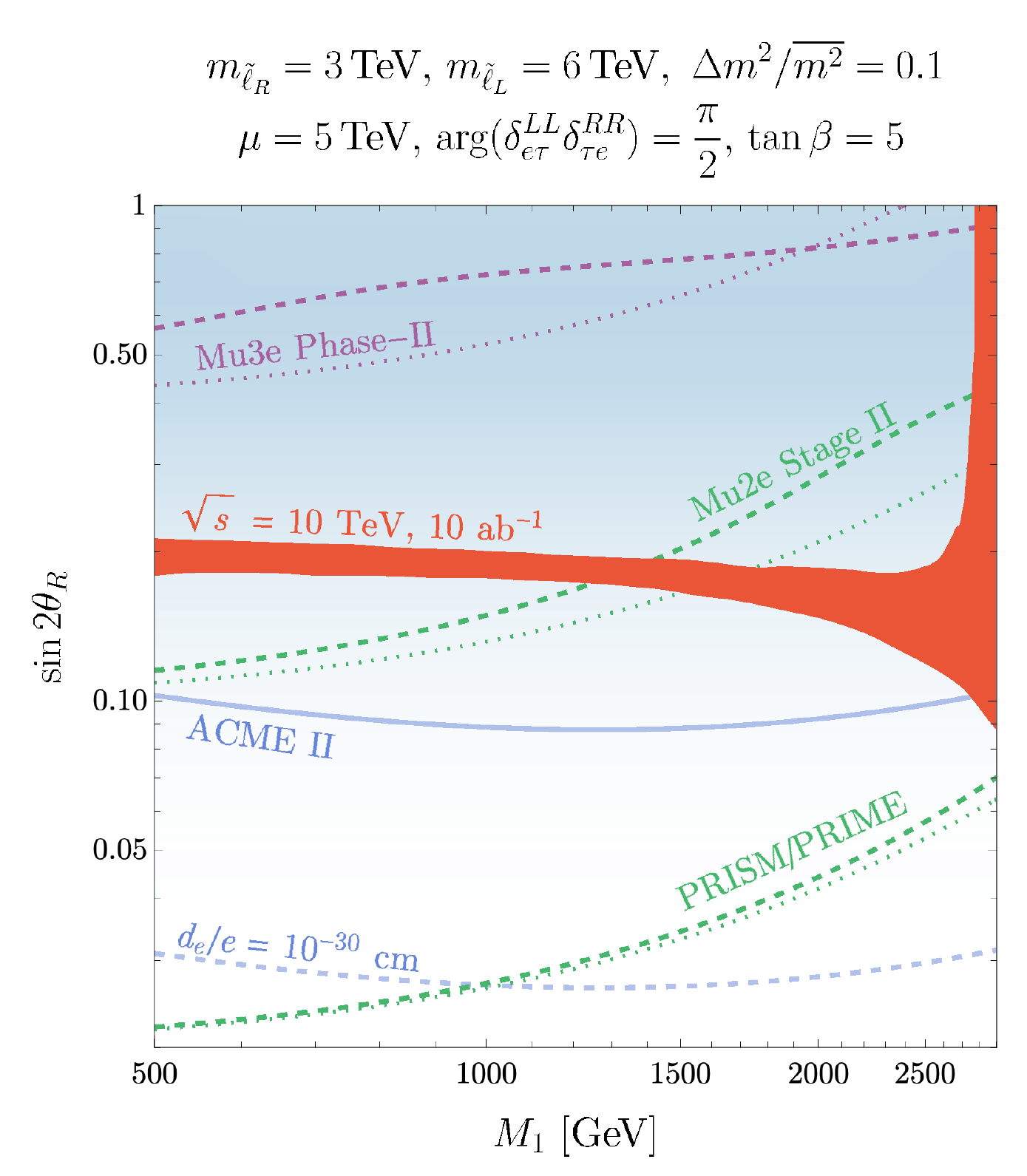}
    \caption{The reach for a 10 TeV muon collider (thick red lines) to discover a flavor-violating signal from mixed slepton production compared to the complementary constraints from LFV experiments (purple and green lines) and EDM limits (blue lines). More details can be found in Ref.~\cite{Homiller:2022iax}.}
    \label{fig:lfv_mssm_reach}
    \end{figure}

As a second example, consider the lepton flavor violating processes $\mu \to e\gamma$, $\mu \to 3 e$ and $\mu \to e$ conversion in atomic nuclei. In motivated extensions of the SM, these processes arise from loops involving new states that are not flavor eigenstates. For instance, in models of supersymmetry, the slepton mass eigenstates will in general be mixtures of different flavors, but may be nearly degenerate, leading to a ``super-GIM'' mechanism that suppresses the LFV signatures and allows the new states to lie at the few TeV scale. As discussed in Ref.~\cite{Homiller:2022iax}, this situation is a highly-motivated target for the next generation of low-energy LFV experiments, but the physics responsible for any signature can also be studied in detail at a muon collider. Once again, a high-energy lepton collider would not only pair-produce the new states responsible, but would also allow for detailed studies of the flavor violation by measuring e.g., $\mu^+\mu^- \to \mu^+ e^- + $ missing momentum mediated by intermediate sleptons. For motivated models of supersymmetry breaking, a 3 TeV muon collider could probe all of the parameter space that will be explored by the Mu2e experiment, while a 10 TeV collider would have reach comparable to the most optimistic future low-energy experimental proposals, as shown in Fig.~\ref{fig:lfv_mssm_reach}. Alternatively, a muon collider could study lepton flavor-violating contact interactions, such as the $\tau 3\mu$ operator studied in Ref.~\cite{AlAli:2021let}. In either situation, the importance of searching for LFV interactions at the TeV scale is clear.
    
Another suite of indirect probes, which already shows some hints of a deviation from the SM, are tests of lepton flavor universality, e.g., in heavy meson decays. The most recent measurements of $R_{K^{(*)}}$, the $B$-meson decay ratios to muons over electrons, at LHCb~\cite{LHCb:2021trn}, show an interesting discrepancy with the SM prediction. While lepton flavor universality is not a fundamental property of the SM, it is respected by the gauge interactions, and any deviation at the level currently being tested would be an unequivocal signal of new physics at the few to $10$s of TeV scale, depending on the precise structure of the new interactions.  A number of studies~\cite{Chakrabarty:2014pja, Capdevilla:2020qel, Buttazzo:2020ibd, Capdevilla:2021rwo, Dermisek:2021ajd, Dermisek:2021mhi, Capdevilla:2021kcf, Huang:2021biu, Asadi:2021gah, Huang:2021nkl, Casarsa:2021rud, Han:2021lnp, Liu:2021akf, Dermisek:2020cod}, summarized in Refs.~\cite{MuonCollider:2022xlm} and \cite{Aime:2022flm}, demonstrate that a muon collider with energies at least $10\,\textrm{TeV}$ would decisively test the most plausible explanations for this anomaly, whether it be a new gauge force, a model involving leptoquarks, or some other new physics leading to $bs\mu\mu$ contact interactions at the $10$s of TeV scale.

As these examples exemplify, while the indirect reach of precision probes can naively extend to very high scales, there are well-understood, natural ways of suppressing these processes that bring these targets to the TeV scale. Furthermore, given the hierarchy problem and the unexplained patterns of masses and mixing angles that exist even in the Standard Model, it is reasonable to expect that there might be new physics with interesting flavor structure lying near the TeV scale. A broad program in precision physics is vital in searching for these new states, and a muon collider that extends the energy reach as high as possible is an ideal tool to supplement this strategy.

%% file: physics-gminus2.tex
As an example of how muon colliders provide complementary sensitivity to existing anomalies, we investigate the overlap with the recent muon $g-2$ measurement. The persistent discrepancy between the theoretical and experimental values for $(g-2)_\mu$ has recently increased to 4.2$\sigma$ between the recent measurement from the Fermilab $g-2$ experiment~\cite{Muong-2:2021vma,Muong-2:2021ojo,Muong-2:2021ovs,Muong-2:2021xzz} and the consensus value from the Muon $g-2$ Theory Initiative~\cite{Aoyama:2020ynm,Aoyama:2012wk,Aoyama:2019ryr,Czarnecki:2002nt,Gnendiger:2013pva,Davier:2017zfy,Keshavarzi:2018mgv,Colangelo:2018mtw,Hoferichter:2019mqg,Davier:2019can,Keshavarzi:2019abf,Kurz:2014wya,Melnikov:2003xd,Masjuan:2017tvw,Colangelo:2017fiz,Hoferichter:2018kwz,Bijnens:2019ghy,Colangelo:2019uex,Pauk:2014rta,Danilkin:2016hnh,Jegerlehner:2017gek,Knecht:2018sci,Eichmann:2019bqf,Roig:2019reh,Colangelo:2014qya}. This observable may point to new physics beyond the SM, perhaps specific to the muon sector. A muon collider offers unique opportunities to probe the new physics of $(g-2)_\mu$ that no other experiment can offer, for two related reasons. 

\begin{figure}
  \centering
    \includegraphics[width=0.45\textwidth]{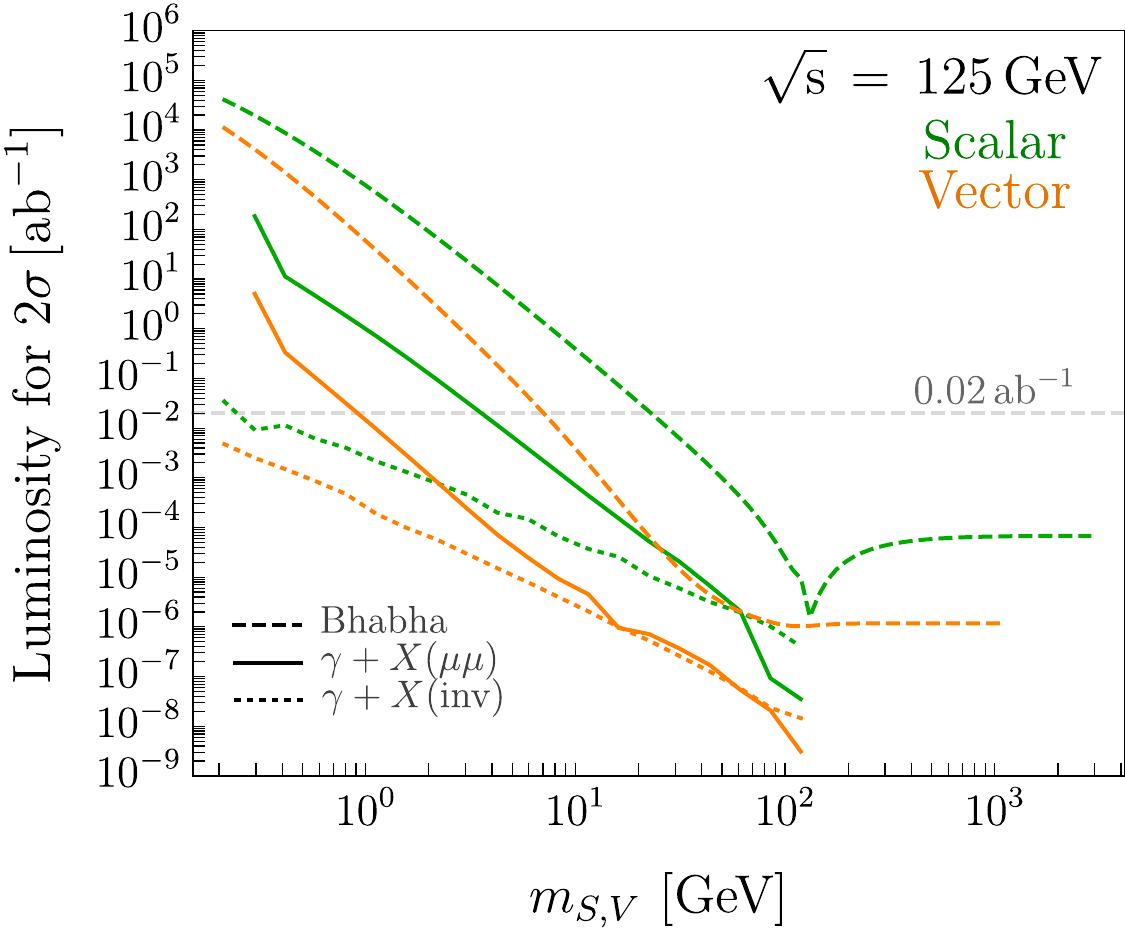} \hfill
    \includegraphics[width=0.45\textwidth]{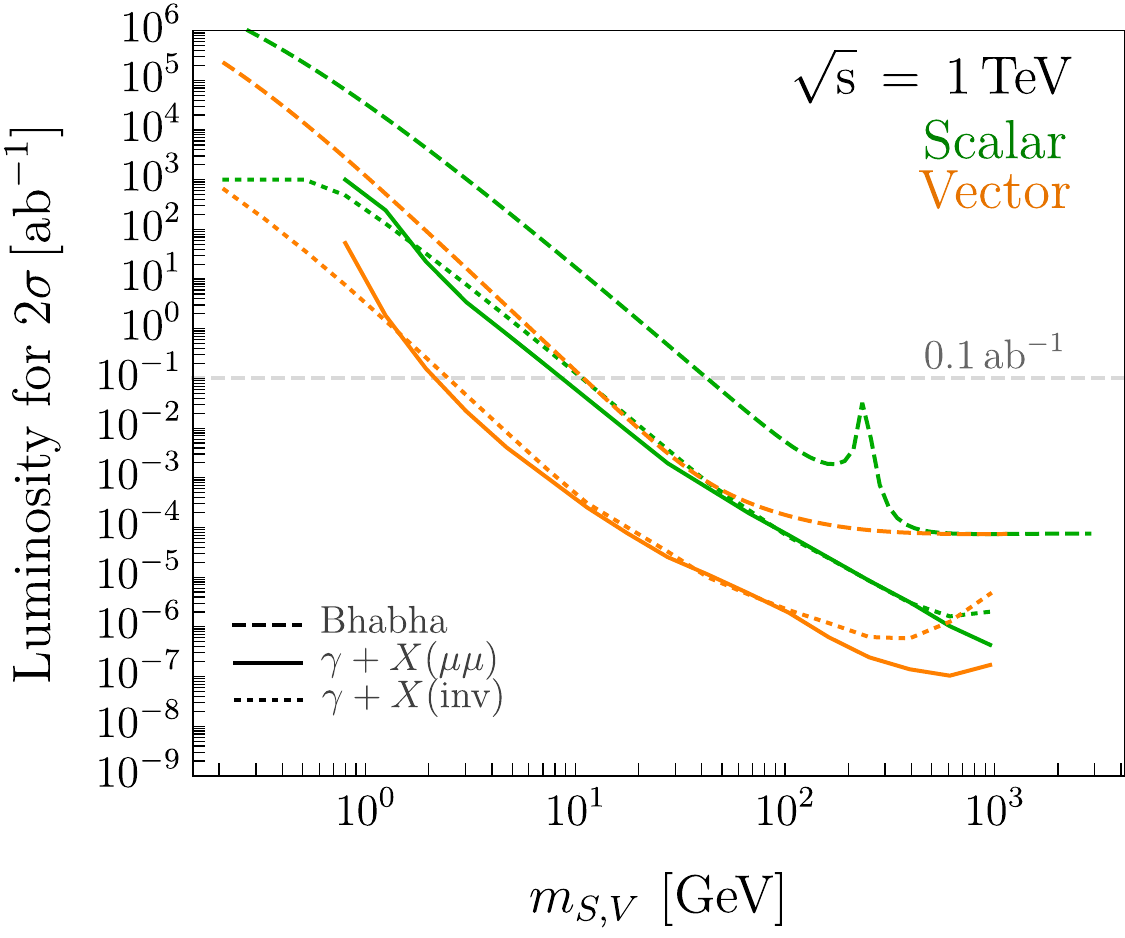}
    \caption{Required luminosity to yield a 2$\sigma$ exclusion of singlets responsible for $(g-2)_\mu$. A 125 GeV muon collider (left) has superior sensitivity to 1 TeV (right) and to 10 TeV (not shown) through the combination of direct singlet production when kinematically accessible and deviations from muonic Bhabha scattering for heavier singlets, even considering the lower luminosity. Based on the analysis of Ref.~\cite{Capdevilla:2021rwo}.}
    \label{fig:gm2Singlet}
\end{figure}

Most obviously, if a new SM singlet (either scalar or vector) resolves the $(g-2)_\mu$ anomaly, in the worst-case scenario where the singlet \emph{only} couples to the muon, only a muon beam can probe the singlet coupling responsible for $(g-2)_\mu$. The maximum singlet mass is at the TeV scale, and to match the observed deviation in $(g-2)_\mu$, the coupling must increase with mass; thus, even if the singlet is too heavy to produce on-shell, its contribution to muonic Bhabha scattering may be probed directly at a muon collider (Fig.~\ref{fig:gm2Singlet}). In a preliminary analysis, Ref.~\cite{Capdevilla:2021rwo} provides sensitivity estimates for both mono-photon searches for on-shell singlets (independent of the singlet decay channel) and deviations in Bhabha scattering. Ref.~\cite{Capdevilla:2021kcf} persuasively shows that only a \emph{low-energy} muon collider can fully probe the $(g-2)_\mu$ singlet parameter space above 1 GeV, providing a strong BSM physics motivation for a staged muon collider program.

If the new physics of $(g-2)_\mu$ comes instead from 1-loop contributions from electroweak-charged states, there must be at least two new particles (either two fermions and a scalar, or two scalars and a fermion), and a careful consideration of the possible SU(2) representations shows that there must always be at least one new charged particle~\cite{Capdevilla:2020qel}. A muon collider is once again the ideal vehicle for discovery of such states because of its strong sensitivity to electroweak physics and the availability of the full center-of-mass energy. The maximum mass scale of the new physics, assuming only perturbative unitarity, is around 100 TeV, but such corners of parameter space require extreme fine-tunings in both the muon mass and the flavor sector, let alone an additional fine-tuning of the Higgs mass. Ref.~\cite{Capdevilla:2021rwo} phrased this optimistically as a ``no-lose theorem'' for $(g-2)_\mu$ at a TeV-scale muon collider: if no new charged particles are discovered, we will learn something profound about the role of fine-tuning in the SM. The situation is even better, though, because for new physics above the energy of the muon collider, the effective operator which yields $(g-2)_\mu$ will contribute a large excess rate for $\mu^+ \mu^- \to h \gamma$~\cite{Buttazzo:2020ibd}. This process is essentially background-free in the SM, and thus a detailed analysis of the reducible background rate from e.g. $Z$'s faking Higgses at a muon collider will be needed to determine the true sensitivity. Regardless, the prospects for confirming the new physics of $(g-2)_\mu$ at all energy scales above 1 GeV are extremely strong.

In addition to $\mu^{+} \mu^{-}\to h \gamma$, any model which generates the effective dipole operator through a chiral-enhancement of the Higgs coupling to new particles also generates di- and tri-Higgs production at a muon collider as discussed in Ref.~\cite{Dermisek:2021ajd}. Relating this to $(g-2)_\mu$ in models of  vector-like leptons which have tree-level couplings to the Higgs boson and muon leads to predictions for di- and tri-Higgs cross sections without a free parameter which are 3-4 orders of magnitude above the SM background. This correlation is predicted in any model which generates $(g-2)_{\mu}$ via chiral enhancement through a coupling of the SM Higgs to new particles~\cite{Dermisek:2022aec}, and thus the di- and tri-Higgs signals can efficiently test this class of solutions to $(g-2)_\mu$ mediated by particles above the running energy of a muon collider.

%% file: physics-HNL.tex
The origin of neutrino masses is one of the fundamental puzzles of Nature. The latest measurements show that at least two of the neutrino masses are at around $O(0.1)$ eV ~\cite{ParticleDataGroup:2020ssz}, which is far lighter than the rest of matter contents of the SM. Why does the neutrino carry such a tiny mass? One possible approach is the seesaw mechanism, like the simple Type I seesaw model, which requires heavy Majorana states. The more testable scenarios, such as the inverse seesaw and linear seesaw models, can realize the observed light neutrino with additional potentially detected heavy fermions. From the phenomenological view, we can introduce heavy fermions, which can be either Dirac or Majorana, that are common testable components of various seesaw models. It can be parametrized in terms of its mass $m_N$ and mixing angle $U_l \ll 1$ with the SM neutrino flavor~\cite{Abdullahi:2022jlv}. Although we know that there are at least two massive neutrinos, thus requiring two HNLs, we only turn on one flavor for simplicity. The relevant Lagrangian say for Dirac is as follows
\begin{equation}
\begin{split}
\mathcal{L} \supset & \bar{N} i \slashed{\partial} N - m_N \bar{N} N  + 
\dfrac{g U_l}{\sqrt{2}} \left(W_\mu \bar{l}_L \gamma^\mu N + h.c. \right) -\\
&\dfrac{g U_l}{2 \cos\theta_w} Z_\mu\left(\bar{\nu}_L \gamma^\mu N + \bar{N} \gamma^\mu \bar{\nu}_L \right)
 -g U_l m_N h \left(\bar{\nu}_L  N + \bar{N} \nu_L \right)
\end{split}
\end{equation}

\begin{figure}[ht]
\centering
\includegraphics[width=8cm]{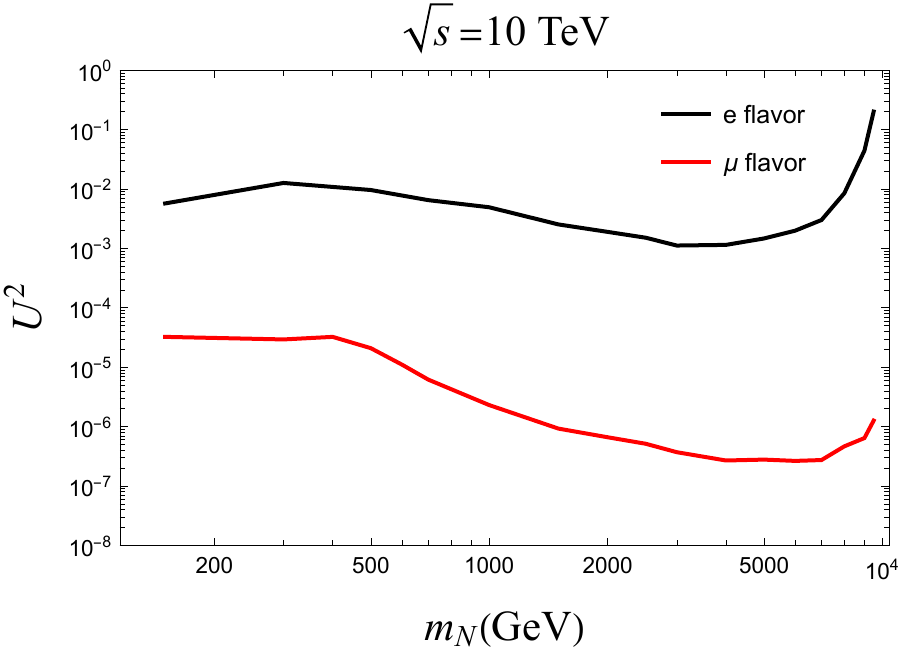}
\caption{The projected sensitivities for probing HNL on the $\sqrt{s} = 10$ TeV muon collider. The black curve refers to the electron-flavored HNL, while the red curve is for the muon-flavored case. The results for the tau-flavored HNL will be similar to the electron-flavored case. More details of the analysis can be found in Ref.~\cite{ZhenLiuTeamHNLMuC}.}
\label{fig:HNL}
\end{figure}

In this current study, we focus on the mass range $m_N > m_h$ and leave the lower mass range for future study. The muon-flavor HNL can be produced in abundance at the muon collider by the $t$-channel diagram $\mu^+ \mu^- \rightarrow N_\mu \bar{\nu}_\mu$. The signal rate for this channel can be as high as $O(10)$ pb on the 3 TeV and 10 TeV muon collider. The heavy neutrino $N_\mu$ will promptly decay via charged or neutral current or the Higgs boson. We select the decay channel $N_\mu \rightarrow W^+ + \mu^-$  and let $W^+$ subsequently decay to dijets. Assuming that W boson can be well reconstructed from the dijets combination by considering the faking from Z boson, the background is $\mu^+\mu^- \rightarrow W^+ \mu^- \bar{\nu}  (l l)$ or $\mu^+\mu^- \rightarrow Z \mu^- \mu^+ (l l)$ in which the leptons in the brackets are required not to be observed. After imposing the basic cuts, eventually, we can focus on the W boson. Then we made optimized cuts in the plane of the $W$ boson $E-p_T$ and selected the events in the specific region. The upper limit for $U_l^2$ at 95\% exclusion C.L. is then given in Fig~\ref{fig:HNL}.

We show four sensitivity curves in Fig~\ref{fig:HNL}, the $\mu$ flavor and $e/\tau$ flavor at 10 TeV muon colliders. Since the background mostly accumulates at low $m_N (m_{W+\mu})$ regime, as $m_N$ increases from $O(100)$ GeV, we can get better sensitivity on $U_\ell^2$.   The best constraint on the squared mixing angle is at the intermediate $m_N$ regime, around $\sqrt{s}/2$. As the HNL mass approaches a higher value, we gradually lose the sensitivity on the mixing angle due to the signal rate reduction. For the muon case, we can probe $U_\mu^2$ down to like $O(10^{-7})-O(10^{-5})$ due to the t-channel enhancement. While for the electron case, the constraint on $U_e^2$ can be probed in a wide mass range for around $10^{-3}$.

%% file: physics-theorysimulations.tex
A multi-TeV muon collider will provide the unique opportunity of performing precision measurements, similar to those foreseen at $e^+e^-$ colliders and to explore very high-energy phenomena like at a $pp$ collider. Therefore, not only high-precision predictions will be needed for specific observables, but also fully exclusive simulations that can account for multi-particle QCD final state radiation and electroweak radiation from the initial as well as  final states. Multi-TeV muon collisions will give the possibility to explore, for the first time in an accelerator context, SM interactions in a regime where EW symmetry is basically restored, i.e., where the weak scale $v$ is negligible with respect to the energies involved and a new set of phenomena, such for example multi weak  boson emissions, may become dominant. Such SM yet novel effects will have an impact in the search and sensitivity to New Physics. 

As discussed above, the physics at muon colliders at high energy can be schematically divided into two broad classes. In $s$-channel annihilation all collider energy flows into the high $Q^2$ interaction, making it possible to create new heavy states (with a reach in mass that is given by the kinematical limit) or very energetic SM final states. From the prediction/simulation point of view, this regime offers challenges qualitatively similar to usual $e^+e^-$ experiments, at least for what concerns QED and QCD. On the other hand, in the multi-TeV regime virtual and real emissions of soft and soft-collinear EW radiation can significantly  impact cross-section predictions and therefore affect measurements. Production of multi-boson final states is enhanced due to large logs, leading to multi-jet/lepton signatures that can be used to reconstruct the original "weak partons" entering in the hard interaction, similarly to how QCD radiation patterns are used to associate jets to gluons or quarks~\cite{Chen:2022msz,Chen:2016wkt}. 

In weak-boson fusion,  the emission low-virtuality vector bosons from the initial state muons, becomes dominant turning the muon collider into a high-luminosity vector bosons collider. This regime poses novel theoretical challenges both for precision as well as for Monte Carlo event generation. Even at the lowest order in pertubation theory, the large scale separation between $\sqrt{s}$ and $v$, the necessity of including both QED and weak radiation respecting EW gauge invariance, and the relevance of multi-boson final states, make  fixed-order event generation challenging for standard tools such as {\texttt{WHIZARD}}~\cite{Kilian:2007gr} or  {\texttt{MadGraph5\_aMC@NLO}}~\cite{Costantini:2020stv,Ruiz:2021tdt}. In addition, with EW radiation becoming dominant makes  a systematic theoretical reformulation of problems in terms of EW PDF's form the initial state and possible fragmentation functions for the final states necessary. Preliminary studies, see e.g. Refs.~\cite{Han:2020uid,Han:2021kes,Han:2022laq}, show that EW resummation effects can be significant at a multi-TeV muon collider. 

Considering initial state and final state effects together, and noting the non-abelian nature of EW interactions, one could be lead to think that techniques developed over decades for QCD resummation would provide a suitable path to correctly account for these phenomena. This is far from being the case. At low energy, EW symmetry is broken and physical states display weak charges (at low energy physical states are QCD neutral) giving rise to very different experimental signatures, e.g., a $Z$ vs a $W$ boson.  This fact also impacts the very definition of IR safety and inclusive observables that should be employed in EW calculations and the matching between an EW-symmetric forward/backward evolution to EW-broken initial/final states.  In this context, even though far from having achieved an accurate understanding and implementation of the corresponding physics, significant progress has been made over the years.  From the seminal work on the effective vector boson approximation~\cite{Kane:1984bb,Dawson:1984gx,Chanowitz:1985hj,Kunszt:1987tk} to the more recent progress on resummation~\cite{Chiu:2007yn,Chiu:2007dg,Chiu:2009ft,Manohar:2014vxa,Chen:2016wkt,Manohar:2018kfx}, including the formal definitions and proof of factorization ~\cite{Bauer:2017isx,Fornal:2018znf,Bauer:2018xag,Borel:2012by,Wulzer:2013mza,Cuomo:2019siu}. The first implementations of EW showering have also become available with different degrees of accuracy~\cite{Christiansen:2014kba,Christiansen:2015jpa,Brooks:2021kji}.
The muon collider project will bring further motivation to the theoretical community towards developing a systematic understanding of EW radiation.

%% file: general.tex
\subsection{General Introduction}\label{gen:accel}

\begin{figure}
        \centering{\includegraphics[width=0.532\textwidth, trim={2.1cm 0 2.1cm 0}]
        {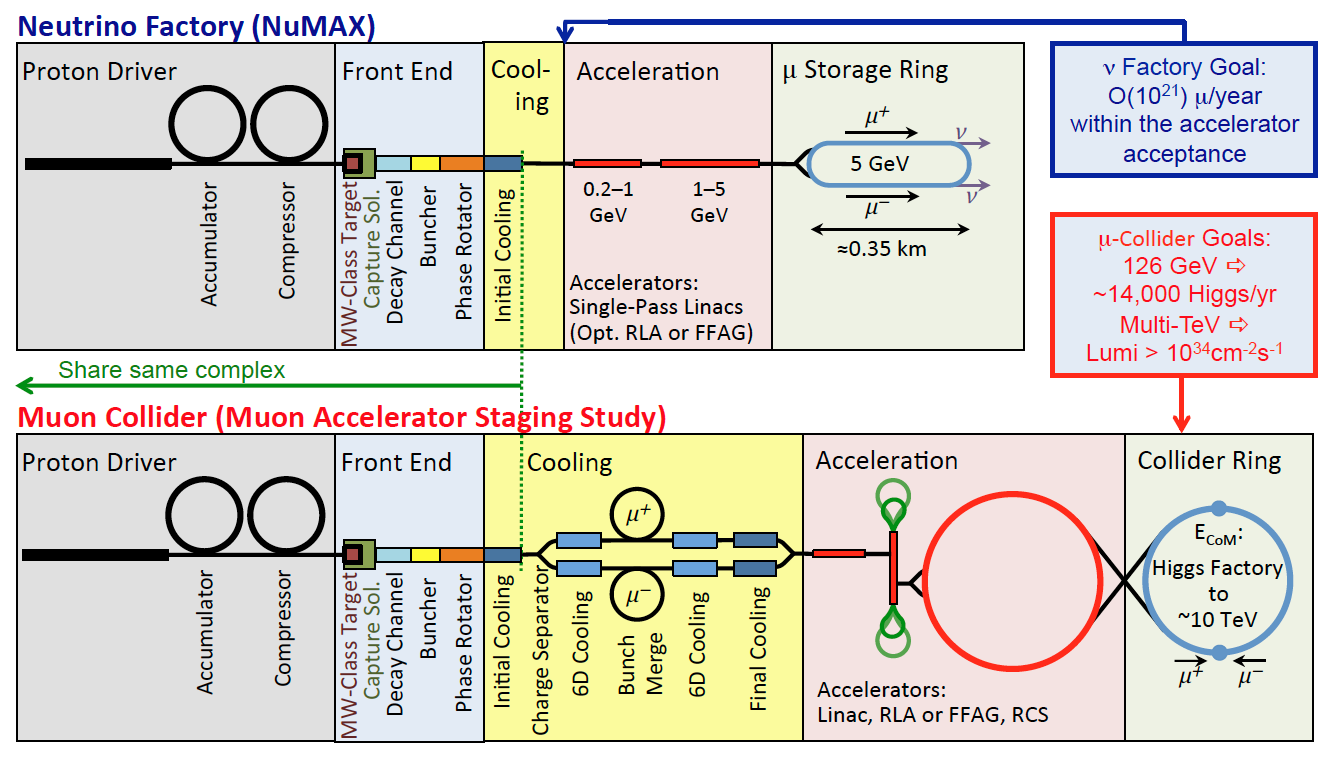}}
        \centering\textbf{}

\caption{A conceptual scheme for the Muon Collider and the NuMax Neutrino Factory. Different stages of the muon production, cooling and acceleration are shown. Various options of accelerating muon bunches to the desired energy (RCS, RLA, etc) are also indicated. Parts of the complex up to initial cooling are common to the Muon Collider and the Neutrino Factory.}
\label{fig_muon:sketch}
\end{figure}

Although muons offer many potential physics benefits, their use brings substantial complications as well. Indeed, if intense muon beams were easy to produce, they would already be available. Firstly, muons are created as a tertiary beam. The commonly proposed production scheme uses a proton beam to bombard a high-Z target. This produces pions, which are captured in a solenoidal decay channel, where they decay to muons. To produce an acceptably large sample of muons, a multi-MW proton beam is required.
%; a typical Neutrino Factory specification is for a MW scale proton driver. 
A target system capable of tolerating such an intense beam is a substantial challenge. The capture and decay process just described gives rise to a muon beam having a large energy spread and a large transverse phase space. The large transverse phase space has several implications: (1) It favors the use of solenoidal focusing in the lower energy portions of the facility, as opposed to the more conventional quadrupole focusing. %A solenoid focuses in both planes simultaneously, avoiding the excessively large beam size in one plane when using an alternating gradient quadrupole channel.
(2) It requires a rapid mechanism for reducing the emittance to more tractable values.
%; (3) it requires a high-acceptance fast acceleration system and decay ring. 
The second major challenge of muon beams is due to the short lifetime of the muon, only 2.2 $\mu$s at rest. Clearly, the short lifetime puts a premium on very rapid beam manipulations. A fast emittance cooling technique, “ionization cooling,” is needed to reduce the transverse emittance of the muon beam, along with a very rapid acceleration system. The ionization cooling technique \cite{Budker1967AnEM,skrinsky,mumu_page,neuffer-yellow} requires high-gradient normal conducting RF cavities due to the need to immerse the cavities in a strong solenoidal magnetic field.
Finally, the decays of the muons lead to potentially severe backgrounds in the detector of a Muon Collider. There are also a number of challenges related to the magnet requirements.
In the target area, the initial capture magnet is a 20 T solenoid design. Lower field target solenoids (15~T) have also been considered and provide a tradeoff between cost/feasibility and performance. In the cooling channel, large aperture magnets having relatively low field, up to 2--3 T and 1.5-m diameter, are utilized. As the beam emittance is reduced through cooling, higher field solenoids with lower diameter bores are needed. In the final cooling stages of a Muon Collider, very high strength solenoids, up to $\sim$~35--50 T, are required. %In the acceleration system, solenoids with very low fringe fields are needed to permit operation of nearby superconducting RF cavities.
In the acceleration system and collider ring, shielded dipoles are needed to accommodate the high heat load from muon decay electrons.

MAP developed the concept shown in Fig.~\ref{fig_muon:sketch}. The proton complex produces a short, high intensity
proton pulse that hits a target and produces pions. The decay channel guides the pions and
collects the muons produced in their decay into a buncher and phase rotator system to form a muon
beam. Several cooling stages then reduce the longitudinal and transverse emittance of the beam using
a sequence of absorbers and RF cavities in a high magnetic field. A linac and two recirculating linacs
accelerate the beams to 60 GeV. One or more rings accelerate the beams to the final energy. As the
beam is accelerated, the lifetime in the lab frame increases due to relativistic time dilation so later stage
accelerators have proportionally more time for acceleration, so that rapid-cycling or fast pulsed synchrotrons can be used.
Fixed-Field Alternating-gradient (FFA) accelerators are an interesting alternative. Finally the two single bunch
beams are injected at full energy into the collider ring to produce collisions at two interaction
points.

The MAP feasibility assessment demonstrated sufficient progress in several key areas to suggest that muon colliders are a viable technology for a high-energy collider. Several issues were identified as part of the MAP Feasibility Assessment that had the greatest risks for the realization of a muon collider concept. These included: operation of RF cavities in high magnetic fields in the front end and cooling channel; development of a 6D cooling lattice design consistent with realistic magnet, absorber, and RF cavity specifications; a direct demonstration and measurement of the ionization-cooling process; development of very-high-field solenoids to achieve the emittance goals of the Final Cooling system; and demonstration of fast-ramping magnets to enable RCS capability for acceleration to the TeV scale.

While other machine design and engineering conceptual efforts were pursued to develop the overall
definition of a muon collider facility, research in the above feasibility areas received the greatest attention as part of the MAP effort.
An important outcome of MAP was that progress in each of the above areas was sufficient to
suggest that there exists a viable path forward.

Parameter sets for the primary collider energy options considered by the Forum are derived from the MAP and IMCC studies and are summarized in Table~\ref{tab:accparameters}.

\begin{table}[]
    \centering
    \begin{tabular}{c|c|c|c|c}
     \hline
                  Parameter & Unit & Higgs Factory & 3 TeV & 10 TeV \\
    \hline
    COM Beam Energy & TeV & 0.126 & 3  &  10\\
    Collider Ring Circumference & km & 0.3 & 4.5 & 10\\
    Interaction Regions &   & 1 & 2 & 2\\
    Est.~Integ.~Luminosity & ab$^-1$/year & 0.002 & 0.4 & 4\\
    Peak~Luminosity & $10^{34}$ cm$^{-2}$ s$^{-1}$ & 0.01 & 1.8 & 20\\
    Repetition rate & Hz & 15 & 5 & 5\\
    Time between collisions & $\mu s$ & 1 & 15 & 33\\
    Bunch length, rms & mm & 63 & 5 & 1.5\\
    IP beam size $\sigma^{*}$, rms & $\mu m$ & 75 & 3 & 0.9\\
    Emittance (trans), rms & mm-mrad & 200 & 25 & 25\\
    $\beta$ function at IP & cm & 1.7 & 0.5 & 0.15\\
    RF Frequency & MHz & 325/1300 & 325/1300 & 325/1300\\
    Bunches per beam  & & 1 & 1 & 1\\
    Plug power & MW & $\sim 200$ & $\sim 230$ & $\sim 300$\\
    Muons per bunch & $10^{12}$ & 4 & 2.2 & 1.8\\
    Average field in ring & T & 4.4 & 7 & 10.5\\
    \hline
    \end{tabular}
    \caption{A summary of parameters for the primary muon collider options considered by the Forum.}
    \label{tab:accparameters}
\end{table}

%% file: feasibility.tex
\subsection{Feasibility statement}

%% file: proton_linac.tex
\subsubsection{Proton Source}

A muon collider requires a high intensity proton source that produces intense pulses of protons that can be focused onto a target to produce pions that decay into muons. The general parameter requirements developed by MAP are a multi-MW source of multi-GeV protons producing pulses at a 5--15 Hz rate. Multi-MW proton sources have been and are being produced for 
spallation neutron sources and neutrino sources (SNS~\cite{SNS}, ESS~\cite{ESS}, J-PARC~\cite{JPARC}, Fermilab~\cite{Fermilab}). Proton sources of 
the required intensity are within the capabilities of existing technology. The R$\&$D challenge is to adapt and extend existing facilities to muon collider requirements, or design an optimized new source. 

In the MAP program, an 8 GeV CW SRF linac capable of 4 MW was presented as the baseline proton source, which accelerated H$^{-}$ beam into an accumulator ring using charge exchange injection. 
This was based upon the Project X design~\cite{ICD2}. The initial part of that design is being constructed at Fermilab as PIP-II \cite{pip2pdr}, an 0.8 GeV linac capable of 1.6 MW in CW operation. As discussed in Snowmass white papers, PIP-II can be extended to higher energies using a linac-based upgrade \cite{Belomestnykh2021} or a linac plus RCS system \cite{Ainsworth22}. These upgrades could be developed into proton drivers for a muon collider. 

%% file: accumulation.tex
\subsubsection{Accumulation and Compression}

The high power beam from the proton source must be collected into short intense bunches on target for a collider.
In the MAP scenario, 8 GeV H$^-$ beam from a 4 MW Linac was accumulated in a small number of bunches in an Accumulator Ring (AR) at 15 Hz rep rate \cite{Alexahin:2012ipac}. The bunches are then transferred to a Compressor Ring (CR) for compression to $\sim 1$ m rms bunch
lengths. Simultaneous delivery of all of the
bunches from the CR onto the target can be obtained by directing each bunch through separate
transport lengths in a ``trombone" transport configuration onto the target. In Ref.~\cite{Alexahin:2012ipac}, separate lattices of $\sim 300$~m circumference for the Accumulator and Compressor are presented, along with simulations of bunch compression. A single ring combining accumulation and compression functions could also be considered.

The general concept can be implemented with other high power proton source configurations.
Output beam from a suitable RCS or FFA could also be bunched and compressed in a ring transport. An essential part of the proton source R$\&$D will be adapting the source to obtain the compressed bunches needed for the collider scenario.

%% file: targets.tex
\subsubsection{Targets}

%LBNF DUNE -- 1.2MW, T2K -- 1MW -- new \\
%Solenoid: ITER solenoid comparable -- new  \\
%With the US magnet development program \\
%Material R\&D 

%%% Introduction for the MC target system
%%% 
%The MAP demonstrated the proof-of-principle of the muon collider target system. %cite MERIT
%However, the system lifetime will be short due to severe radiation damages. It will be a very expensive device. 
The proposed R\&D criteria for the target system for muon colliders are high pion/muon production yields and high tolerances of the thermal stress caused by beam impact. 
This is the extended goal of high power target (HPT) systems for future neutrino facilities, such as the LBNF at Fermilab~\cite{DUNE:2016hlj} and the upgraded Tokai to Kamioka (T2K) facility at J-PARC~\cite{T2K:2011qtm}. 
Neutrino target technology has matured to the 1 megawatt (MW) level which will be in operation at Fermilab and J-PARC in the next decade. 
The real challenge for the neutrino community is to produce the next-generation 2.4 MW LBNF target system.
% which will be the  target system. 
A key difference between MC and neutrino target systems is in the thermal stress on the system, because of the differing time structures of the incident proton beams. 
The proton beam for the neutrino target is
multi-bunched within a microsecond (the bunch length is a few to a few hundred nanoseconds) and the beam repetition rate is around 1 Hz so as to generate a high integrated neutrino flux at a neutrino detector, while the proton beam for the MC will be single bunched with a 1-m-long rms bunch length and a beam repetition rate of 5--15 Hz to generate a high luminosity at an interaction region. 
As a result, the instantaneous thermal stress in the MC target system is several orders of magnitude higher than in the neutrino target system. 
Moreover, the repetition rate of the thermal stress cycle on the MC target is an order of magnitude higher than that of the neutrino target. 
%The advanced target technology R\&D is required for the MC target system which has higher mechanical stress and damage tolerances than the neutrino target. 

%%% Material science 
The international collaboration, Radiation Damage In Accelerator Target Environments (RaDIATE), has been carrying out R\&D on target materials for high power beam facilities since 2012~\cite{radiate}. %cite "Radiation Damage In Accelerator Target Environments," [Online]. Available: https://radiate.fnal.gov.
Various sorts of material response  (mechanical, chemical and physical) have been studied using high energy beams and radiation. 
They propose to extend the study of novel materials for multi-MW beam facilities~\cite{kavin2022}. %cite Kavin Ammigan, Sujit Bidhar, Frederique Pellemoine, Vitaly Pronskikh, et al. ”Novel Materials and Concepts for Next-Generation High Power Target Applications”, arXiv:2203.08357 [physics.acc-ph] (pdf).
These materials include high-entropy alloys (HEA), electrospun nanofiber materials, SiC-coated graphite and SiC-SiC composites,  toughened fine-grained recrystallized (TFGR) tungsten, dual-phase titanium alloys, and advanced graphitic materials. 
In addition, liquid and highly granular targets, such as fluidized tungsten powder, will be studied. 
The MC target design effort will be part of the HPT material science. 

%%% Simulation
Previously observed particle production measurements have been compiled as a ``big data" set. 
This is used for calibrating particle simulation codes such as MCNP~\cite{Briesmeister:1993zz}, FLUKA~\cite{Ferrari:2005zk}, MARS~\cite{Mokhov:1995wa}, \GEANT 4~\cite{Agostinelli2002hh}, and PHITS~\cite{Sato:2018imy}. 
These simulation codes compute a radiation map and a precise production yield of secondary and tertiary particles and their time evolution in phase space in the beam transport system.
The code is also used to estimate the instantaneous thermal stress due to beam impact. Time domain heat propagation in the target material is modeled by applying another simulation code, such as Finite Element Analysis (FEA) or a fluid dynamics program. 
Often, the interface among these different types of code is an issue. 
Further development of  particle simulation codes is proposed in the white paper~\cite{charlotte2022}. %cite Charlotte Barbier, Sujit Bidhar, Marco Calviani, Jeff Dooling, et al. ”Modeling Needs for High Power Target”, arXiv:2203.04714 [physics.acc-ph] (pdf).
In additon, the application of artificial intelligence (AI) is considered for optimization of the dimensions and shape of the target~\cite{kavin2022}. % cite Kavin Ammigan, Sujit Bidhar, Frederique Pellemoine, Vitaly Pronskikh, et al. ”Novel Materials and Concepts for Next-Generation High Power Target Applications”, arXiv:2203.08357 [physics.acc-ph] (pdf). 
The MC target design group will be part of this simulation R\&D effort. 

%%% High field magnet R&D
%%% Sasha, please add your sentence below
The target area and the front end consist of a series of large-aperture high-field solenoids. The (most challenging) 20 T target solenoid is composed of a 15 T field, 2 m aperture superconducting outsert and a 5 T field, 0.3 m aperture normal conducting insert. It is followed by 12 m long decay channel solenoids with tapered apertures from 2 m to $\sim0.6$ m and field from 20 T to 2.5 T. The SC solenoid design  faces several challenges including the severe radiation environment, high field, large stored energy and magnetic forces, etc. Given the field strengths, the outer SC solenoid may use hybrid Nb$_3$Sn inner and Nb-Ti outer coils. Detailed studies may be needed to develop reinforced cryogenically stabilized superconducting cables and validate a magnet cooling design that can withstand the high heat deposition. Experience with large Nb-Ti detector solenoids and the Nb$_3$Sn 13 T ITER~\cite{ITER} central solenoid may be used. Possibilities for use of HTS cables need also to be studied.

%A pion/muon collection magnet will be optimized. 
%The baseline design of the collection magnet is a solenoid. 
%Because the target is immersed in a solenoid field, the pion/muon collection efficiency is very high. 
%The magnet needs to be a large bore and a high peak field. 
%ITER has a Nb3Sn central solenoid (CS) model coil. %cite https://www.usiter.org/us-hardware/central-solenoid
%The test CS has demonstrated which is 1.6 meter inner diameter and 3.6 meter outer diameter and generate a 13 Tesla field. 
%We will collaborate with the US high field magnet development program. 

%%% Alternate pion/muon collection design 
%Alternate pion/muon collection design was discussed in the muon collider forum workshop. 
%It is based on the magnetic horn system, which is widely used for the high power neutrino facilities, like LBNF and T2K. 
%In order to focus the both charges in the horn system, a FODO horn lattice was discussed in the muon collider forum meeting. 
%We will develop the concept. 

%%% Future target R&D facility
%Fermilab proposes the hot cell facility at the Target System Integration Building (TSIB) for the post-irradiation examinations (PIE)~\cite{frederique2022}. %cite F. Pellemoine, C. Barbier, Y. Sun, K. Ammigan, S. Bidhar, B. Zwaska, et al. ”Irradiation Facilities and Irradiation Methods for High Power Target”, arXiv:2203.08239 [physics.acc-ph] (pdf).
There is strong need for a new beam irradiation facility for target material R\&D on the Fermilab site. 
 PIP-II is a versatile machine that 
generates sufficient beam power to serve several HEP programs as well as  beam irradiation tests~\cite{jeff2022}. % cite Jeffrey Eldred, Sergei Nagaitsev, Vladimir Shiltsev, Alexander Valishev, Robert Zwaska, Michael Syphers. ”Design Considerations for Fermilab Multi-MW Proton Facility in the DUNE/LBNF era”, arXiv:2203.08276 [physics.acc-ph] (pdf). (also under NF09, RF05)
A possible alternative is an upgraded Booster, which has sufficient beam power to serve both the beam irradiation facility as well as an ionization-cooling demonstrator~\cite{john2022}. 
%cite "Physics Opportunities for the Fermilab Booster Replacement" John Arrington, Joshua Barrow, Brian Batell, Robert Bernstein, Nikita Blinov, S. J. Brice, Ray Culbertson, Patrick deNiverville, Vito Di Benedetto, Jeff Eldred, Angela Fava, Laura Fields, Alex Friedland, Andrei Gaponenko, Corrado Gatto, Stefania Gori, Roni Harnik, Richard J. Hill, Daniel M. Kaplan, Kevin J. Kelly, Mandy Kiburg, Tom Kobilarcik, Gordan Krnjaic, Gabriel Lee, B. R. Littlejohn, W. C. Louis, Pedro Machado, Anna Mazzacane, Petra Merkel, William M. Morse, David Neuffer, Evan Niner, Zarko Pavlovic, William Pellico, Ryan Plestid, Maxim Pospelov, Eric Prebys, Yannis K. Semertzidis, M. H. Shaevitz, P. Snopok, M.J. Syphers, Rex Tayloe, R. T. Thornton, Oleksandr Tomalak, M. Toups, Nhan Tran, Yu-Dai Tsai, Richard Van de Water, Katsuya Yonehara, Jacob Zettlemoyer, Yi-Ming Zhong, Robert Zwaska, arXiv:2203.03925

%% file: ionization_cooling.tex
\subsubsection{Ionization Cooling}

%Muons for neutrino factories and muon colliders are produced as a tertiary beam: protons are directed onto a target to yield a beam of pions. Those pions are then captured in a high-field solenoid and allowed to drift and decay into muons. The resulting muon beam has a very large phase space. Tightly focused (``cooled'') muon beams are desired for neutrino factories and muon colliders. Given the muon relatively short life span, ionization cooling~\cite{budker,skrinsky,mumu_page} is deemed to be the only technique fast enough to cool beams within the muon lifetime.

%Various aspects of muon ionization cooling were addressed over the last few decades, first by the Neutrino Factory and Muon Collider Collaboration (NFMCC)~\cite{nfmcc} and later by the Muon Accelerator Program (MAP)~\cite{map} described above.

A complete scheme for cooling a muon beam sufficient for use in a Muon Collider has been previously defined by MAP. The proposed scheme consists of a sequence of steps: first,  an ionization cooling channel  reduces the 6D emittance of the incoming bunch train until it can be injected into a bunch-merging system. The single muon bunches, one of each sign, are then sent through a second 6D cooling channel in which the transverse emittance is reduced as much as possible and the longitudinal emittance is cooled to a value below that needed for the collider.  If necessary, the beam can then be sent through a final 4D cooling channel using high-field solenoids that further reduces the transverse emittance while allowing the longitudinal emittance to grow. 

An important outcome of MAP was that sufficient progress was made in each of the above areas to
suggest that there exists a viable path forward. For instance, a 6D cooling lattice was designed that incorporated reasonable physical assumptions
\cite{bib_muon:MAP_Rectilinear}; a final cooling channel design, which implemented the constraint of a 30 T maximum
solenoid field, came within a factor of two of meeting the transverse emittance goal for a high energy
collider and current development efforts appear poised to deliver another factor of 1.5 improvement.
In parallel, the Muon Ionization Cooling Experiment (MICE) collaboration built and operated a section of a solenoidal cooling channel and demonstrated
the ionization cooling of muons using both liquid hydrogen and lithium hydride absorbers~\cite{MICE_nature}. This demonstration of ionization cooling is an important advance in the development of high-brightness muon beams.

Together with these MAP Studies, since the end of MAP some relevant technology R\&D has continued to progress, enhancing the promise of muon colliders as an avenue of study. In particular, new studies are required to leverage the now increased limits of RF gradient in a strong solenoidal magnetic field, which theory suggests should give an improved cooling channel performance.

%% file: rf.tex
\subsubsection{RF in Magnetic Field}
Because muons have a finite lifetime and are produced within a very large phase space, making a compact and efficient muon ionization cooling channel is essential, in which high gradient RF cavities must be placed in a strong magnetic field to compensate quickly for the muons' ionization energy loss. 
However, early RF experimental studies at the MuCool Test Area (MTA) at Fermilab showed that the probability of  RF breakdown events is significantly enhanced when cavities are operated in  external magnetic fields. As a result,  the maximum achievable RF gradients were seriously degraded. The impact on RF system performance is mainly caused by enhanced effects of multipacting (MP) and field emission due to the external magnetic field, where field emission electrons are focused by the magnetic field into beamlets that have current density high enough to damage the cavity surface material and eventually lead to RF breakdown~\cite{diktys2010}. % cite D. Stratakis, J. C. Gallardo, and R. B. Palmer, Effects of external magnetic fields on the operation of high-gradient accelerating structures, Nucl. Instrum. Methods Phys. Res., Sect. A 620, 147 (2010).
To mitigate breakdown in external magnetic fields, two types of RF cavities were proposed and experimentally studied in MAP: vacuum and high pressure gas filled cavities. 

A low-Z, refractory,  high electrical and thermal conductivity material---beryllium\,---is chosen to terminate a conventional open iris RF cavity, taking advantage of the penetrating nature of muons. As a result, the cavity has nearly a factor of two higher cavity shunt impedance---a factor of two saving in peak RF power; moreover, beryllium can withstand high peak surface field without surface damage from the magnetically focused field emission electrons. In addition, the cavity inner surface was post-processed using the techniques developed for SRF cavities, which include electropolishing (EP), dry-ice cleaning, and assembly in clean rooms. 
%The 805 MHz Beryllium wall modular cavity is demonstrated the peak gradient 50 MV/m in a 3 Tesla solenoid field. 
% cite Bowring
The experimental RF program included cavities with beryllium windows at two frequencies, 201 MHz and 805 MHz. The required RF performance was demonstrated in the available experimental conditions at the MTA, where a 201 MHz prototype cavity with thin beryllium windows for MICE achieved 14 MV/m in the fringe field of a 5 T solenoid field, limited by available peak RF power~\cite{MICE_cavity,Li:2008af}; and an 805 MHz beryllium-wall modular cavity achieved 50 MV/m within a 3 T solenoid field~\cite{modular_cavity}. 

The success of these experimental demonstrations indicates the importance of cavity design (taking external magnetic field into consideration ), cavity surface finish, and understanding of RF breakdown in magnetic field.  %Many experiences have been learnt from the R\&D of these two cavities on RF design, RF commissioning, mechanical design, and manufacturing techniques.
Although significant progress was made in understanding  RF breakdown in magnetic field, there are still open issues that need to be further explored, such as the RF gradient limit versus resonant frequency and cavity stored energy, cavity frequency versus external magnetic field, etc. Our RF breakdown physics model predicts that the breakdown probability depends on the temperature and material strength of the cavity. With the recent advancement in C$^{3}$ technology~\cite{Dasu:2022nux}, a cryogenically cooled copper cavity may have potential as the cavity option for muon ionization cooling channels in cases in which the required performance outweighs system electrical efficiency. Nevertheless, design, engineering, and high-power testing of a practical muon ionization cooling channel with RF remains  one of the most important R\&D tasks for a future muon collider. 

A dense hydrogen-gas filled RF cavity is proposed as the second method to overcome electrical RF breakdown in strong magnetic fields. Field emission electrons in the RF cavity are diffused by the dense gas via Coulomb scattering and are slowed down and prevented from cascading. Experimental studies confirmed that the maximum accelerating gradient in a gas filled RF cavity is indeed independent of the external magnetic field~\cite{hanlet2006}. % cite P. Hanlet, M. Alsharo’a, R. E. Hartline, R. P. Johnson, M. Kuchnir, K. Paul, C. M. Ankenbrandt, A. Moretti, M. Popovic, D. M. Kaplan, and K. Yonehara, in Proceedings of the 10th European Particle Accelerator Conference, Edinburgh, Scotland, 2006, (EPS-AG, Edinburgh, Scotland, 2006), p. 1364.
Moreover, dense hydrogen is an ideal ionization cooling medium because of its long radiation length and large ionization energy loss rate and has already been chosen as the preferred energy-absorber material~\cite{moses2013,ben2016}.
% cite M. Chung et al., Pressurized H2 rf Cavities in Ionizing Beams and Magnetic Fields, Phys. Rev. Lett. 111, 184802 (2013).
% cite Pressurized rf cavities in ionizing beams B. Freemire, A. V. Tollestrup, K. Yonehara, M. Chung, Y. Torun, R. P. Johnson, G. Flanagan, P. M. Hanlet, M. G. Collura, M. R. Jana, M. Leonova, A. Moretti, and T. Schwarz Phys. Rev. Accel. Beams 19, 062004 – Published 20 June 2016
In addition, the gas--plasma--beam interactions have been studied experimentally by using an intense proton beam at Fermilab and numerically by a particle-in-cell plasma simulation~\cite{kwangmin2017}.  This physics model of beam-induced-plasma RF loading has been developed and verified by experiments~\cite{ben2016}. 
A  plasma is created by the incident beam.
The studies show that this beam-induced plasma is thermalized by interactions with neutral gas on a picosecond time-scale, and the electron density %and temperature 
in the cold plasma can be minimized by adding a small amount of electronegative dopant, in order to reduce plasma loading of the cavities. The collective dynamics neutralizes the space charge in the incident beam. As a result, the incident beam is focused by a self-induced azimuthal field. This is a new plasma focusing mechanism that can be used to bring about extra focusing of muons \cite{Chung-IPAC2014,Freemire-2018}.

%% file: final_cooling.tex
\subsubsection{Reaching the Final Collider Emittances}

The cooling channels described thus far cannot reach the transverse emittance of around 25~$\mu$m (normalized) required for the beams in the collider ring. A ``final cooling" system first described by R.~Palmer involving an alternation between uniform, high-field solenoids and acceleration was designed and simulated during the MAP program~\cite{prstab-18-091001}. The beamline does not so much cool the beam as ``exchange'' emittance to reduce the transverse emittance at the cost of a longitudinal emittance increase. The design demonstrated the feasibility of the concept, but its performance was somewhat lower than expected. There are a number of  research directions that could lead to improvements in that design. One direction likely to yield some improvement would be to perform more extensive studies of the optimization of that channel. Due to the nonadiabatic nature of the system and the low energies it reaches, maintaining the beam emittance even in the conservative parts of the system is challenging, particularly in the longitudinal plane.

In the final cooling stage, a series of 50 mm aperture solenoids with magnetic field above 30 T (ideally in the range of 50 to 60 T) will be needed. These solenoids require hybrid coils with HTS and LTS sections. The key challenges of this ultrahigh-field solenoid are related to choices of the HTS material and cable, forces and stresses in the coil, large stored energy, and quench management. The parameters of this solenoid go significantly beyond available technology, and it will require considerable R\&D and demonstration, including studies and measurements of effects that are specific to ultrahigh fields.

The design in~\cite{prstab-18-091001} assumed a maximum solenoid field of 30~T; a higher field would be expected to produce lower emittances. Hybrid superconducting--resistive solenoids exist with fields as high as 45~T~\cite{ieee-tasc-13-1385}, purely superconducting solenoids exist with fields as high as 32~T~\cite{ieee-tasc-26-1}, and an R\&D project toward a 40~T purely superconducting solenoid is funded~\cite{ieee-tasc-30-1}. Advances in high-temperature superconducting magnet design lead to hopes of achieving even higher fields. We should examine the impact that such higher field magnets would have on the performance of this channel. Such  magnets can also be used to improve the performance of the cooling channel upstream of this system~\cite{summers-2000}. The impact of improved input emittances on this design should be studied, both on the channel's overall performance and to provide input on any properties of the input distribution that would improve performance.

There are other ideas for replacing all or part of this last solenoid channel, or improving the beam characteristics entering that channel. One of the most frequently discussed is parametric resonance ionization cooling (PIC)~\cite{maloney-2013}, which relies on enhancing the beam divergence by inducing a half-integer resonance at an ionization absorber. Therefore, it does not require a strong focusing field. % cite final cooling white paper
While in principle this channel can provide very small transverse emittances, the practical implementation faces challenges in managing chromatic and nonlinear aberrations~\cite{maloney-2013}. We believe there are still possibilities for managing these aberrations, and the potential performance of PIC warrants further theoretical studies. 

There are other ideas for reaching the required collider emittances that are less well-developed, such as plasma focusing~\cite{jinst-15-p03004} and the slicing and recombining of beam distributions at higher energies~\cite{arxiv-1504-03972}.

%A near-term research program could focus on
%\begin{itemize}
%    \item Further improvements in the high-field solenoid cooling channel, incorporating both more extensive simulation studies and expected improvements from more advanced magnet technologies, and
%    \item Further simulation studies of PIC to achieve acceptable performance or rule out its utility.
%\end{itemize}
% Magnetic field, potential for improvement of\\
% Tallahassee 32T magnets in regular ops with 50 mm

%% file: acceleration.tex
\subsubsection{Acceleration}

Acceleration to moderate energies (10s of GeV) can be accomplished by the relatively conventional means of linacs and recirculating linear accelerators (RLAs). The MAP program proposed concrete designs for accelerating to 5~GeV, and suggested how an RLA could accelerate to 63~GeV \cite{jinst-2018-p02002,arxiv-1308.0494,bnl-105417-2014-ir,bogacz-2021}. What was not studied in detail was acceleration from the final cooling systems to an energy of around 200~MeV. We expect this to use systems similar to those used in the cooling system, but without the requirements of high magnetic fields.

Reaching higher energies would ideally be done via pulsed synchrotrons. Rapid cycling synchrotrons are extensively utilized in existing accelerators, but the requirements for acceleration in muon colliders differ in two important aspects: short (ms-scale) acceleration cycles with a relatively low repetition rate (around 10~Hz), thus ``pulsed'' rather than ``rapid cycling''; and the requirement for a high average bend field. To achieve a high average bend field, fixed field superconducting magnets are interleaved with bipolar pulsed iron dipoles, denoted a ``hybrid'' pulsed synchrotron. Pulsed iron magnets can achieve fields of 1.5~T with manageable power consumption~\cite{aipcp-1777-100002}; they could reach somewhat higher fields at the cost of increased power losses. Conventional copper coils can be used and make a relatively small contribution to the power losses, and small magnets have been built demonstrating the feasibility of the required short pulsed times~\cite{ipac2012-3542}. HTS coils can be used instead if that should prove more cost-effective, and have demonstrated significant ramping rates~\cite{ieee-tas-32-1}. If that technology could be adapted to create pulsed fields well beyond what could be achieved with iron-dominated magnets, that could significantly improve the achievable energy and reduce the number of stages of pulsed synchrotron acceleration. Providing power to the pulsed magnets can be challenging and costly~\cite{brauchli-2021}. A single power supply can likely drive only a small number of magnets. The peak power delivered is high and, to have reasonable average power requirements, a very large fraction of the energy delivered to the magnets in a pulse must be recovered. To maintain good beam properties, the pulse shape must be well-controlled and reproducible. No detailed design for such a power supply currently exists.

The impact of collective effects in acceleration systems must be assessed~\cite{metral-2021}. Accelerators are not duplicated for each bunch sign, but rather the two charges counter-rotate in each system, thereby colliding twice per turn. Those beams also pass through RF cavities, leaving significant wakes due to the high bunch charges. Since high-frequency RF is desirable for cost and efficiency, the energy extracted by a bunch from a cavity on each pass can be a significant fraction of the cavity's stored energy, which would be a new way of operating an RF cavity.

%A near-term research program focused on understanding the capabilities of a muon collider could study
%\begin{itemize}
%    \item Acceleration from a final cooling stage to the initial energy of the MAP acceleration scenario, with goals of having a more general description of the acceleration scenario, an approximate idea of achievable muon transmission, and studying emittance preservation, particularly in the longitudinal plane.
%    \item A detailed pulsed acceleration lattice design that would fit on the Fermilab site. The goal would be to have sufficient detail in the design to be able to specify reasonable confidence the maximum energy that could be achieved for a muon collider on the Fermilab site given certain technology choices.
%    \item FFA acceleration designs, both for moderate and maximum energies, with particular emphasis on advanced designs (such as nonlinear and vertical FFAs), to understand if and for what parameter ranges they may be more favorable than other types of accelerators.
%    \item A study of the collective effects in acceleration systems, to asses their impact on the beam distribution, and identify any required mitigates and their impacts.
%\end{itemize}

% MAP proof of principle: Lattice 1.5\,T 400\,Hz \\
% HTS gives alternatives 0.5\,T 300\,T/s, conductor development \\
% Power Supply technically considered feasible, question power eff/cost/control

%% file: collider_ring.tex
\subsubsection{Collider Ring}

This section briefly summarizes the submitted white papers~\cite{Alexahin:2022tav,alexahin:2022ztv} and our previous studies of MC SR and Interaction Regions (IR)~\cite{Ankenbrandt:1999cta,Mokhov:2018fyn,Zlobin:2018ayy,Kashikhin:2012zza,Mokhov:2011zzd} which brought together in a coherent form the results of the previous studies on a Muon Collider and presented a design concept of the 3 and 6 TeV MC optics, the conceptual design and parameters of large-aperture high-field superconducting (SC) magnets, and a preliminary design and analysis of the protection system to substantially reduce radiation loads on the magnets as well as particle backgrounds in the collider detector. The SC magnets and detector protection considerations impose strict limitations on the lattice choice, hence the designs of the collider optics, SR and IR magnets and Machine-Detector Interface (MDI) are closely intertwined.

For a 3~TeV Muon Collider with $\beta^*$ of 5 mm it was possible to achieve the target design parameters with either a triplet or a quadruplet Final Focus (FF) with moderate strength quadrupoles. The momentum compaction factor for a standalone arc cell is $\alpha_c = -0.004$. Each arc consists of six such cells and two dispersion suppressors. The betatron phase advance is 300$^\circ$ in both planes to ensure cancellation of spherical aberrations. This phase advance includes the aforeemntioned 6 cells and the two dispersion suppressors.  Though the sextupoles for chromaticity correction are interleaved they are too weak to noticeably affect the dynamic aperture. In the context of this paper, dynamic aperture refers to as the minimum amplitude boundary where unbounded particles become unstable and are getting lost. 

\begin{figure}
\begin{center}
\includegraphics[width=0.7\textwidth,trim={0 0 0 5}]{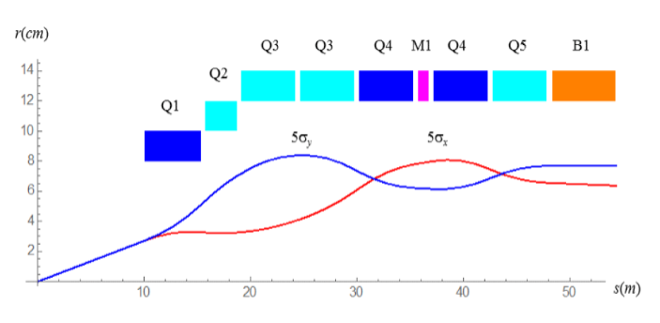}
\includegraphics[width=0.7\textwidth]{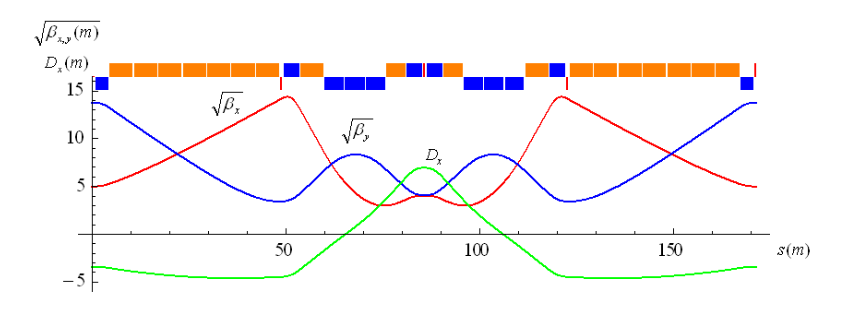}
\caption{Top: IR layout of the 6 TeV COM collider with 5$\sigma$ envelopes and quadrupole apertures. In cyan are shown the quadrupoles with up to 5 T dipole component. Bottom: 3 TeV arc cell concept.}
\label{mc_ring_Fig1}
\end{center}
\end{figure}

In the 6 TeV COM MC, as a first approximation we use the IR design with $\beta^*$ of 3 mm described in \cite{Alexahin:2018svu}, whereas for the arcs we rescale the arc cell design of the 3 TeV collider~\cite{Alexahin:2018svu}. The concepts of the 6 TeV IR layout and the 3 TeV arc cell are shown in Fig.~\ref{mc_ring_Fig1}. The next step toward a consistent design is to build the complete ring including utility sections for $\beta^*$ tuning and RF cavities. Chromaticity correction, crucial for achieving sufficient dynamic aperture and large momentum acceptance in the presence of very small $\beta^*$, will be based on the concepts developed in the previous studies~\cite{Alexahin:2018svu,Wang:2015yyh}.

The baseline 3 TeV collider is based on 150-
mm aperture dipoles. In the arcs combined function magnets are used. More specifically, the focusing magnets include 8 T field with a gradient of 85 T/m and defocusing magnets with 9 T field with a gradient of -35 T/m as well as pure dipoles of 10.4 T.   
In the IR the quadrupoles start at a gradient of 250 T/m with 90 mm aperture and proportionally lower gradient for larger aperture
quadrupoles. The magnets are based on traditional cos-theta coil geometry and Nb$_3$Sn superconductor and were used to provide realistic field maps for the analysis and optimization of the arc lattice and IR design, as well as for studies of beam dynamics and magnet protection against radiation. 

The 6 TeV IR design assumed use of HTS technology to achieve 20 T nominal operation fields in dipoles and 16 T pole tip fields in quadrupoles. The magnet inner bore diameter is required to be at least 5$\sigma_{x,y}$, where $\sigma$ is the rms beam size, plus 3 cm for absorbers or 24 cm. The flexible momentum compaction arc cells compensate the large positive contribution of the IR to $\alpha_c$. To mitigate neutrino induced radiation as well as energy deposition by decay electrons, the arc cell quadrupoles are combined-function magnets with dipole fields of about 9 T.

An important issue to address is the stress management in SC magnet coils to avoid substantial degradation or even damage of the brittle Nb$_3$Sn and/or HTS superconductor. Stress management concepts for shell-type coils are being studied experimentally for high-field accelerator magnets based on LTS (Nb$_3$Sn)~\cite{Novitski:2022hbw} and HTS (Bi2212 and ReBCO) cables~\cite{2056715,Kashikhin:2019uee}.

In the assumed IR designs, the dipoles close to the Interaction Point (IP) and tungsten masks in each interconnect region (needed to protect the magnets) help in reducing background particle fluxes in the detector by a substantial factor. The tungsten nozzles in the 6 to 600 cm region from the IP, assisted by the detector solenoid field, allow trapping of most of the decay electrons close to the IP as well as most of the incoherent $e^+e^-$ pairs generated in the IP. Analysis shows that with sophisticated tungsten, iron, concrete and borated polyethylene shielding in the MDI region, a total reduction of background loads between 2 and 3 orders of magnitude can be achieved.

\begin{figure}
\begin{center}
\includegraphics[width=0.8\textwidth]{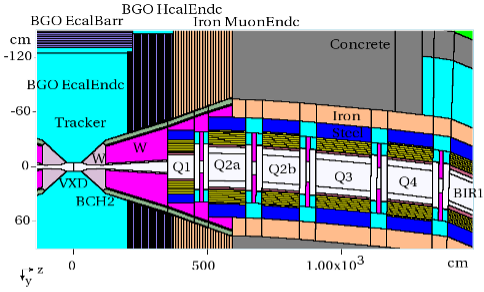}
\caption{MARS15 MDI model with tungsten nozzles on each side of IP, tungsten masks in interconnect regions and tungsten liners inside each magnet.}
\label{mc_ring_Fig2}
\end{center}
\end{figure}

An optimized MDI from the thorough MARS15~\cite{mars2018} Monte Carlo simulation study of \cite{Mokhov:2018fyn} is shown in Fig.~\ref{mc_ring_Fig2} with a sophisticated tungsten nozzle clad with borated polyethylene on each side of the interaction point, massive steel/concrete shielding around, tungsten masks in  magnet interconnect regions, and steel and tungsten liners inside each IR magnet.

All the most critical concerns of the MC optics design, magnets, and radiation protection have not only been conceptually resolved but also addressed in sufficient detail for a 3--6 TeV COM muon collider which can be considered either for the International Muon Collider Collaboration or as a Fermilab site-filler. The R\&D program would include extension of these studies to 10 TeV and higher energies.

A low-energy medium-luminosity MC has  also been studied \cite{alexahin:2022ztv,Mokhov:2018fyn} and discussed as a possible Higgs Factory (HF). It was shown that electrons from muon decays will deposit more than 300 kW in the HF SC magnets. This imposes significant challenges on SC magnets used in the HF SR and IR. Magnet design concepts were proposed that provide high operating gradient and magnetic field in a large aperture to accommodate the large size of muon beams (due to the expected large transverse emittance), as well as a cooling system to intercept the large heat deposition from the showers induced by decay electrons. The distribution of heat deposition in the MC SR lattice elements requires large-aperture magnets to accommodate thick high-Z absorbers to protect the SC coils. Based on the developed MARS15 \cite{mars2018} model and intensive simulations, a sophisticated radiation protection system was designed for the collider SR and IR to bring the peak power density in the superconducting coils below the quench limit and reduce the dynamic heat deposition in the cold mass by a factor of 100. The system consists of tight tungsten masks in the magnet interconnect regions and elliptical tungsten liners in the magnet aperture optimized individually for each magnet. These also reduce the background particle fluxes in the collider detector.

%% file: neutrino_flux.tex
\subsubsection{Neutrino Flux} 
%Talk about annual doses\\
%Magnets on movers 10-100\\
%Minimize straight sections (MAP -feasible)\\
%Beam Wobble\\
%Depth\\
%Big picture: Physics is pushing us towards higher energy, collider design also pushes towards higher energy. 
%Neutrino beams are more ``intense” at higher energies.

One of the important issues~\cite{Alexahin:2022tav}to consider when designing a muon collider is a potentially elevated flux of charged particles generated by neutrino interactions with soil and building materials at very large distances from MC~\cite{Ankenbrandt:1999cta,Johnstone:1997zu,King:1999rr,Mokhov:1999rk,Mokhov:2000gt}. In this section we describe the problem and present various ways to mitigate it. According to the latest studies, these mitigation strategies can reduce the dose to below the Fermilab annual off-site regulatory limit.

Intense highly collimated neutrino beams, created from muon decays in the ring and various straight sections of a high-energy MC, can cause elevated radiation doses even at very large distances from the machine. The more energetic decay neutrinos emanate radially outward from the collider ring at angles with respect to the muon direction of order $\theta_\nu=1/\gamma_\mu=m_\mu/E_\mu\simeq10^{-4}/E_\mu\text{[TeV]}$.
\begin{figure}[!htbp]
    \centering
    \includegraphics[width=0.5\linewidth]{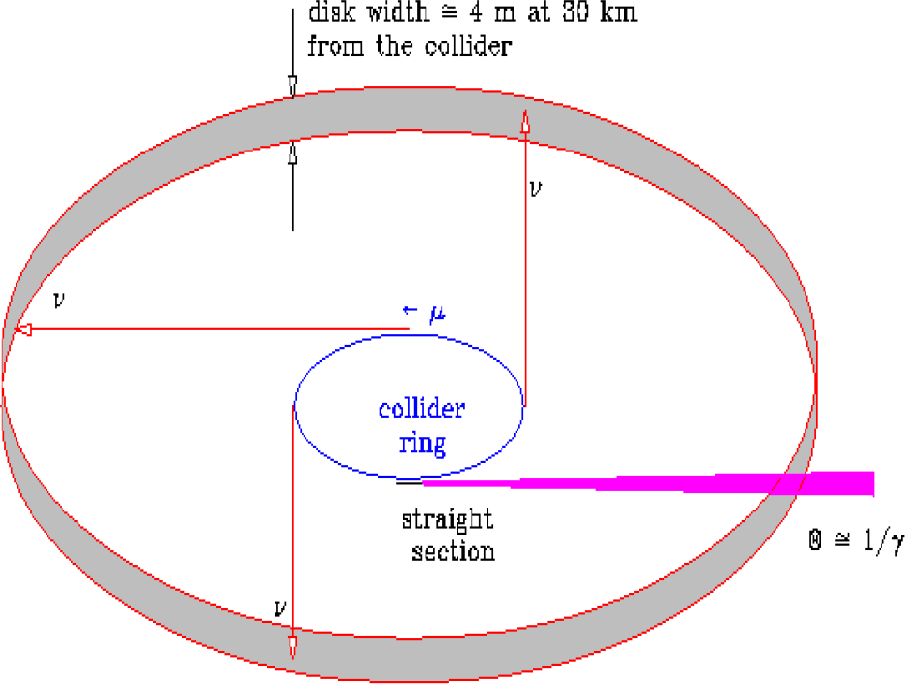}\\
    \includegraphics[width=0.5\linewidth]{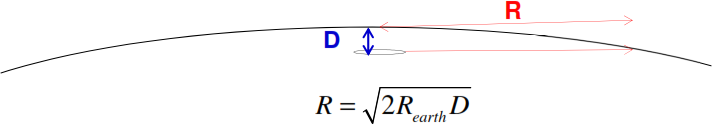}
    \caption{Intense highly collimated neutrino fluxes around MC~\cite{King:1999rr}}
    \label{fig:Mokhov1}
\end{figure}

Neutrino flux and dose per neutrino at a given location from muon colliders grow with muon energy roughly as $E_\mu^3$ due to increase with energy of the neutrino cross section, growth of the total deposited energy, and collimation of the decay neutrinos (each responsible for a factor of $E_\mu$).

This will strongly impact siting issues and cost of a high energy muon collider and needs to be taken seriously in evaluating long-term averaged neutrino flux and resulting dose. Developed in~\cite{Mokhov:2000gt}, a weighted neutrino-interaction generator for the MARS Monte Carlo code~\cite{mars2018} permitted detailed simulations of the interactions with matter of neutrinos and of their progeny in and around an MC, capable of modeling neutrinos in the energy range from 10~MeV to 10~TeV.

The model~\cite{Mokhov:2000gt} serves to represent energies and angles of the particles emanating from a simulated interaction. These particles, along with the showers initiated by them, are then further processed by the MARS15 code~\cite{mars2018} which calculates, e.g., energy deposition, absorbed, and effective dose as a function of location in a user specified geometry model. Effective dose, caused by charged particles from neutrino interactions, is calculated with particle- and energy-dependent quality factors taken into account. Muon and electron neutrinos and their antiparticles are included and distinguished throughout, which are represented in the decays from MC in roughly equal amounts. The MARS15 model identifies charged and neutral current deep inelastic neutrino and antineutrino interactions with nuclei as the dominant channels forming the main contributions to the dose from neutrino interactions. Besides that, the model accurately describes neutrino-nucleon elastic and quasi-elastic scattering, interactions with atomic electrons, and coherent elastic scattering. In the latter, a Pauli form-factor of quarks---topological fluctuations of the QCD vacuum---is included (as a weight) to discourage small $\lvert q^2\rvert$ insufficient to liberate a nucleon or promote the nucleus to an excited state.

Extremely low interaction and scattering probabilities mean that neutrinos travel essentially in a straight line and survive over enormous distances. Much like neutrons and gammas, neutrinos by themselves cause little or no biological damage but instead create charged particles which in turn deposit their energy in tissue to be interpreted as dose ``due to neutrinos.'' ``Neutrino'' dose is thus by charged particles generated by neutrinos upstream of a human. Therefore, radiological neutrino-induced effects around MC cause negligible effects for anyone above the ground and/or in an above-ground building, but can create unacceptably high radiation levels for a person lying stationary in a basement room for a year.

Dose to a human body strongly depends on neutrino energy and presence of material immediately upstream of that body. Total whole-body effective dose in a bare seated person (non-equilibrium) is lower than that in one embedded in infinite soil (equilibrium). The whole-body dose is a factor of~2 lower than the maximum dose, because a neutrino flux footprint could be smaller than typical human dimensions. The equilibrium dose is achieved after 3--4~m of soil or concrete at all neutrino energies considered here.

\begin{figure}[!htbp]
    \centering
    \includegraphics[width=0.5\linewidth]{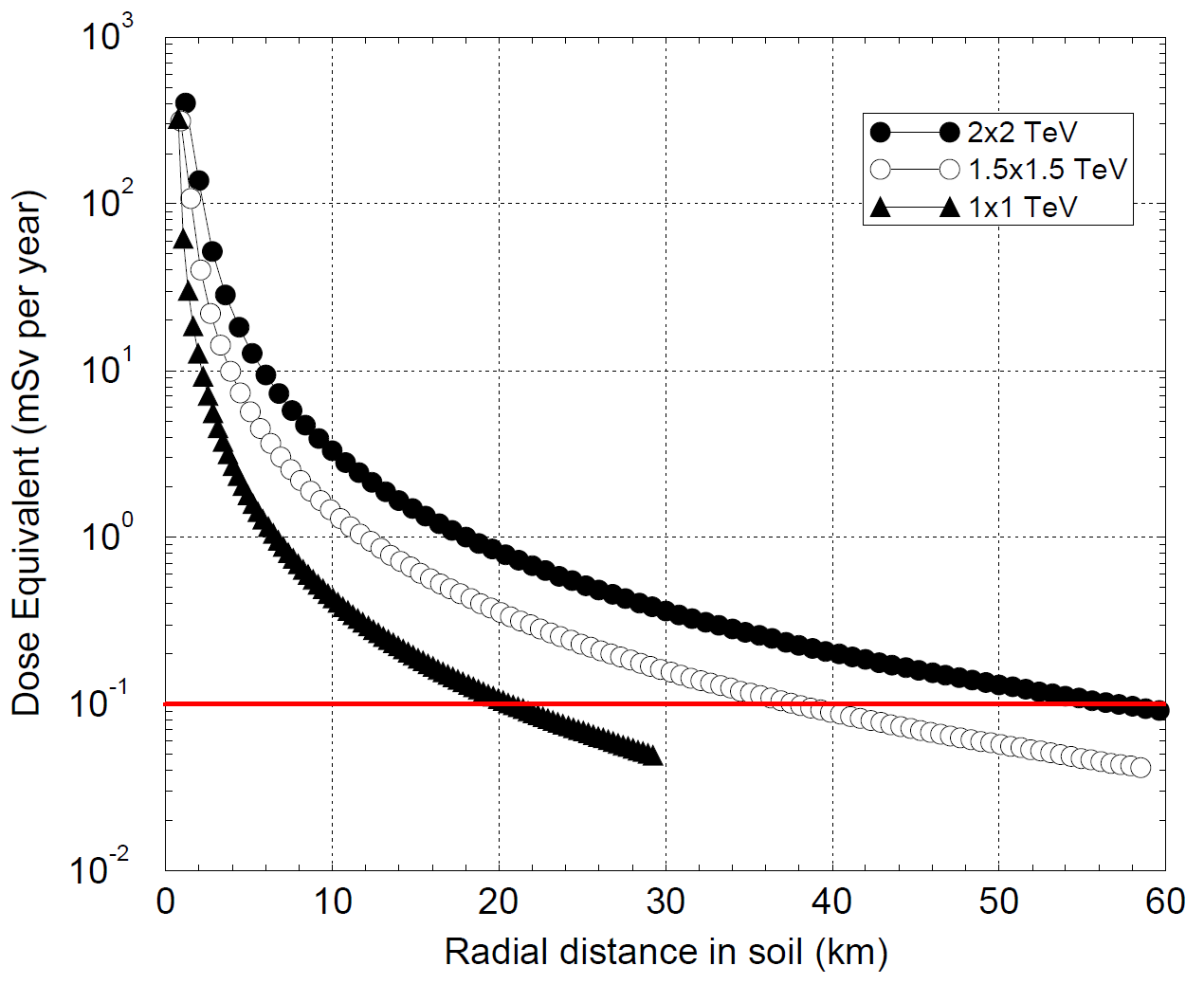}%
    \includegraphics[width=0.5\linewidth]{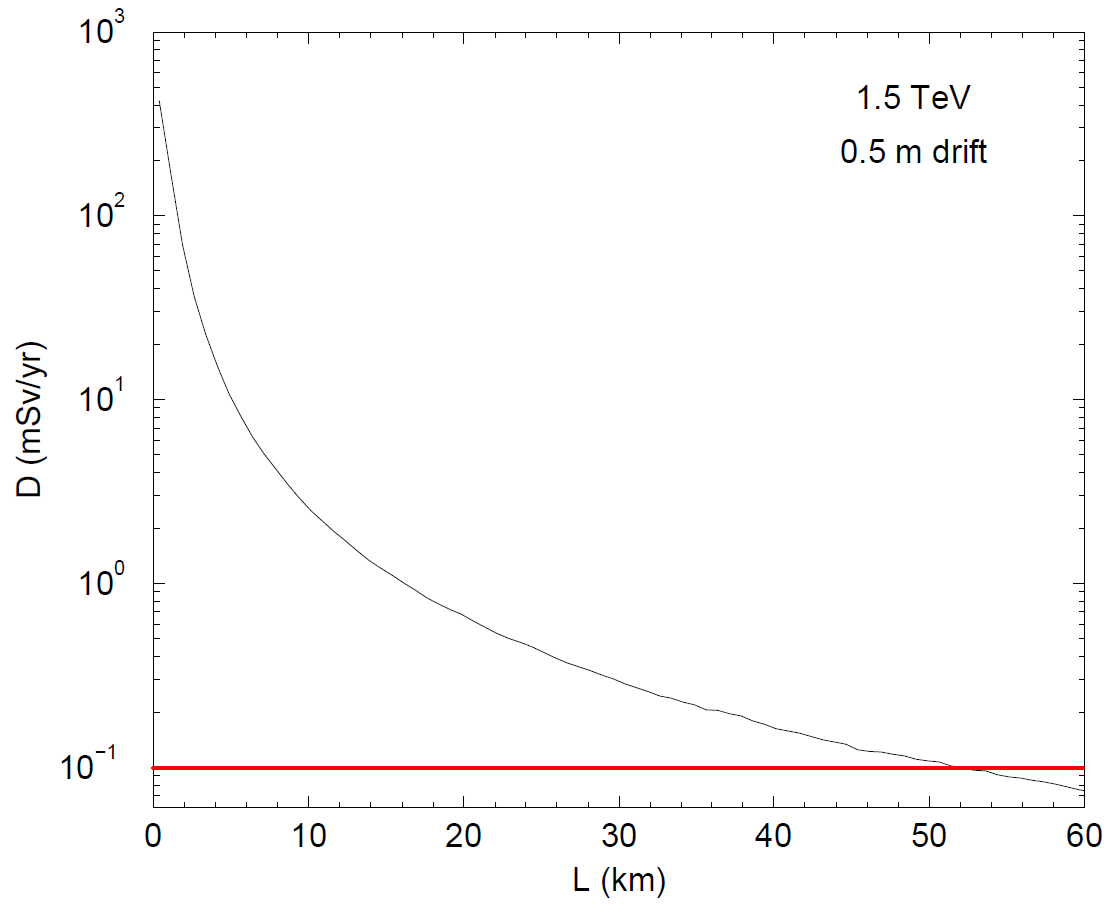}
    \caption{Left: Dose in the orbit plane vs radial distance from the ring center of 2, 3 and 4~TeV COM MC. Right: Dose downstream of a 0.5~m drift with $2.6\times10^{16}$~decays/yr of a 1.5~TeV muon beam there vs. distance in soil downstream of the drift. Red line shows  annual off-site regulatory limit at Fermilab, that is, $10\text{ mrem/year} = 0.1\text{ mSv/year}$~\cite{Mokhov:2000gt}.}
    \label{fig:Mokhov2}
\end{figure}

Instead of providing shielding, the presence of soil/concrete upstream enhances the dose by a factor of up to 1000 in the TeV region as compared to the case with no shielding. The dose can be higher than the annual off-site regulatory limits, e.g., 10~mrem/year at Fermilab. Contrary to conventional irradiation, the use of  high-Z shielding in front of a ``human'' can increase neutrino-related dose by a factor of ten compared to  low-Z shielding at low neutrino energies, while the values converge in the TeV energy range. Fig.~\ref{fig:Mokhov2} (left) shows MARS15 calculated dose around the 2, 3 and 4~TeV COM MC rings in the orbit plane with $1.2\times10^{21}$~decays/yr vs distance in soil from the ring center, while Fig.~\ref{fig:Mokhov2} (right) shows the annual dose as a function of soil thickness downstream of a 0.5~m drift with $2.6\times10^{16}$~decays/yr of a 1.5~TeV muon beam. One sees that for the 4~TeV COM MC, the dose drops below the regulatory limit at a radial distance of 60~km from the ring center, and at 55~km downstream of the 0.5~m drift with 1.5~TeV muon beam decaying there.

The most obvious way to reduce neutrino fluxes and mitigate radiological issues at large distances from the collider ring is to place the ring deep underground. MARS-calculated depths D along with distances R (see Fig.~\ref{fig:Mokhov1}) are shown in Table~\ref{tab:Mokhov-t1}~\cite{Mokhov:2000gt} for five COM energies from 0.5 to 4~TeV and two annual off-site limits $1\text{ mSv}=100\text{ mrem}$ and $0.1\text{ mSv} =10\text{ mrem}$ (Fermilab), all for N muon decays/yr. Results here were obtained assuming suppressed contribution from the field-free regions. %There is also the regulatory question whether delivering an off-site dose above the limit at any depth underground or height above it is permissible.

\begin{table}[htbp]
    \centering
    \begin{tabular}{c|c|ccccc}
    &E COM TeV&0.5&1&2&3&4\\\hline
Limit&$N\times10^{21}$&0.2&0.2&1.2&1.2&1.2\\
1 mSv/yr&R (km)&0.4&1.1&6.5&12&18\\
1 mSv/yr&D (m)&$\leq$1&$\leq$1&3.3&11&25\\
0.1 mSv/yr&R (km)&1.2&3.2&21&37&57\\
0.1 mSv/yr&D (m)&$\leq$1&$\leq$1&34&107&254\\
0.01 mSv/yr&D (m)&--&--&--&300&--
    \end{tabular}
    \caption{Required depths D, radial distances R and off-site dose limits for several scenarios of muon colliders~\cite{Mokhov:2000gt}}
    \label{tab:Mokhov-t1}
\end{table}
Note that simplified expressions derived in~\cite{Ankenbrandt:1999cta,King:1999rr} give noticeably more conservative results compared to those from MARS full Monte Carlo~\cite{Mokhov:2000gt}. For example, for the 3~TeV case, depth to stay within 0.1~mSv/yr (1\% of the DOE limit) is 300~m (see Table~\ref{tab:Mokhov-t1}) compared to the analytical depth of 500~m~\cite{Ankenbrandt:1999cta}.

\begin{figure}[!htbp]
    \centering
    \includegraphics[width=0.8\linewidth]{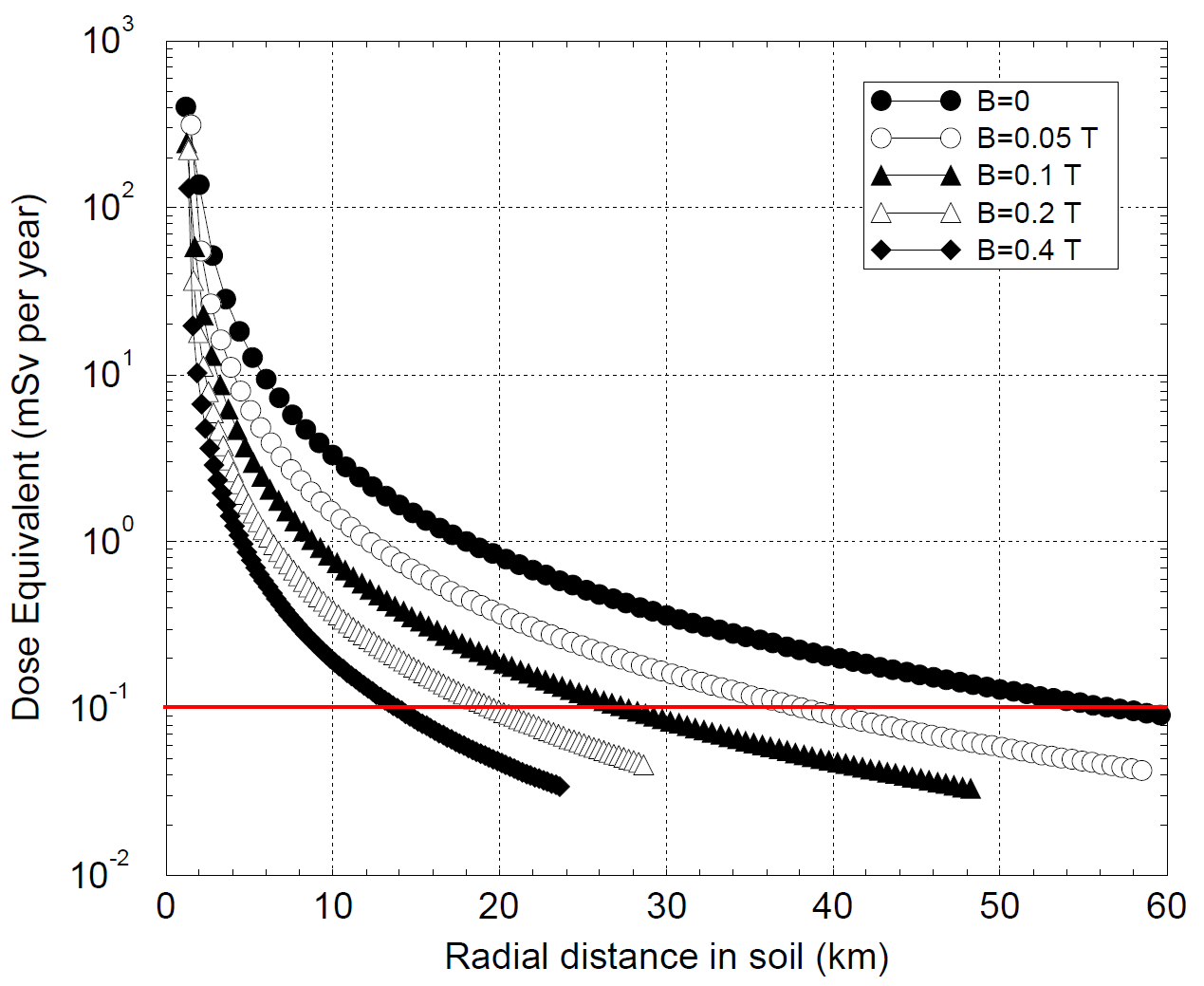}
    \caption{Neutrino flux and dose reduction around the 4~TeV COM muon collider using beam wobbling induced by wave field of 0, 0.05, 0.1, 0.2 and 0.4~tesla. The red line is the Fermilab offsite annual limit which is reached at 57~km with no wave field and at 14~km with 0.4~T wave field.}
    \label{fig:Mokhov3}
\end{figure}

The second mitigation technique was proposed in~\cite{Johnstone:1997zu} and studied in great detail in~\cite{Mokhov:2000gt}. Beam wobbling by a systematic time-varying vertical wave field in the ring would drastically disperse the strongly-directed neutrino flux in the orbit plane outside the MC ring. The dose reduction is quite impressive as shown in Fig.~\ref{fig:Mokhov3} for the wave fields of 0.05, 0.1, 0.2 and 0.4~T for a 4~TeV COM energy MC. Without such a wave field, the radial distance in the orbit plane to reduce the neutrino induced dose to the offsite limit of 0.1~mSv/yr is 57~km, while a modest wave field of 0.4~T reduces that distance to 14~km, and the effect rapidly increases with the field value. A recently proposed alternative technique with large-stroke high-resolution mechanical magnet movers (being refined at CERN) has even higher mitigation potential, as it can provide larger beam wobbling amplitude. 

Preliminary studies indicate that placing the collider ring at the depth of 200~m alone is sufficient to keep the dose well below the target level of 10~mrem/year for COM energies of up to 3 TeV. At 10 TeV, a combination of 200~m tunnel depth and beam wobbling is necessary to reduce the dose to below the target. The results bolster confidence in feasibility of a high-energy muon collider. 

Further improvements are certainly possible and should be studied. For example, one of the strategies investigated in MARS simulations~\cite{Mokhov:2000gt}, was to minimize the field-free regions in the collider. It was found that the presence of a field of even a fraction of 1~tesla is enough to reduce the dose at large distances from an MC to a below-limit level. The application of such a field over all RF and other components seems possible. The straight sections could also be shortened by using continuous combined function magnets. Furthermore, improved performance of the muon cooling might significantly reduce the emittances, thus greatly reducing the muon beam currents and neutrino fluxes, while keeping the luminosity the same.

%The focusing strength could be further increased by the use of plasma or other exotic focusing methods at the IP.
%The ``exotic," mitigation looked at in the early days of high-energy muon collider studies was to build it at an isolated site such as a desert or mountain region or remote island with no earth or rock immediately after a muon vector exits the Earth's surface.

%% file: imcc.tex
\subsection{European Accelerator Roadmap}

In 2020, the European LDG formed a muon beam panel and charged it with delivering input to the European Accelerator R\&D Roadmap covering the development and evaluation of a muon collider option. In parallel, CERN formed IMCC to assess feasibility of building a high energy muon collider, identify critical challenges, and develop an R\&D program aimed at addressing them~\cite{IMCC}. Besides the accelerator, the effort includes development of the MDI, detector concepts, and an evaluation of the physics potential. 

The IMCC held four "community meetings" in 2020 and 2021 to develop the scope and the plan of work to be done between now and the next ESPPU. R\&D objectives have been identified in several key areas, including muon production and cooling, neutrino induced radiation mitigation, MDI studies and optimization, and the high energy complex. Technologically, the design imposes challenging requirements on the high power targets where short proton bunch length and frequency may compromise the target's lifetime and integrity, on the high-field solenoidal magnets used in the production, collection and cooling of the muons, as well as on the specs of fast-ramping and fixed-field magnets used in the accelerator and collider rings. The ionization cooling system is a novel concept and requires careful studies for optimal integration of the absorber and RF stations inside of high magnetic fields. Successful demonstration of a partial muon cooling system is therefore crucial for the design verification. Currently rough dimensions of the facility have been identified and siting at CERN is being actively explored. An alternative siting at Fermilab is possible and is mentioned in~\cite{Bhat:2022hdi}.

\begin{figure}
        \includegraphics[width=0.8\textwidth, trim={2.1cm 0 0 0}]
        {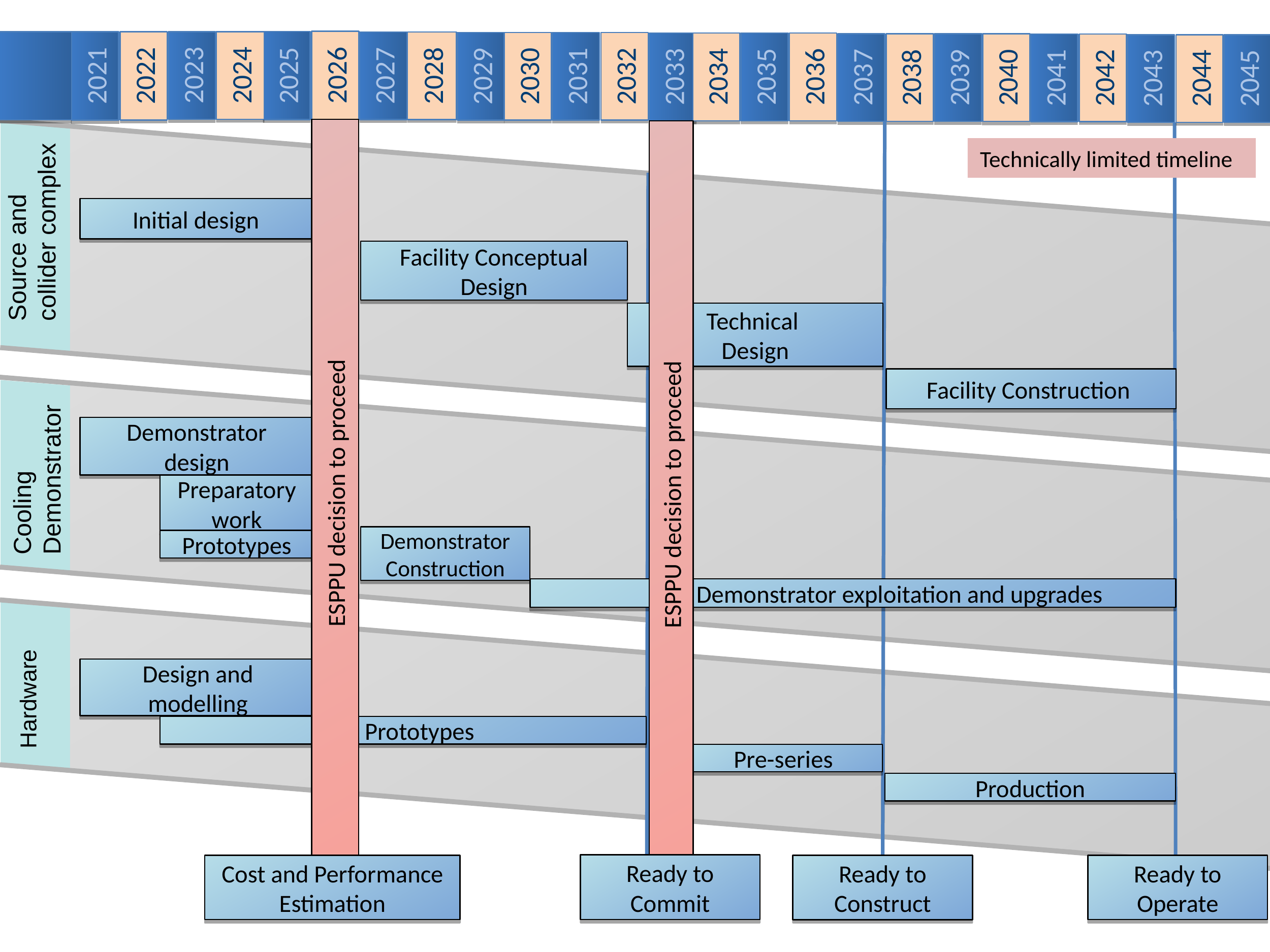}
        \centering\textbf{}

\caption{A technically limited IMCC timeline for the Muon Collider R\&D program}
\label{fig_muon:RDtimeline}
\end{figure}

A technically limited time line developed by IMCC is shown in Fig.~\ref{fig_muon:RDtimeline} and discussed in greater detail in \cite{bib_muon:LDGReport}. A muon collider with a center-of-mass energy around 3 TeV could be delivered on a time scale compatible with the end of operation of the HL-LHC. The muon collider R\&D program will consist of an initial phase followed by the conceptual and the technical design phases. The initial phase will establish the potential of the muon collider and the required R\&D program for the subsequent phases. 

The performance and cost of the facility would be established in detail. A program of test stands and prototyping of equipment would be performed over a five-year period, including a cooling cell prototype and the possibility of beam tests in a cooling demonstrator. This program is expected to be consistent with the development of high field solenoid and
dipole magnets that could be exploited for both the final stages of cooling and the collider ring development. A technical design phase would follow in the early 2030s with a continuing program focusing on prototyping and preserves development before production for construction begins in the mid-2030s, to enable delivery of a 3 TeV collider by 2045. The program is flexible, in order to match the prioritization and timescales defined by the next ESPPU, P5, and equivalent processes.

%\begin{table*}
%\caption{Tentative target %parameters for a muon collider at different
  %energies based on the MAP design with modifications. These values are only to give a first,
  %rough indication. The study will develop coherent %parameter sets of its own.}
  
%\label{MC:t:param}
%\begin{center}
 % \begin{tabular}{|*6{c|}}
  %  \hline
  %  Parameter & Symbol & Unit &\multicolumn{3}{c|}{Target value} \\
   % \hline
    %center-of-mass energy & $E_\textnormal{cm}$ & \si{\tera\electronvolt} & 3 & 10 & 14 \\
    %Luminosity & $\cal L$ & \SI{e34}{\per\square\centi\meter\per\second} & 1.8 & 20 & 40 \\
    %Collider circumference& $C_\text{coll}$ & \si{\kilo\meter} & 4.5 & 10 & 14 \\
    %\hline
    %Muons/bunch & $N$ & \num{e12} & 2.2 & 1.8 & 1.8 \\
    %Repetition rate & $f_\text{r}$ & \si{\hertz} & 5 & 5 & 5 \\
    %Beam power  & $P_\text{coll}$ & \si{\mega\watt} &5.3  & 14.4 &20 \\
    %Longitudinal emittance& $\epsilon_\text{L}$ & \si{\mega\electronvolt\meter} & 7.5 & 7.5 & 7.5 \\
    %Transverse emittance& $\epsilon$ & \si{\micro\meter} & 25 & 25 & 25 \\
    %\hline
    %IP bunch length& $\sigma_z $ %& \si{\milli\meter} & 5 & 1.5 & 1.07 \\
    %IP beta-function& $\beta $ & \si{\milli\meter} & 5 & 1.5 & 1.07 \\
    %IP beam size& $\sigma $ & \si{\micro\meter} & 3 & 0.9 & 0.63 \\
   % \hline
  %\end{tabular}
%\end{center}
%\end{table*}

Based on the MAP design, target parameter sets have been defined for the collider as a starting point. The parameter sets have a luminosity-to-beam-power ratio that increases with energy. They are based on using the same muon source for all energies and a limited degradation of transverse and longitudinal emittance with energy. The design of the technical components to achieve this goal are a key element of the study. It is important to emphasize that a 10~TeV lepton collider poses a number of key challenges. The collider can potentially produce a high neutrino flux that might lead to increased levels of radiation far from the collider; this would need to be mitigated and is a prime concern for the high-energy option. The MDI might limit the physics reach due to beam-induced background, and therefore the detector and the machine need to be optimized simultaneously. The collider ring and the acceleration system that follows the muon cooling have not been studied for 10~TeV or higher energy and can limit the energy reach. Finally, the production of a high-quality muon beam is required in order to achieve the desired luminosity. Optimization and improved integration of various sub-systems are required to achieve the performance goal, while maintaining low power consumption and cost. 

%Despite strong interest and expertise, U.S. participation in IMCC has been mainly limited to the work done in the context of Snowmass. As mentioned above, the design strategy taken by IMCC relies heavily on the concepts developed by the MAP collaboration. The European muon beam panel included two representatives (including the co-chair) from the U.S., and a large number of scientists helped to organize the IMCC working group activities. U.S. scientists made key contributions to most areas of the IMCC design development and planning, including magnets, RF cavities, muon production and cooling, muon acceleration, beam dynamics, machine-detector interface, and the high-energy complex. Besides the accelerator design, the Energy and Theory Frontier communities in the U.S. provided strong contributions in the areas of physics studies and detector design. 

%% file: rd_priorities.tex
\subsection{R\&D Priorities and possible US contributions}

Below we list possible areas  of research with emphasis on areas wherein the U.S. can contribute:

\begin{itemize}
\item \textbf{Proton driver:} Existing facilities can be upgraded to provide multi-MW beams for a muon collider or neutrino factory. Fermilab’s PIP-II program will be capable of delivering beam power up to 1.2 MW. Several proposals are under development for either expanding the superconducting proton linac (PIP-III) or combining the existing linac with an RCS to increase the beam power to $> 2$ MW. The ESS MW proton linear accelerator can also be upgraded and extended to demonstrate the generation of nanosecond-scale beams with very high-charge (10$^{15}$) proton pulses that can be used for the generation of the muon pulses required for a muon collider. The R$\&$D program will explore these and other source options.

\item \textbf{Target:}
The design criteria of the target system are high pion/muon production yields and high tolerances of the thermal stress caused by the beam impact.
\begin{itemize}
    \item Extending the RaDIATE effort to study high stress tolerant material. Optimizing dimensions of the target system to produce high pion/muon yields as well as minimizing thermal stress on the target material by use of a new target material. 
    Applying AI algorithms to find the best mix of target compounds to fulfill the design criteria. 
    \item Collaborating with the MagLab and other national institutions to develop a radiation robust high field magnet system for capturing pions/muons and transporting them to the cooling channel. 
    \item Utilizing the TSIB hot lab and other PIE facilities to develop target material science. Propose the needed beam irradiation facility: intense proton beam facilities will be considered for the beam irradiation facility, such as PIP-II and the upgraded-Booster accelerator complex. These accelerators are also good candidates for the cooling demonstrator. 
\end{itemize}

%Fermilab has an active target development program, including targets for Mu2e-II (100 kW) and LBNF (1.2-2.4 MW).  The Mu2e-II geometry is a simpler version of the MC target system, with targets within high field large-bore solenoids.  The field strength of Mu2e-II solenoids is lower and the target length is shorter than the MC target system. However, making the Mu2e-II target system is still extremely challenging. Fermilab also hosts RaDIATE to explore targets for LBNF at 2.4 MW operation. The Fermilab research can collaborate with the Mu2e-II target group and with RaDIATE to synergetically develop the target technology for the MC. 

\item \textbf{Cooling:} %Demonstrations of the performance of RF cavities in magnetic fields are crucial. 50 MV/m at 3 T has been demonstrated at the MTA. Further tests are needed to establish performance at the parameters of cooling scenarios. Higher field magnets are needed at the cooling aborbers. 
%MAP considered cooling with magnetic fields up to 30 T. (Commercial MRI magnets are now available at 29 T and the record field demonstrated is 32 T with bores similar to those needed for cooling; these could be extended to MC parameters.)
Improving the cooling performance is a primary goal of the cooling design R\&D. Depending upon the future target system, decay, bunching, and phase rotation (called the  ``front end") systems, the subsequent 6D cooling channel must be optimized. Improving cooling can significantly relax the beam requirements, reducing the primary proton beam power, the beam induced background at the collider detector, and the neutrino flux. Research on integration of AI techniques can aid in making the channels shorter and perhaps identify new parameter sets for improved cooling. The FOFO Snake~\cite{FOFOSnake}, which cools both muon charge signs simultaneously, should be further explored. Further simulation studies should be made of PIC~\cite{maloney-2013,PIC} to determine whether it can achieve acceptable performance or to rule out its utility. Further improvements should be considered in the high-field solenoid cooling channel to reach the final collider emittances, incorporating both more extensive simulation studies and expected improvements from more advanced magnet technologies. Other research areas of importance are:
\begin{itemize}
    \item Investigating the space charge effect in the cooling channel, especially in the later cooling channel. 
    \item Engineering efforts to integrate high gradient RF cavities into high field magnet coils. 
    Designing modular beam components including  beam instrumentation.
    \item Investigating the cooling performance with imperfect beam optics, especially in the later cooling channel.  
    \item Investigating the influence of ionizing particles in the cooling channel, especially for the initial cooling channel.
    \item Investigating the beam-plasma interactions, especially the plasma focusing effect. 
\end{itemize}
\item \textbf{Acceleration:} To better understand acceleration for a muon collider, a near-term research program should include: 
\begin{itemize}
    \item Creating a first design of acceleration from a final cooling stage to the initial energy of the MAP acceleration scenario, with goals of having a more general description of the acceleration scenario, an approximate idea of achievable muon transmission, and studying emittance preservation, particularly in the longitudinal plane.
    \item Desiging a detailed pulsed acceleration lattice design that would fit on the Fermilab site. The goal would be to have sufficient detail in the design to be able to specify with reasonable confidence the maximum energy that could be achieved for a muon collider on the Fermilab site given certain technology choices.
    \item Study FFA acceleration designs, both for moderate and maximum energies, with particular emphasis on advanced designs (such as nonlinear and vertical FFAs), to understand if and for what parameter ranges they may be more favorable than other types of accelerators.
    \item A study of the collective effects in acceleration systems, to assess their impact on the beam distribution, and identify any required mitigations and their impacts.
    \item Designing and testing an approximately full-scale pulsed magnet similar to a magnet that would be used for pulsed muon acceleration, along with its power supply. It should reach fields required for muon acceleration (at least 1.5~T) and ramp at the desired rate (around 1~ms). The goals would be to confirm the expected magnet performance, our ability to control the field variation with time, and to understand the engineering and cost issues for the system, particularly the power supply.
\end{itemize}
\item \textbf{Collider Ring:} 
%The collider ring requires 16 T arc dipoles with a 15 cm bore. Moreover, neutrino flux mitigation is a concern. In addition, MAP only has studied lattices up to 6 TeV. The US-MDP program will have ID  120 mm, 12-15 T dipole demonstrators with Nb3Sn coils within the next 3-4 years. 
The TF and EF groups can investigate the physics cases at 600 GeV, 3 TeV, and 10~TeV center of mass energy. A new collider lattice must be designed. Possible solutions to mitigate neutrino flux are to situate the collider at $\sim100$ m depth, add magnets, or move the lattice over time. Improving cooling can significantly relax these requirements.   

\item \textbf{Magnets:}
The following key MC magnet systems will require focused R\&D efforts:
\begin{itemize}
    \item Muon cooling. MAP considered muon beam cooling with magnetic fields up to 30 T. This field level has been demonstrated with commercial 29 T MRI magnets. The record field of 32 T achieved in a superconducting solenoid with  bore diameter similar to that needed for cooling allows extending its design and parameters to the MC level. Experimental demonstration of the final cooling solenoid will be needed.
    \item Muon acceleration. Recent tests of the HTS-based 0.5-m long two-aperture SC dipole at Fermilab have shown record-high field ramp rate of ~300 T/s at 10 Hz repetition rate and 0.5 T field amplitude. Based on this result, a possible upgrade of this magnet design to MC requirements of 2 T amplitude in a 10 mm beam gap with dB/dt up to 1000 T/s for the rapid cycling acceleration dipoles has been proposed and needs to be supported.
    \item Collider ring. The collider ring for 6 TeV or higher COM energy MC requires 24 cm aperture, 20 T nominal field arc dipoles with large operating margin in a high radiation environment. The US-MDP is working on designs and technologies of accelerator magnets based on low temperature and high temperature superconductors. In 4--5 years the program plans experimental demonstration of a 120-mm aperture, 10--12 T Nb$_3$Sn dipole, and development of conceptual and engineering designs of a 20 T hybrid HTS/Nb$_3$Sn dipole with 50-mm aperture, which will be fabricated and tested. Combining these two results would pave the way towards the large-aperture ($>$ 150 mm diameter) high-field ($>$ 20 T with large margin) dipoles and quadrupoles needed for the HE MC.
\end{itemize}    
\item \textbf{RF cavity:}
Significant achievements have been made on the RF cavities in magnetic field for muon ionization cooling under MAP. However, there is still a considerable amount of R\&D needed to make and test realistic cavities for the current cooling channel designs, which includes:
\begin{itemize}
    \item
    Design a compact multiple-cavity module, with practical frequency tuning and RF power feed systems. In the current cooling channel design, the voltage in each segment requires multiple cavities or one linac with multiple cells. To achieve the required strong solenoidal fields, the cavity transverse dimension should be kept as compact as possible. The RF tuning mechanism and RF power coupling feed must fit into the tight space, and more importantly be immune to the strong magnetic field background. Novel ideas, such as distributed coupling structure~\cite{distributed_coupling}, should be explored and studied. The RF module design is an integral part of the cooling channel,  and comprehensive system engineering efforts are required to include all accessory components, such as RF coupler, RF feed-through probe, cryogenic system and diagnostics for the cooling channel during the early design phase.
    \item
    Study of the beryllium windows for RF cavities at various cooling stages: The thickness of beryllium windows is determined by the acceptable RF frequency tuning range due to window deformation from the thermal stress induced by RF heating and scattering effects. The Be window should be kept as thin as possible, and especially becomes extremely challenging at the later stage of the cooling channel where the scattering effect from beryllium may become one of the dominant heating effects on muon emittance. 
    \item
    Explore RF cavities at other frequencies: RF breakdown scales with  frequency, as expressed by the empirical Kilpatrick limit; optimization study of cavity frequency choice, in consideration of RF breakdown limit (performance), cavity dimensions, and surrounding magnets may provide additional benefit for a compact cooling channel design.
    \item
    Cavity operation at cryogenic (LN$_2$) temperature: A recent experiment shows that a copper cavity operated at cryogenic temperature is more resilient to RF breakdown than at room temperature ~\cite{cryogenic_cavity}. Although the experiment was carried out without a strong external magnetic field, it is expected that it will withstand higher RF breakdown even in the presence of a strong magnetic field.  In addition, both copper and beryllium have a factor  2--3 improvement in electrical conductivity at LN$_2$ temperature, a significant and potential peak RF power savings for a given RF gradient.  
    \item
    Investigation of beam loading and wakefield effects in vacuum RF cavities: Theoretical and experimental study to include  collective beam loading and wakefield effects and mitigation methods and implementation of higher-order-mode (HOM) damping schemes if needed.
    \item 
    Further investigation of the gas--plasma--beam interactions in gas-filled RF cavities: Study the plasma chemistry and learn how to control the cold plasma dynamics. In order to take advantage of the extra plasma focusing effects, a new type of plasma focusing experiment should be conducted by using intense beams, and simulated using particle-in-cell codes. 
\end{itemize}
\end{itemize}

%% file: fermilab_site.tex
\subsection{Fermilab Site Option}

\begin{figure}
\begin{center}
\includegraphics[width=0.9\textwidth]{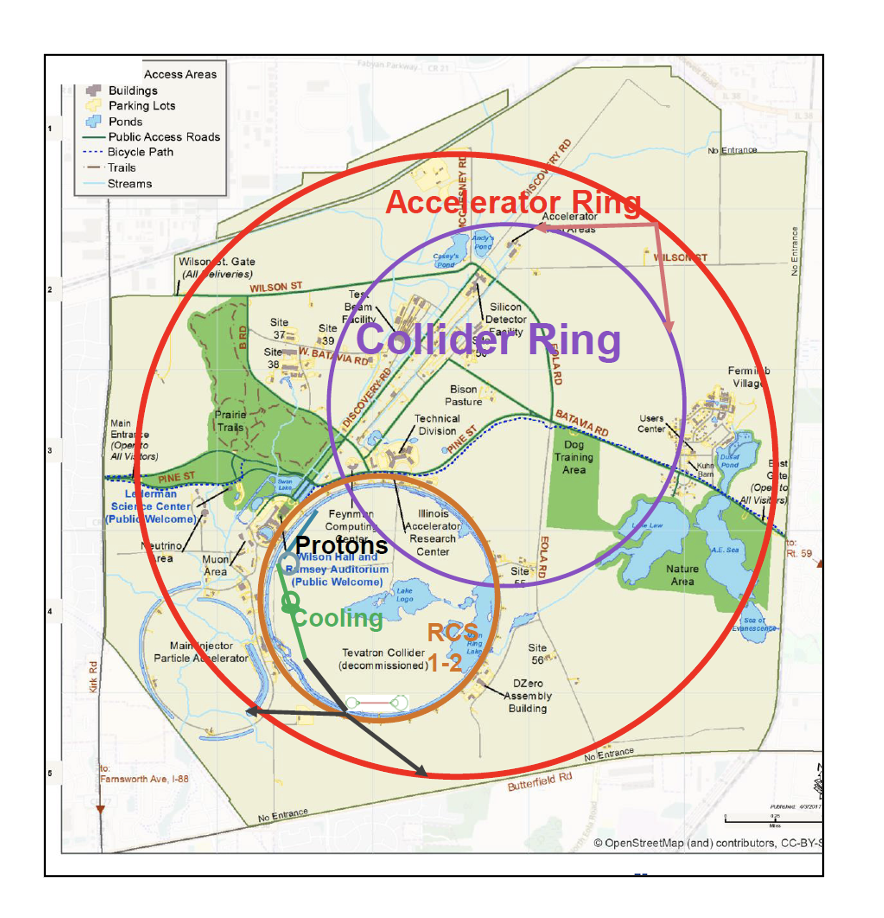}
%/\includegraphics[width=0.6\textwidth]{125-600collider.png}
\caption{A schematic view of the Fermilab site and the layout of a possible complex for the Muon Collider. The protons start at PIP-II and are accelerated, bunched and pulsed onto a high power target. Muon cooling chain is indicated in green. Acceleration happens in stages with the final stage taking place inside the large Accelerator Ring. Muons at the nominal energy are injected into the Collider Ring, where  the experiment(s) are located.}
\label{fig:FnalMC}
\end{center}
\end{figure}

The idea of having a Muon Collider as a potential ``site filler" for Fermilab dates back to the early 2000’s, when parameters for a 4 TeV machine were presented. More recently, using higher field magnets and higher-gradient acceleration, the parameter space towards a 10 TeV Muon Collider concept that would fit within the Fermilab site has been identified and a first design concept has been developed. A schematic layout of this configuration is shown in Fig.~\ref{fig:FnalMC}. 
%g 
The concept begins with use of PIP-II as the initial part of the proton source. The PIP-II linac would be extended to higher energy and followed by either a higher energy linac leading into  proton accumulation and bunching rings or a rapid-cycling synchrotron or FFA (fixed-field accelerator) ring. The goal would be to produce intense $\approx$\,10 GeV proton pulses at $\sim5$ Hz and $\sim2$ MW onto a pion production target. This is followed by muon collection (from $\pi$ decay) and bunching that leads into 6D muon cooling channels, obtaining minimal emittance beams. The collection and cooling channel would be $\sim$\,1--2 km long.

Muon acceleration is achieved in three stages: (1) A Linac (up to 5 GeV) first that is followed by a Recirculating Linac (up to 65~GeV). This energy would be sufficient for a Higgs Factory~\cite{Neuffer:2013wrd}. (2) This is followed by a set of two Rapid Cycling Synchrotrons that can fit into the Tevatron ring tunnel and are capable of delivering an energy up to 1 TeV. (The first RCS would accelerate to $\sim$\,300 GeV, using normal-conducting magnets. The second would be a hybrid high-field RCS.) (3) A final RCS ring that has a radius of 2.65 km and can bring the energy up to 5 TeV.
(This is a hybrid RCS ring with $\sim$\,16 T magnets interlaced with cycling $\pm$4 T magnets.
If cycling is limited to $\pm\sim$2 T, two rings are required.)
The acceleration will use superconducting RF cavities at frequencies of 650~MHz and 1300 MHz. 

The 5 TeV beams would be injected into a 10 TeV collider ring using high field bending magnets. Based on extrapolations from Ref.~\cite{Wang:2015xoa} the 10 TeV collider is expected to have a radius of 1.65 km. It is important to note that, given the 3 accelerator stages, staging is possible and operation at 125 GeV, 1 TeV, and 3 TeV can be envisioned as intermediate states. Figure~\ref{fig:FnalMC} shows a schematic view of the collider in its various stages. 

%% file: det-intro.tex
\subsection{General Introduction}

A circular \mumu collider is a particularly attractive option for the future of energy frontier exploration. Such a machine has the potential to deliver a vast physics program in a relatively compact and power efficient accelerator complex. However, on the experimental side, a great physics potential is accompanied by unprecedented technological challenges, due to the fact that muons are unstable particles. Their decay products interact with the machine elements and produce an intense flux of background particles that eventually reach the detector and may significantly degrade its performance.

The physics program includes a precision component consisting of the exploration of the electroweak sector of the SM and a broad spectrum of searches for BSM physics. In order to achieve these physics goals, experiments at a Muon Collider need to be able to reconstruct products of \mumu collisions with required performance. The performance demands identification and reconstruction of charged leptons, photons, and jets with high efficiency and good energy/momentum resolution. This in turn places stringent requirements on the performance of tracking and calorimeter reconstruction. Identification of displaced vertices originating in decays of heavy flavour mesons is also of major interest, in particular for the Higgs program and for BSM signatures involving heavy flavor. 

The unstable nature of muons (lifetime $\tau_{\mu} \approx$~2 microseconds at rest) results in a significant fraction of muons to decay, with their decay products producing a large flux of secondary and tertiary particles after interacting with accelerator and detector elements. Such a flux of particles is referred to as Beam-induced background (BIB). This results in a unique experimental environment~\cite{Mokhov:2011zzd}, with a related set of challenges. One of the biggest challenges for a muon collider detector is to successfully disentangle the products of the \mumu collisions from an intense BIB coming primarily from the muon decay and shower products. The detectors and event reconstruction techniques need to be designed to cope with the presence of the BIB. In this sense, considerations and challenges related to the detector design at a muon collider are somewhat distinct from these at $pp$ and \ee colliders. 

The goal of this section is to demonstrate that high quality physics is achievable despite challenges originating from the BIB. This is possible due to significant advancements in detector technologies during the last decade, motivated in large by requirements for the HL-LHC experimental upgrades. A comprehensive review of  promising technologies for a muon collider can be found in Ref.~\cite{Jindariani:2022gxj}. Major breakthroughs have also been made in development of dedicated BIB suppression and physics object reconstruction techniques. This work has been carried out within the context of IMCC and Snowmass studies and is summarized in Ref.~\cite{MuonCollider2022ded}.

Here we first discuss the environment in which the detectors are expected to operate and draw some comparisons with High Luminosity LHC. We present the current simulated detector configuration and highlight the expected performance. We also briefly describe technologies that have a potential to match challenging specifications of a muon collider detector. The current detector configuration was adopted from CLIC and is optimized for 3 TeV center-of-mass energy; the design needs to be updated for future higher energy studies. We then present two very distinct physics studies ($H\rightarrow bb$ and Dark Matter with disappearing track) and compare results at the full and fast simulation levels. The goal of these comparisons is to demonstrate that impact of the BIB can be mitigated and the residual differences between full- and fast- simulation are small. Finally, we outline a path forward for future improvements.

%% file: det-environment.tex
%% Simulation of BIB
The expected characteristics of BIB depend on the beam properties, accelerator lattice, interaction region as well as detector design. 
Detailed simulation studies have been performed~\cite{Mokhov:1996tq} using the MARS15~\cite{mars15} software and, more recently~\cite{Collamati:2021sbv,MuonCollider:2022ded,Jindariani:2022gxj}, using a combination of Linebuilder~\cite{Mereghetti:1481554} and FLUKA~\cite{Ferrari:2005zk, BOHLEN2014211}. The two simulations are found to give compatible results.
Muons decays are simulated within $\pm 200$ meters from the interaction point (IP) in a collider ring with the parameters summarised in Table~\ref{tab:collider_parameters}. In particular, these studies were performed for a collider with center-of-mass energy of 1.5 TeV.
The expected distance between bunches is such that the effect of nearby bunches is negligible.

\begin{table}[h]
    \centering
    \begin{tabular}{l|c|c}
        Parameter & Symbol & Value \\
        \hline\hline
        Center-of-mass energy & $\sqrt{s}$ &  $1.5$~TeV\\
        Muons per bunch      & $N_\mu$ & $2\cdot 10^{12}$\\
        Normalised transverse emittance & $\epsilon_{TN}$ & 25 $\pi \,\mu$m rad\\
        Normalised longitudinal emittance &$\epsilon_{LN}$ & 7.5 MeV m \\
        IP relative energy spread & $\delta_E$   & 0.1 \%\\ 
        IP beta function &$\beta^{*}_{x,y}$ & 1 cm \\
        IP transverse beam size & $\sigma_{x,y}$ & 6 $\mu$m \\
        IP longitudinal beam size & $\sigma_{z}$ & 10 mm   \\
    \end{tabular}
    \caption{Representative set of muon collider parameters used in the detailed simulation presented in~\cite{MuonCollider:2022ded}.}
    \label{tab:acc-params}
    \label{tab:collider_parameters}
\end{table}

\begin{figure}
  \centering
    \includegraphics[width=0.45\textwidth]{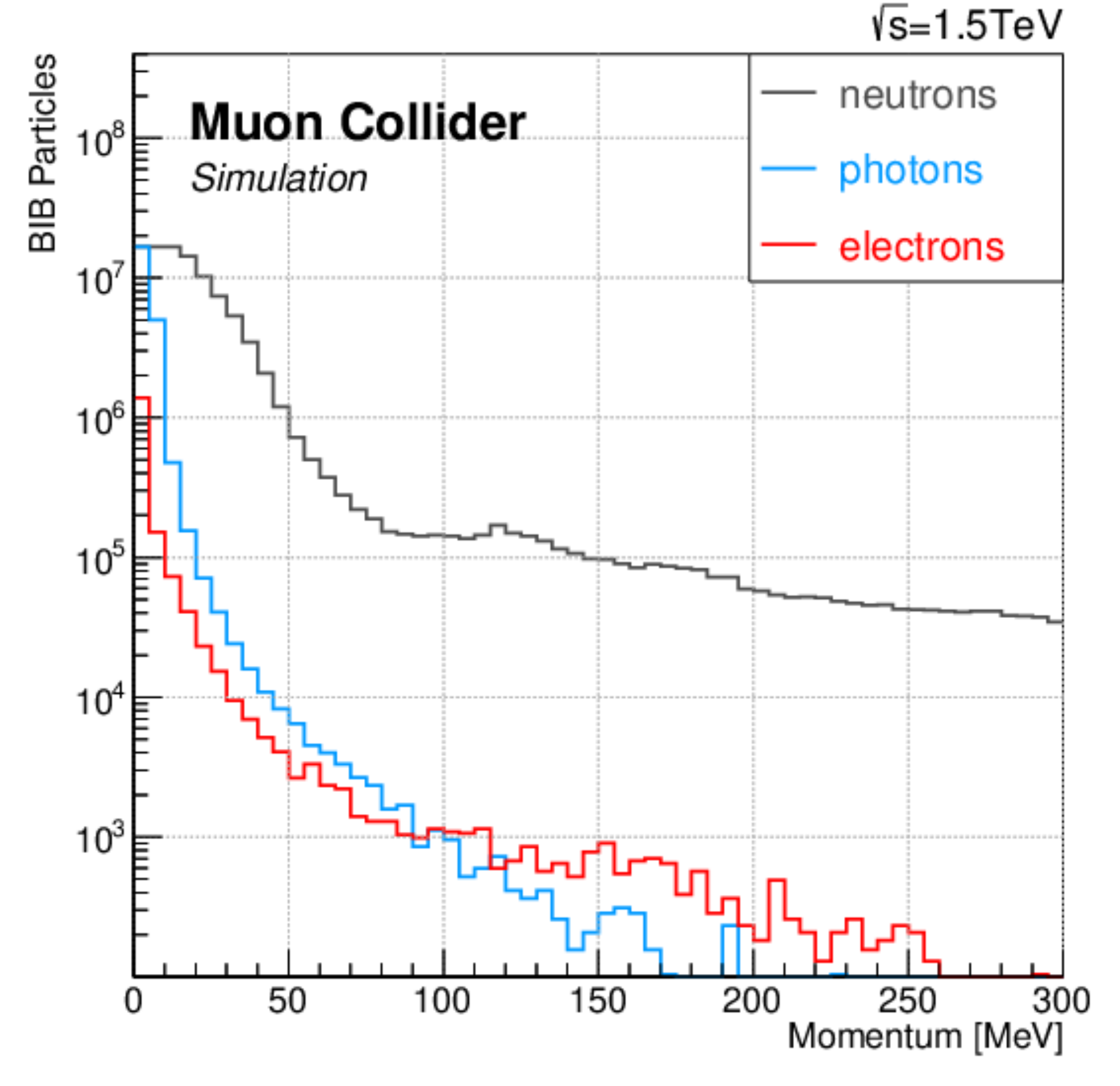} \hfill
    \includegraphics[width=0.45\textwidth]{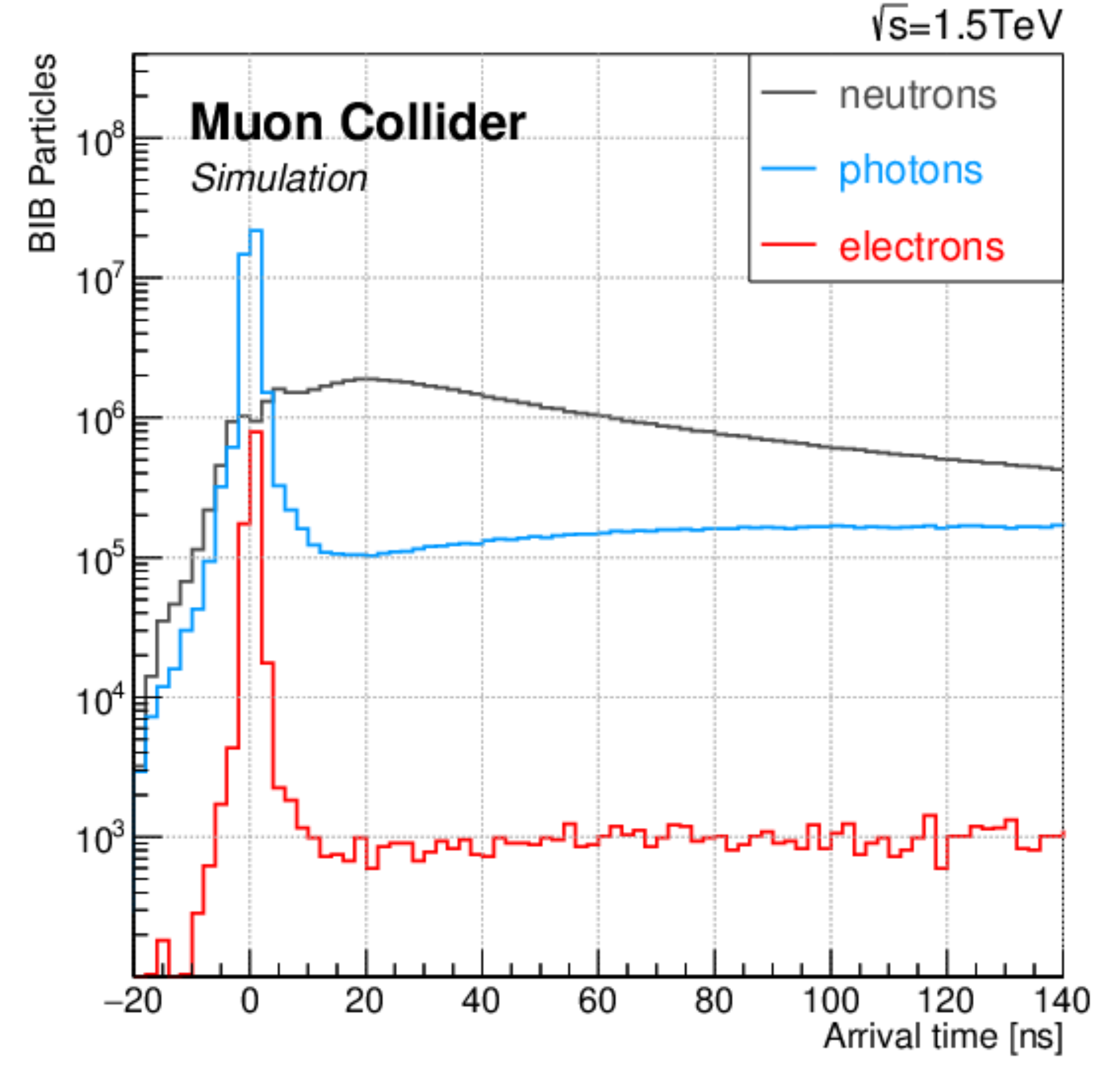}
    \caption{Kinematic properties of BIB particles entering the detector region: momentum (left) and arrival time with respect to the bunch crossing (right).}
    \label{fig:bib-props}
\end{figure}

%% BIB composition, introduction of nozzle, composition and position source of BIB after nozzle
Particles from BIB can deposit a huge amount of energy in the detector, if not shielded properly.
For this reason an essential part of the machine detector interface at a Muon Collider is a pair of tungsten (W) nozzles cladded with borated polyethylene (BCH), which reduce the rate and energy of BIB particles reaching the detector by several orders of magnitude. Such a nozzle also limits the acceptance of the detector to polar angles $\theta > 10^\circ$.
The flux of particles surviving the shielding and entering the detector arise partially through shower products of BIB particles exiting the nozzle and through back-scattering of particles from one beam into the nozzle on the opposite side of the IP with respect to the direction the beam is arriving from. The result is a diffuse background of mostly low-momentum and out-of-time photons, neutrons, and electrons/positrons; as shown in Figure~\ref{fig:bib-props}, their flux reduces by 2 to 5 orders of magnitude for energies above 100 MeV and the expected time structure extends to several hundreds of ns. A smaller flux of muons and charged hadrons is expected but their flux is smaller by several orders of magnitude.

% Radiation levels, comparison with HL-LHC, FCC-hh
The amount of expected radiation from BIB has been estimated using the FLUKA simulation mentioned above, using a simplified detector geometry. Detailed maps are available in Ref.~\cite{Jindariani:2022gxj} for both expected dose and 1~MeV neutron equivalent fluence. Table~\ref{tab:dose_fluence} provides an approximate comparison of the expected doses and fluences in the muon collider environment when compared to high-radiation environments of high-energy colliders such as HL-LHC.
The expected HL-LHC doses and fluences are taken from Ref.~\cite{CERN-LHCC-2017-021}; such calculations include a very detailed simulation of detector elements and a safety factor that is not included in the other numbers of this table, but should not change qualitatively the comparison. We also re-normalized the expected dose and fluence to a single year of data-taking at the ultimate performance of the HL-LHC accelerator.
%%TODO: x-check numbers
%%TODO: decide if to add a map from the detector paper

\begin{table}[]
    \centering
    \begin{tabular}{c|cc|cc}
     \hline
                  &\multicolumn{2}{c|}{Maximum Dose (Mrad)} & \multicolumn{2}{c}{Maximum Fluence (1~MeV-neq/cm$^2$)}\\
                  & R$=22$~mm & R$=1500$~mm & R$=22$~mm & R$=1500$~mm \\
    \hline
    Muon Collider & 10 & 0.1 & $10^{15}$ & $10^{14}$\\
    HL-LHC        & 100 & 0.1 & $10^{15}$ & $10^{13}$\\
    \end{tabular}

    \caption{Rough comparison of expected dose and 1~MeV neutron equivalent per cm$^2$ fluence per year of data-taking for a muon collider environment as well as for HL-LHC. The numbers are reported for different radii (R), loosely corresponding to possible positions for a first tracking layer, and before the entrance of an electromagnetic calorimeter. See text for details and caveats.}
    \label{tab:dose_fluence}
\end{table}

% Further BIB reduction with in-detector capabilities and design choices -- just mention, fwd reference appropriate sections
The flux of BIB particles can be further reduced exploiting the different timing, direction and energy spectrum compared to the products of the main \mumu collisions. The effect of such requirements is explored in
Ref.~\cite{MuonCollider2022ded,Jindariani:2022gxj,https://doi.org/10.48550/arxiv.2203.06773} and summarized in the following Sections~\ref{sec:det:current-configuration},~\ref{sec:det:det-new}.

% Particle multiplicity at critical detector positions (1st tracking layer, entrance of calo, muon system barrel / endcap)
% if time allows, might also go into the detector chapters?

% Scaling at higher energy
FLUKA simulations of higher center-of-mass energy muon collider configurations is still under development. A very preliminary comparison of the multiplicity of particles entering the detector volume using FLUKA and MARS15 and for beam energies up to 5.0 TeV (corresponding to 10 TeV COM energy) is shown in Tab.~\ref{fig:bib-comps}. The numbers are obtained with MDI optimized for 750 GeV beam energy. These and further preliminary results reported in Ref.~\cite{MuonCollider:2022ded} show energy evolution of two main effects: muon lifetime becomes longer in the laboratory frame, leading to fewer decays per meter, while the average initial energy of muon decay products is larger, leading to a higher multiplicity of particles generated by shower in material. It turns out that the two effects roughly cancel out resulting in similar multiplicity, spatial, and time distributions of BIB particles entering the detector. Optimization of MDI at 3 and 10 TeV should results in further reduction of the BIB.

%\begin{figure}
%  \centering
%    \includegraphics[width=0.9\textwidth]{Figures/bibComparisonNew.png} \hfill
%    \caption{Comparison of multiplicities of different types of particles produced in each bunch crossing by the BIB at different energies. The uncertainties are about 20\%. The MDI optimized for 750 GeV is used for the 1.5 and 5.0 TeV columns. Vast majority of these particles have very low momenta.}
%    \label{fig:bib-comps}
%\end{figure}

\begin{table}[h]
    \centering
    \begin{tabular}{l|c|c|c|c|c}
    Simulation Source & \multicolumn{2}{c|}{MARS15} & \multicolumn{3}{c}{FLUKA}\\
    \hline
    Beam Energy [GeV] & 62.5 & 750 & 750 & 1500 & 5000\\
    MDI Optimization & yes & yes & yes & no & no\\
    Muon decay length [m] & $3.9\times10^{5}$ & $46.7\times10^{5}$ & $46.7\times10^{5}$ & $93.5\times10^{5}$ & $311.7\times10^{5}$\\

    Muon decays/m per beam  & $51.3\times10^{5}$ & $4.3\times10^{5}$ & $4.3\times10^{5}$ & $2.1\times10^{5}$ & $0.64\times10^{5}$\\

    $\gamma$/BX (E$_\gamma>0.1$ MeV) & $170\times10^{6}$ & $86\times10^{6}$ & $51\times10^{6}$ & $70\times10^{6}$ & $116\times10^{6}$\\

    n/BX (E$_n>1$ MeV) & $65\times10^{6}$ & $76\times10^{6}$ & $110\times10^{6}$ & $91\times10^{6}$ & $89\times10^{6}$\\

    e$^{\pm}$/BX (E$_e>0.1$ MeV) & $1.3\times10^{6}$ & $0.75\times10^{6}$ & $0.86\times10^{6}$ & $1.1\times10^{6}$ & $0.95\times10^{6}$\\

    h$^{\pm}$/BX (E$_h>0.1$ MeV) & $0.011\times10^{6}$ & $0.032\times10^{6}$ & $0.017\times10^{6}$ & $0.020\times10^{6}$ & $0.034\times10^{6}$\\

    $\mu^{\pm}$/BX (E$_\mu>0.1$ MeV) & $0.0012\times10^{6}$ & $0.0015\times10^{6}$ & $0.0031\times10^{6}$ & $0.0033\times10^{6}$ & $0.0030\times10^{6}$\\
    \hline
    \end{tabular}
    \caption{Comparison of multiplicities of different types of particles (photons, neutrons, electrons/positrons, charged hadrons, and muons/antimuons) produced in each bunch crossing by the BIB at different energies. The numbers correspond to the nominal bunch intensities of $2\times 10^{12}$ muons/bunch. The uncertainties are about 20\%. The MDI optimized for 750 GeV is used for the 1.5 and 5.0 TeV columns. Vast majority of these particles have very low momenta.}
    \label{fig:bib-comps}
    \label{fig:bib-comps}
\end{table}

%% file: det-currentconf.tex
\begin{figure}
    \centering
    \includegraphics[width=0.9\textwidth]{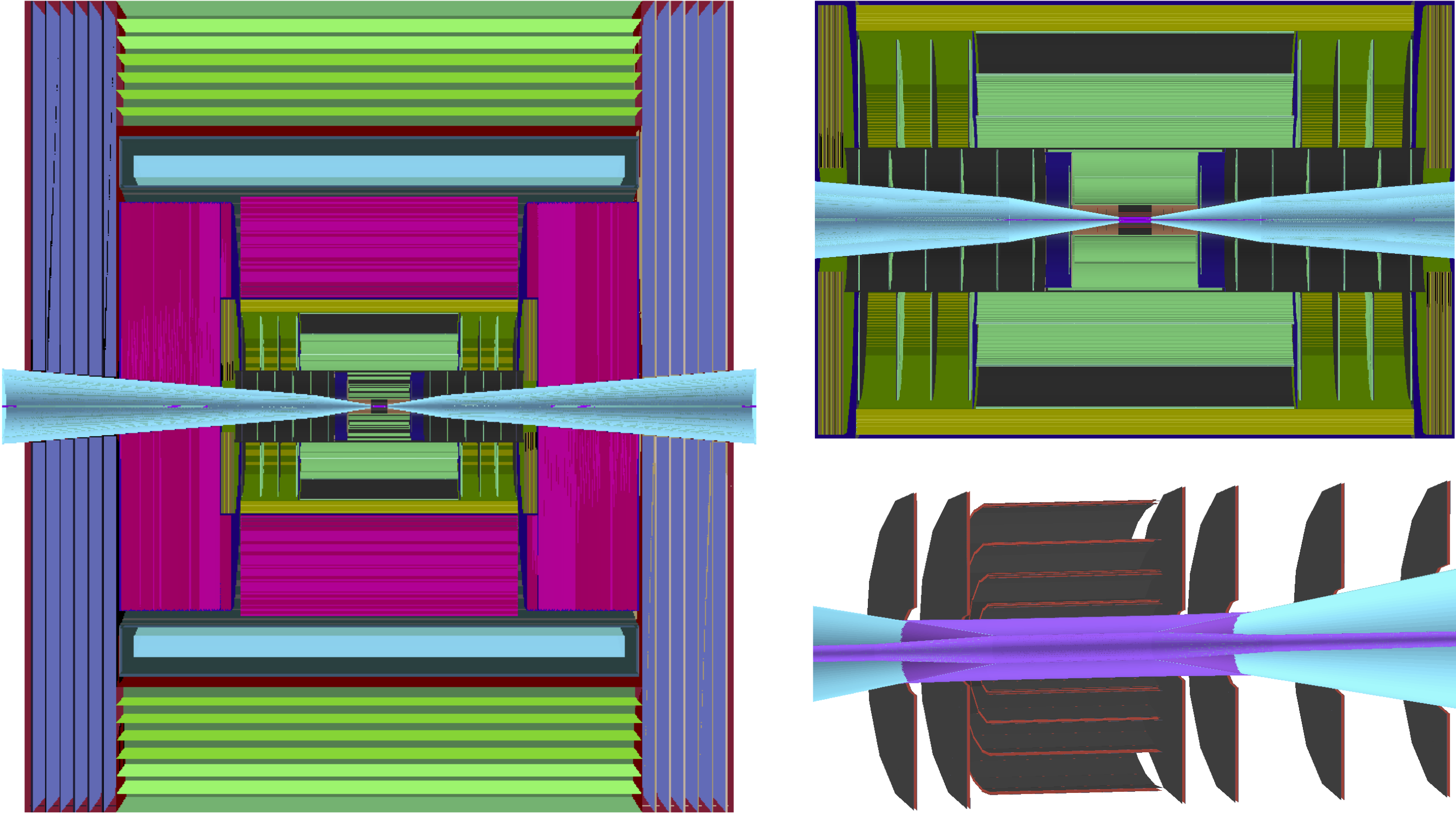}
    \caption{Rendering of the muon collider detector geometry used for the presented simulation studies, including the cone-shaped shielding nozzles (cyan) and the beryllium beampipe (violet). Shown are the cross sections of the full detector geometry (left) and two zoomed-in portions: up to ECAL (top right) and up to Vertex Detector (bottom right). Muon Detector (violet and green) surrounds the solenoid (cyan), which encloses the HCAL (magenta), ECAL (yellow) and the Tracking Detector (green and black).}
    \label{fig:detector-geometry}
\end{figure}

In order to benchmark the realistic physics expectations of a muon collider, a full \GEANT 4 model of a detector based on the CLIC detector has been developed. The detector model was adapted from the post-CDR CLIC detector model, \texttt{CLICdet}~\cite{CLICdet}. It should be of course noted, that since the time scale for the R\&D and construction of a muon collider facility requires several decades it should be assumed that detector technology and performance will naturally evolve during that time period and advancements in detector technology are almost guaranteed to improve the performance that a real detector could obtain in the future. Nonetheless, in order to make progress on understanding the physics potential and background challenges of such a machine a detector with minimal requirements on technology R\&D has been simulated in detail~\cite{MuonCollider2022ded}. The rendering of the detector geometry is presented in Fig.~\ref{fig:detector-geometry}. The detector model uses the basic cylindrical layout of an electron collider detector but makes important modifications to the machine detector interface (MDI) and tracking detectors to shield and optimize the layout for the large beam induced backgrounds.

The detector is comprised of the typical cylindrical layout of a collider detector with a silicon inner tracker, eletromagnetic and hadronic calorimeters enclosed by a strong solenoid magnet with an iron return yoke with muon chambers interleaved. The detector features an all-silicon Tracking Detector with three subsystems: a pixel Vertex Detector from 30 mm to 104 mm radius, a macropixel Inner Tracker from 127 mm to 550 mm in radius, and a microstrip detector from 819 mm to 1486 mm. The spacial resolution is presumed to be $5 \mu \mathrm{m}$ by $5 \mu \mathrm{m}$ for the Vertex Detector and  $7 \mu \mathrm{m}$ by $90 \mu \mathrm{m}$ for the Trackers with timing resolutions of 30 (60) ps for the Vertex (Tracking) Detectors. Such requirements are not achievable by current technology, but are the subject of intense R\&D and are thought to be achievable in the relatively near future, as outlined in Section~\ref{sec:det:det-new}.  Immediately after the Outer Tracker is a silicon-tungsten (SiW) ECAL made of 40 layers of 1.9 mm thick absorber plates with $5 \times 5 $ {\rm mm}$^2$ silicon sensor pads extending to a radius of about 1700 mm. This is followed by the HCAL comprised of 60 layers of iron and plastic scintillating tiles with a total of 7.5 interaction lengths. A 3.57 T solenoid magnet surrounds the previous elements comprised of 3-module 4 layer aluminum coil inside a liquid helium cryostat. The final element are 6 layers of resistive plate chambers (RPCs) inside a large iron return yoke which extends out to 6.5 meters in radius. Design details including the placement of various components, exact material composition, and estimate of the timing and position resolution are given in ~\cite{MuonCollider2022ded} and ~\cite{CLICdet}. 

The most distinctive addition to the CLIC design is modification of the MDI to account and shield the large beam induced backgrounds. The MDI includes the placement of two large tungsten nozzle shields. These are cones of tungsen originating at the center of the detector and limit the forward coverage. In order to improve the shielding borated polyethylene is included in the outer portion of the cone in the inner detector and cylinders of iron, concrete and borated polyethylene surround the beam pipe in the MDI approaching from both sides of the detector with layout and dimensions as shown in ~\cite{MuonCollider2022ded}.

%% file: det-new.tex
Since the original proposals of colliding muons to study physics at high center of mass energies more than fifty years ago, there have been multiple advances in detector technologies that represent a quantum leap forward in experimentalists' ability to perform precision measurements in high occupancy environments . 

In addition, there has been a technological revolution in high throughput data processing, which has been readily integrated into modern particle physics experiments, namely, Graphical Processing Units (GPUs) and Field Programmable Gate Arrays (FPGAs).

%In order to estimate magnitude of the challenge what is possible, it is useful to draw parallels with HL-LHC. The instantaneous luminosity of the HL-LHC will reach up to 7.5× $10^{34}$ cm$^{−2}$s$^{−1}$, 
%a large increase from Run~2 of the LHC. This will cause the occupancy seen by the LHC detectors increase by approximately a factor of 5. LHC detectors cope with this increase in the occupancy by incorporating precision timing detectors. Combining them with silicon tracking and advanced particle flow calorimetry allows for a highly granular, precision measurement. For the muon collider detectors, timing information can be provided directly by the tracker and the calorimeter system. 

\subsubsection{Tracker}
\begin{figure}[t!]
\centering
\includegraphics[width=0.9\linewidth]{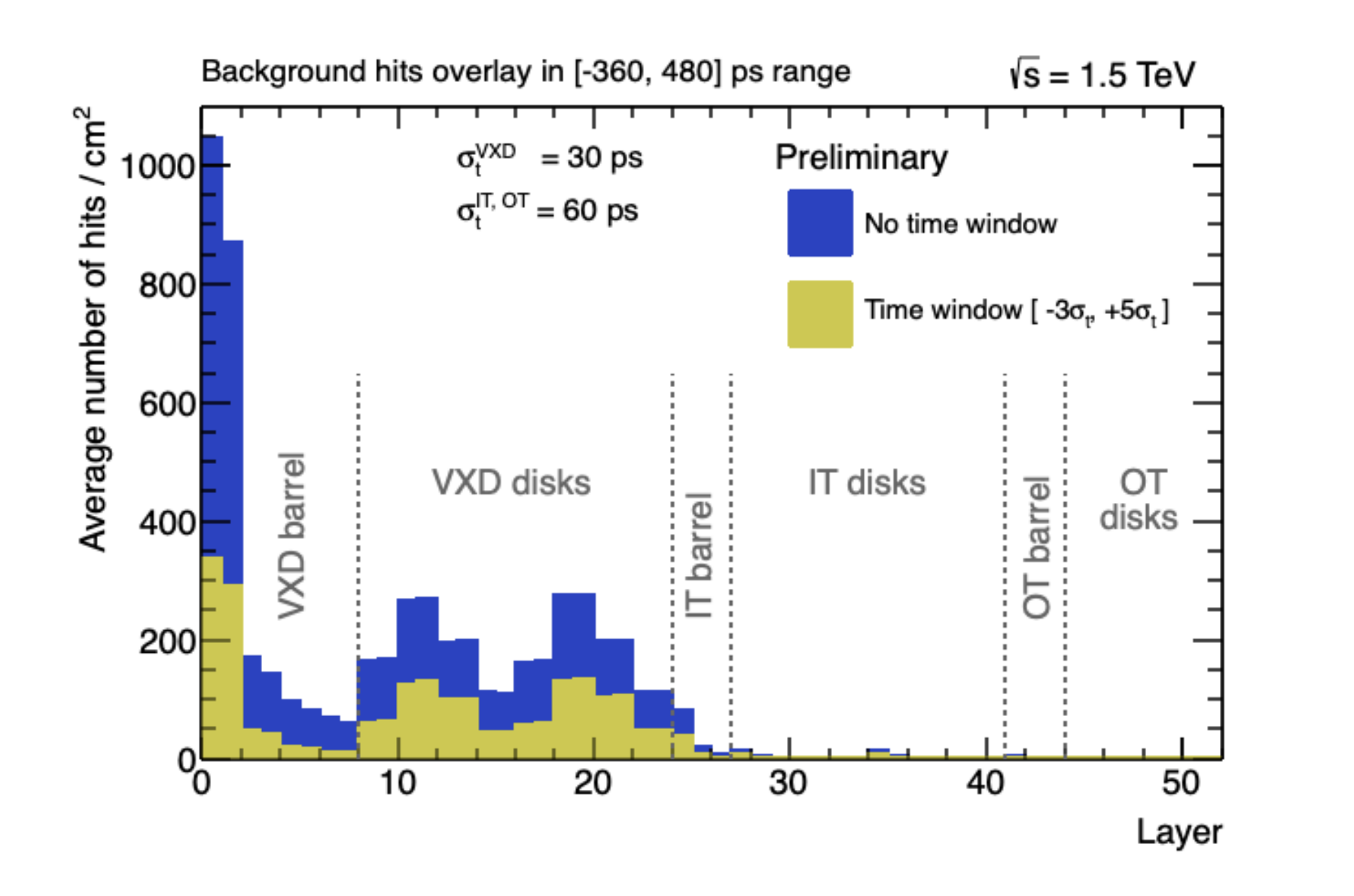}
\caption{Average hit density per bunch crossing in the tracker as a function of the detector layer~\cite{https://doi.org/10.48550/arxiv.2203.07224}.}
\label{fig:trackerhits}
\end{figure}

Closest to the beamline, the tracker suffers from the highest density of BIB. At the track level, the BIB is easy to separate from collision products: it is predominantly low-energy, and does not originate at the interaction point. However, the number of hits produced by the BIB is many orders of magnitude larger than that of the collision products, which creates a challenge for tracker technology, requiring high granularity silicon detectors to avoid saturation. Figure~\ref{fig:trackerhits} shows the hit density per bunch crossing throughout the tracker which reaches 1000 hits/cm$^2$ in the vertex detector. A comparison of the expected hit density (after applying loose timing selection) with ATLAS ITk detector is shown in Table~\ref{tab:hitdensitycomp}. One can see that the density is approximately an order of magnitude larger than in ATLAS in HL-LHC.

\begin{table}[]
    \centering
    \begin{tabular}{c|c|c}
    Detector Layer & ITk Hit Density [mm$^{-2}$] & Muon Col. Hit Density [mm$^{-2}$] \\
    \hline
    Pixel Layer 0 & 0.643 & 3.68\\
    Pixel Layer 1 & 0.22  & 0.51\\
    Strip Layer 1 & 0.003  & 0.03 \\
    \hline
    \end{tabular}
     \caption{Rough comparison of expected hit density at a muon collider detector compared to ATLAS ITk in HL-LHC. The numbers are reported for different radii, corresponding to possible positions for the first and second layers of a pixelated vertex detector and first layer of the silicon strip detector.}
    \label{tab:hitdensitycomp}
\end{table}

Over the past decade several transformative new developments in silicon tracking detector technology have been proposed.  The introduction of Low Gain Avalanche Diodes (LGAD) and similar devices with intrinsic amplification have improved particle tracking time resolutions by a factor of ten, while AC coupled LGADs achieve few-micron position resolution \cite{White_2014, PELLEGRINI201412, CARTIGLIA2015141}. CMOS-based sensors have the potential to lower cost/cm$^2$ by a factor of 5-10 with excellent resolution. In addition, the ability to integrate small pixels with 3D integration or double-sided LGADs can add angular information to each data point. Finally, low power radiation hard extensions of optical transceivers and advanced power delivery systems developed for the HL-LHC will allow for a low mass radiation-hard tracking system.
 
These capabilities are enabling for tracking in the high beam-induced background environment of the Muon Collider. Much of the background is low energy, out of time, and non-pointing. These emerging technologies can reduce the majority of these backgrounds at the sensor level by factors of 10-100 by requiring in-time hits and pixel cluster patterns and energies consistent with tracks emerging from the primary vertex. This would reduce the data load and associated power and mass, simplify triggering, and possibly allow for full tracker event readout every crossing.

Various handles to reduce the BIB can be explored for both on- and off-detector filtering. Possible filtering schemes are described in~\cite{Jindariani:2022gxj} and quickly summarized here:
\begin{itemize}
\item \textbf{Timing}: Removing out of time hits appears to reduce the data load by a factor of 3. Timing information will eventually be needed in the reconstruction, but it makes sense to apply initial filtering on-detector. The time spread expected from the finite size of the interaction region is approximately 20--30 ps.
\item \textbf{Clustering}: Clustering reduces the number of pixel groups read out. This requires more on-detector processing and results in more bits per cluster and a higher power budget, but can reduce the number of hits read out. Selection requirements can also be applied to the cluster shape. The effectiveness needs to be assessed for each BIB cluster type.
\item \textbf{Energy Deposition}: Each of the backgrounds has a characteristic energy deposition signature. For example neutrons have low, localized energy 
deposit. This should be a useful requirement to make on-detector.
\item  \textbf{Correlation Between Layers}: This is a powerful handle for background rejection. However, implementation may be complex and costly, doubling the number of channels. For on-detector filtering, it also requires transfer of data between layers in a very busy environment.
\item \textbf{Local Track Angle}: Track angle measurement can be made in a single detector if the thickness/pitch ratio distributes the signal over several pixels. This avoids the complexity of inter-detector connections and could provide a monolithic solution~\cite{Lipton:2022njd,Lipton:2019drv}.
\item \textbf{Pulse Shape}: Signals from BIB can come with a variety of angles and may not give the deposit profile and pulse shape of a typical MIP.  Appropriate pulse processing, such as multiple sampling, RC-CR filters, zero crossing, or delay line clipping can be used to further reduce the data load.
\end{itemize}

The basic trade-offs are between the complexity, power, and mass needed to implement a on-detector filter, and the benefit of reduced data rate. Currently, the simulated tracker relies primarily on using precision timing information to suppress the BIB. Studies of the other handles outlined above are ongoing and demonstrate good potential for further improvements.

It is evident that developments in 4D tracking are an essential tool for reducing the BIB. New detector technologies, such as the LGADs being used in the ATLAS and CMS Phase-2 upgrades, can provide 20-30 ps per hit timing resolution. Simple timing window requirements that take advantage of this tens of picosecond timing can reduce the background by a factor of three or more ~\cite{https://doi.org/10.48550/arxiv.2203.07224}.

The double-layered tracker strategy currently being employed by the CMS Experiment for the HL-LHC \cite{CERN-LHCC-2017-009} provides extra handles for BIB reduction. Double-layer correlation significantly reduces backgrounds, and can be used to make measurements of local track angles. These measurements can be used to reject particles emanating from the BIB-blocking tungsten nozzles. 

\begin{figure}[t!]
\centering
\includegraphics[width=0.8\linewidth]{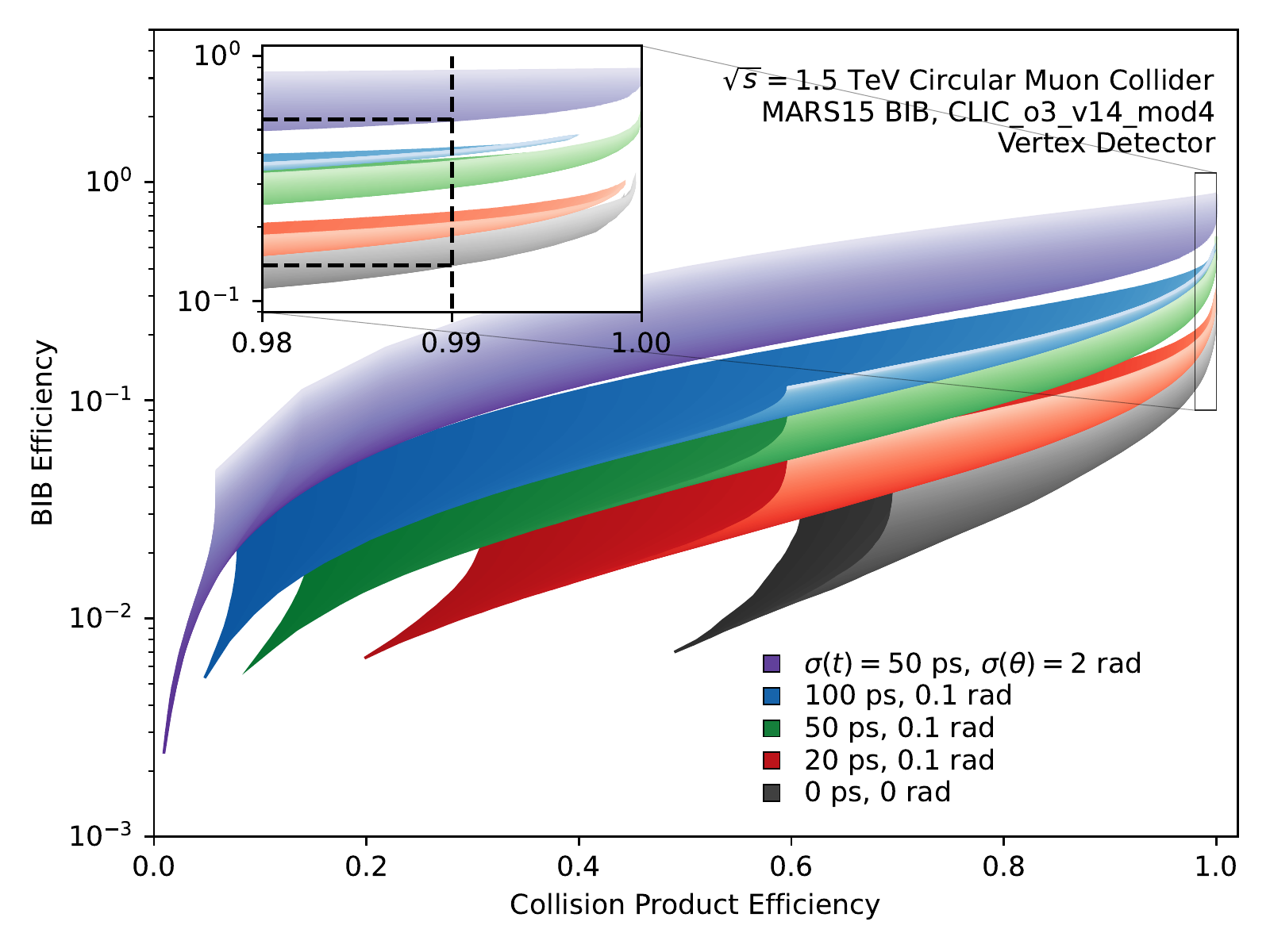}
\caption{Performance of a BIB-reduction algorithm at the hit level of the vertex detector, after a fiducial timing window selection of [-250, +300] ps. An ideal resolution case is shown in black, while the other colors represent a range of timing and angular resolutions. Different hue ribbons represent different detector capabilities, while the surface of the ribbon represents different choices for cut values~\cite{https://doi.org/10.48550/arxiv.2203.06773}.}
\label{fig:bibcuts}
\end{figure}

Figure~\ref{fig:bibcuts} shows potential hit-level BIB reduction from a combination of angular and timing variables designed to isolate collision products. On top of a fiducial timing window that already substantially reduces this background, these requirements have the potential to eliminate another order of magnitude of BIB hits in the vertex detector, before entering the tracking stage, substantially reducing the combinatoric problem introduced by the BIB. More complex analysis of pulse shapes and hit cluster characteristics could potentially reduce this background even further.

\subsubsection{Calorimetry}
%In a multi-TeV muon collider, the calorimeter system must operate in an intense flux of low energy particles arising from the BIB. The BIB in the calorimeter region is mainly formed by photons (96\%) and neutrons (4\%). 
A calorimeter design for a muon collider must be able to disentangle photons and neutrons produced by beam muon decays upstream of the interaction point (IP) from prompt, high-energy electromagnetic and hadronic showers produced by the $\mu^{+}\mu^{-}$ collision at the IP.  The beam-induced background (BIB) consists primarily of soft photons ($p_{\gamma}\,\approx\,2\,\MeV$) and moderately energetic neutrons ($p_{n}\,\approx\,500\,\MeV$)~\cite{Bartosik:2019dzq}. The occupancy of the showers produced by these particles in the calorimeter is high.  A large fraction of the BIB can be rejected with a short readout window, of order a few ns or less, around the nominal beam crossing.  This requires fast signal formation in the front ends. With high lateral granularity, advanced pattern recognition may be used to reject the non-pointing background of low-energy BIB photon and neutron showers.  Finally, the calorimeter active materials and front end electronics need to withstand the 1-MeV-equivalent neutron fluence (a bit less than expected at HL-LHC) and total ionizing dose (worse than expected at HL-LHC) of this BIB.

The amount of ambient diffuse energy in the calorimeters is a quantity that provides a measure of energy deposited by the beam background that needs to be corrected for during reconstruction. To estimate this quantity we deployed a 
grid-median background estimation technique that divides the calorimeter into an arbitrary grid and sums up the energy within the grid~\cite{Soyez:2018opl}. We find that the average ECAL+HCAL ambient energy is 50 GeV per unit area when integrated over the entire detector rapidity region. This number is similar to approximately 40 GeV at HL-LHC.

In the last decade, advances have been made in the areas of precision timing and high granularity that pave the way towards a workable muon collider calorimeter design.  Cell sizes as small as 13~mm$^{2}$ have been realized in prototype silicon sensors for the SiD detector, one of two validated detector designs for the ILC.  These prototypes have been exposed to cosmic rays and test beam electrons and shown to function properly~\cite{Barkeloo:2019zow}.  Similar sensors are being used for the HGCal endcap calorimeter upgrade of the CMS detector for the HL-LHC, but with cell sizes of $\approx\,0.5$ and $\approx\,1.1$ cm$^{2}$.  Prototyping of these sensors is well advanced, with positron test beams indicating that an EM energy resolution of 22\%/$\sqrt{E[\GeV]}\,\bigoplus\ 0.6$\% is achievable~\cite{CMSHGCAL:2021nyx}.  This is not as good as homogeneous crystal calorimeters, but demonstrates reasonable performance with a particle flow technology that is optimized for jet energy resolution and BIB rejection.

For hadronic calorimetry, compact arrays of scintillator or crystal tiles glued directly over PCBs holding silicon photomultipliers (SiPMs) for readout have been shown to be a scalable, cost-effective active material choice. The CALICE Collaboration, whose mission is to develop particle flow calorimetry technologies for any of the linear $\ee$ collider options, has already demonstrated in the early 2000s that hadronic energy resolutions of $\approx$ 60\%/$\sqrt{E[\GeV]}$ are possible with the so-called ``SiPM-on-tile'' technology~\cite{Sefkow:2015hna}.  A 2018 pion test beam exercised the feasibility of constructing and operating a large-scale hadronic calorimeter slice---the $72\,\mathrm{cm}\,\times\,72\,\mathrm{cm}\,\times\,38\mathrm{-layer}$ prototype alone consisted of 22k $3\,\times\,3\,\mathrm{cm}^{2}$ channels~\cite{Krueger:2021abc}.  Even smaller cell sizes of $\approx 1\mathrm{cm}^{2}$ with binary readout have been realized using resistive plate chambers~\cite{Ruchti:2022ixx}.

Important advances have also been made in precision timing and high-speed front ends.  As mentioned above, silicon low-gain avalanche detectors (LGADs) have achieved 30 ps timing resolution for minimum-ionizing particles in test beams.  Discrimination of pulses arriving 5 ns apart and single-cell time resolution of 20 ps has been demonstrated in PbF$_{2}$ crystals with SiPM readout for the Muon g-2 electromagnetic calorimeter~\cite{Kaspar:2016ofv}.  Many scintillators with ultra-fast decay times of less than 1-10 ns, for example BaF$_{2}$, have been identified and characterized for light yield, timing performance, and radiation hardness~\cite{Yeh:2022yog}.  Candidate inorganic scintillators, both in crystal and ceramic form, that survive up to 1-100 Mrad with light loss no worse than 70\% have been identified~\cite{Zhu:2019ihr}.

In~\cite{https://doi.org/10.48550/arxiv.2203.07224}, four key features are identified that will enable good energy measurements at a muon collider experiment:

\begin{itemize}
    \item \textbf{High granularity} to reduce the overlap of BIB particles in the same calorimeter cell. The overlap can produce hits with an energy similar to the signal, making harder to distinguish it from the BIB;
    \item \textbf{Good timing} to reduce the out-of-time component of the BIB. An acquisition time window of about $\Delta t = 300$~ps could be applied to remove most of the BIB, while preserving most of the signal. This means that a time resolution in the order of $\sigma_t = 100$~ps (from $\Delta t \approx 3 \sigma_t$) should be achieved. The CMS BTL based on LYSO:Ce bars with SiPM readout on each end integrated a precision $30$ ps timing layer with a high resolution crystal calorimeter~\cite{Addesa:2022qlt};
    \item \textbf{Longitudinal segmentation}: the energy profile in the longitudinal direction is different between the signal and the BIB, hence a segmentation of the calorimeter can help in distinguishing the signal showers from the fake showers produces by the BIB;
    \item \textbf{Good energy resolution} of $\frac{10\%}{\sqrt{E}}$ in the ECAL system for photons and a jet energy resolution of $\frac{35\%}{\sqrt{E}}$ are expected to be enough to obtain good physics performance. These have been demonstrated for conceptual particle flow calorimeters.
\end{itemize}

Stemming from the ILC R\&D program, two major approaches are being studied to exploit sampling calorimeters and improve upon the current generation of collider experiments: multi-readout (dual or triple)~\cite{Lee:2017shn,dr20} and particle flow~\cite{Thomson:2009rp} calorimetry.
The first approach focuses on reducing the fluctuations in the hadronic shower reconstruction, which are the main responsible for the deterioration in the determination of the jet energy. This goal is achieved by measuring independently the electromagnetic and the non-electromagnetic components of a hadronic shower, thus allowing to correct event-by-event for the different response of the calorimeter to various particle species. Notable examples of of the Dual Readout calorimeters can be found in the work of the DREAM/RD52 collaboration~\cite{Wigmans:2007es} and the proposed IDEA~\cite{FCC:2018evy, CEPCStudyGroup:2018rmc} detector for FCCee/CepC.

The second approach focuses on the reconstruction of the four-momenta of every particle recorded by the detector. This method exploits tracking information and requires a detector with extreme granularity, combined with powerful reconstruction algorithms aimed at resolving each particle's trajectory through the whole detector. Silicon sensors can be used as active elements to achieve a high channel granularity and longitudinal segmentation. State-of-the-art silicon sensors can sustain the high radiation dose of the expected BIB. Analogous technologies are being adopted by CMS HGCal~\cite{CERN-LHCC-2017-023} and considered by the CLIC collaboration. HGCal implements a precise timing measurement in these sensors ($<$100 ps), making the approach usable at a muon collider.

An alternative hybrid approach implements dual-readout methods with a segmented crystal ECAL and fiber HCAL to further improve the particle flow performance of jets with intrinsically high resolution calorimetry.  The fine traverse granularity, longitudinal segmentation and precision timing provide additional handles to suppress BIB, increase particle identification and tracking matching with individual calorimeter hits~\cite{Lucchini:2022vss}. Further development toward test beam evaluations is being organized as part of the CalVision Detector R\&D~\cite{Lucchini:2020bac}.

Comparisons of performance of different calorimeter technologies in the muon collider environment have not yet been done. A comparison for FCC-ee can be found in Ref.~\cite{Aleksa:2021ztd}.

\subsubsection{Readout and Computation}
With a single bunch collider operation scheme, beam crossing frequency is defined by the beam energy and the size of the collider ring. Collisions are expected to happen at the maximum rate of 100 kHz, corresponding to the minimum time between crossings of 10 $\mu$s.

The intense beam induced backgrounds that will be seen at a Muon Collider will present real computational challenges. Recent advances in real-time particle reconstruction techniques for reducing the amount of saved data for offline processing. At any experiment where the bulk of events contain interesting signal, using a trigger system to down-select events is undesirable. In order to reduce the data size two methods could be employed: implementation of selections based on timing windows matching particles to the primary vertex and processing detector signals in real-time and saving only a subset of the high-level reconstructed event information computed in the final software trigger stage is recorded and sent to permanent storage. The former is already implemented in many detectors where signals are only stored within a window of the primary interaction. The latter requires real-time reconstruction and has been employed at LHCb since Run-2.

Initial loose data filtering can happen on-detector.The basic trade-offs are between the complexity, power, and mass needed to implement a on-detector filter, and the benefit of reduced data rate. Overly aggressive front-end filtering schemes can introduce irrecoverable inefficiencies and biases in the data and limit acceptance for certain BSM physics signatures, such as these of long-lived particles. 

Preliminary estimates of data rates based on simulation ~\cite{Jindariani:2022gxj} indicate that a streaming DAQ architecture can provide an attractive solution for a future Muon Collider experiment. With improvements to the tracking speed, such a solution can likely be realized with technologies available today. Future advancements (e.g. higher speed optical links, fast processors, etc) are likely to result in a smaller and/or more performing DAQ system. Work should be invested in improving HLT reconstruction algorithms and exploiting hardware acceleration schemes with the aim to bring per-event processing time down to a few second level. 

%% file: det-perf.tex
\begin{figure}
    \centering
    \includegraphics[width=0.8\textwidth]{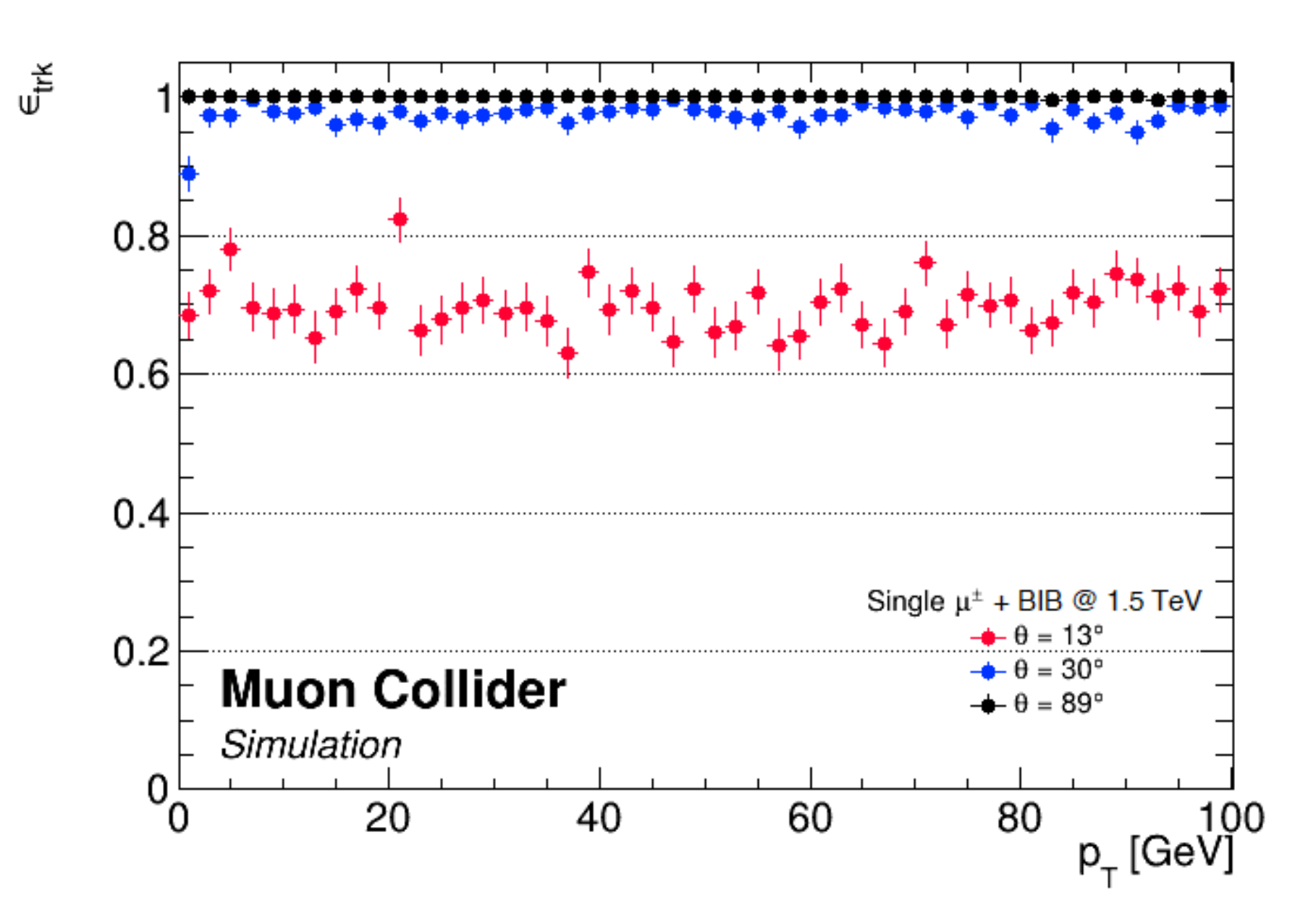}
    \includegraphics[width=0.8\textwidth]{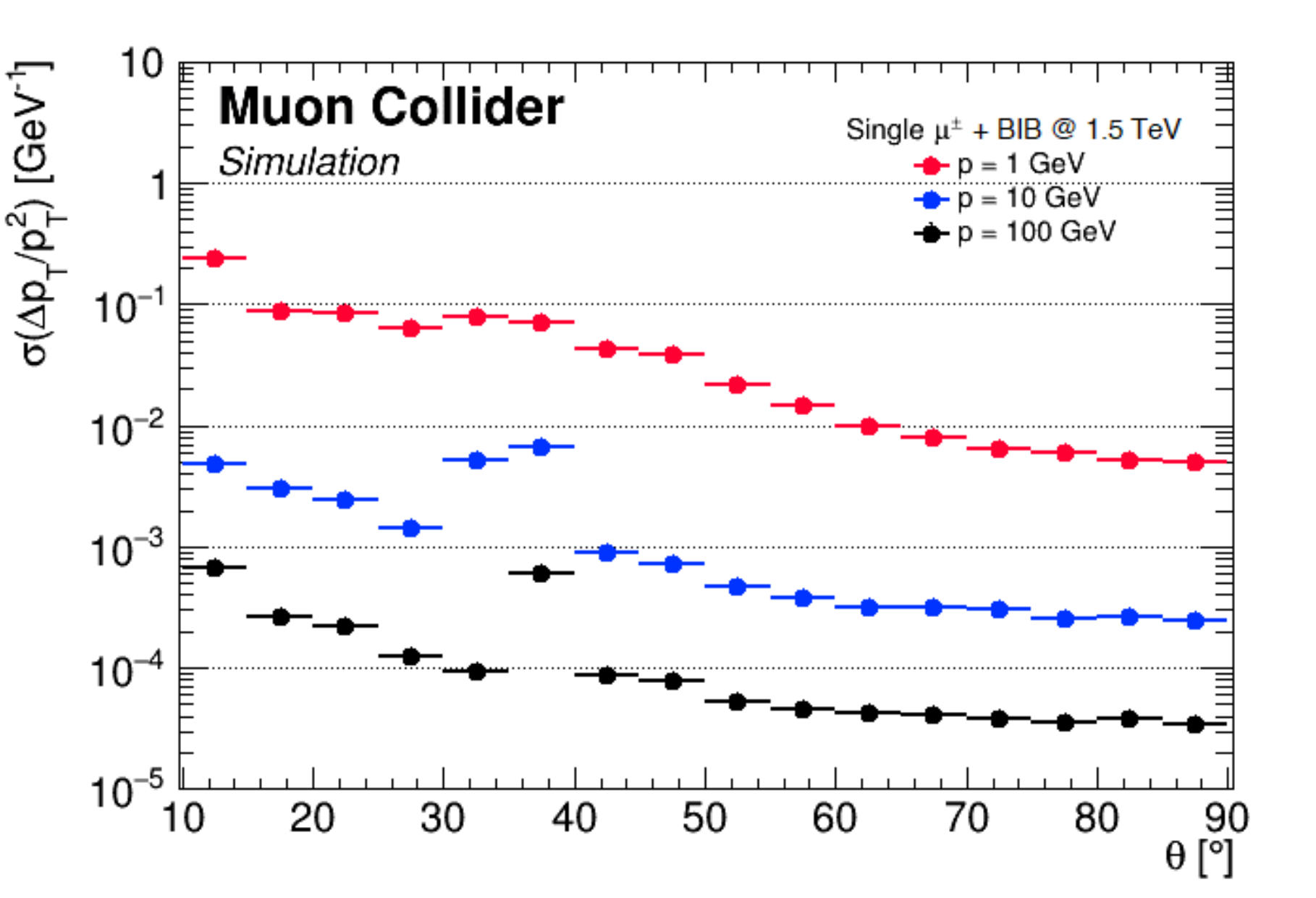}
    \caption{Tracking efficiency for single muon in events with beam-induced background versus transverse momentum (top) and resolution versus $\theta$ (bottom) from ~\cite{MuonCollider2022ded}. The loss of efficiency at small angle comes from the effects of the nozzle used to limit the beam-induced background flux into the detector. Work to improve tracking algorithm and to recover the loss of efficiency is ongoing. }
    \label{fig:tracking}
\end{figure}

A set of reconstruction algorithms were developed and tuned as described in detail in Ref.~\cite{MuonCollider2022ded}. As an example, Fig.~\ref{fig:tracking} shows that charged particles can be successfully reconstructed in the tracker system with momentum resolutions comparable or better than the ones expected for HL-LHC. The developed algorithms also show that running a full track reconstruction in such a busy environment is feasible in a reasonable amount of time (minutes) without huge optimization efforts, proving that the large combinatorics will not be a significant problem in the tracker of a muon collider detector that satisfy the requirements outlined above. It should be noted that time information was used merely in measurement pre-selection and it is not actively used in the track reconstruction at this stage; using that information in track reconstruction (as noted in Section~\ref{sec:det:det-new}) is expected to further improve performance.

\begin{figure}
    \centering
    \includegraphics[width=0.8\textwidth]{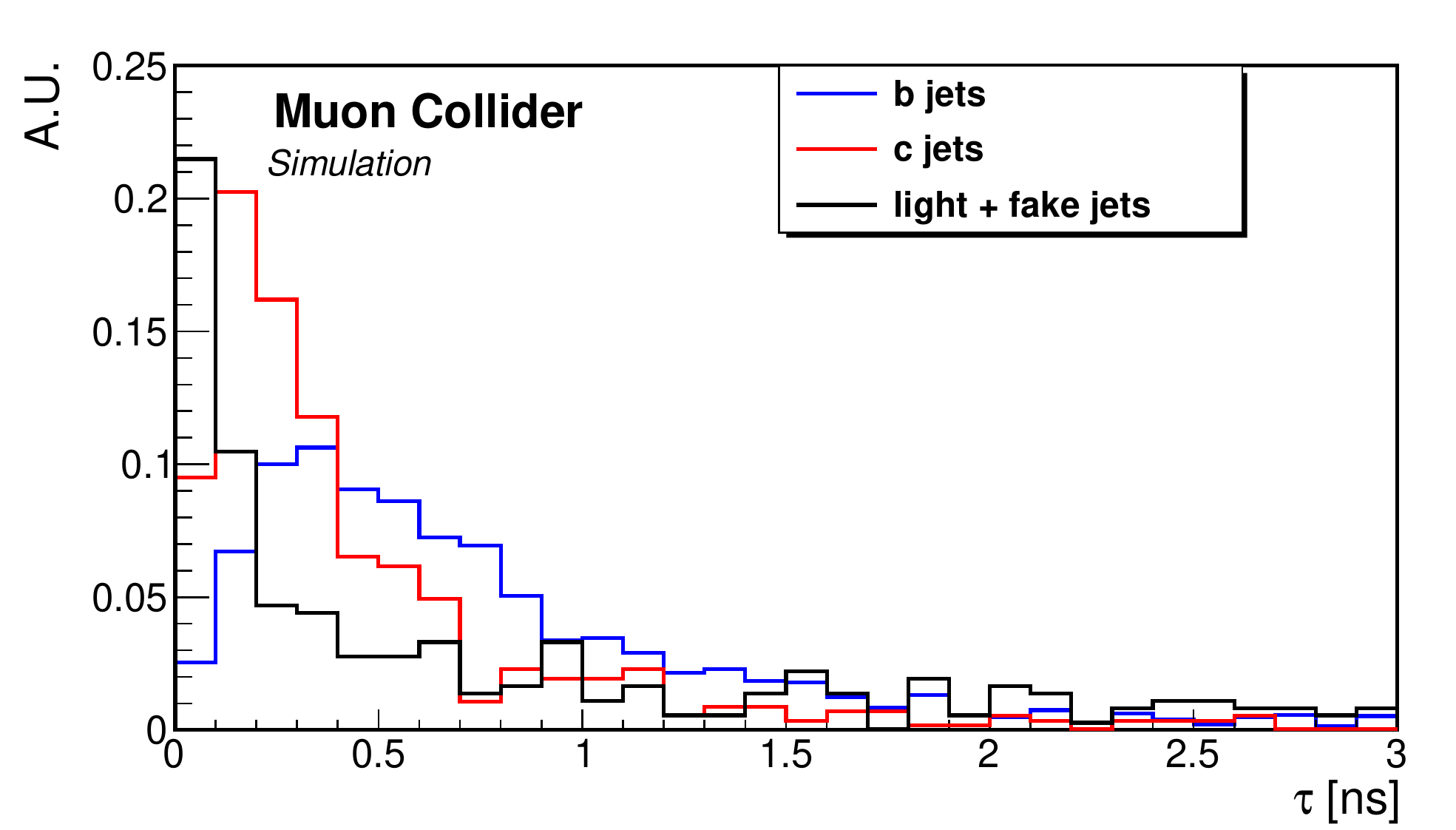}
    \includegraphics[width=0.8\textwidth]{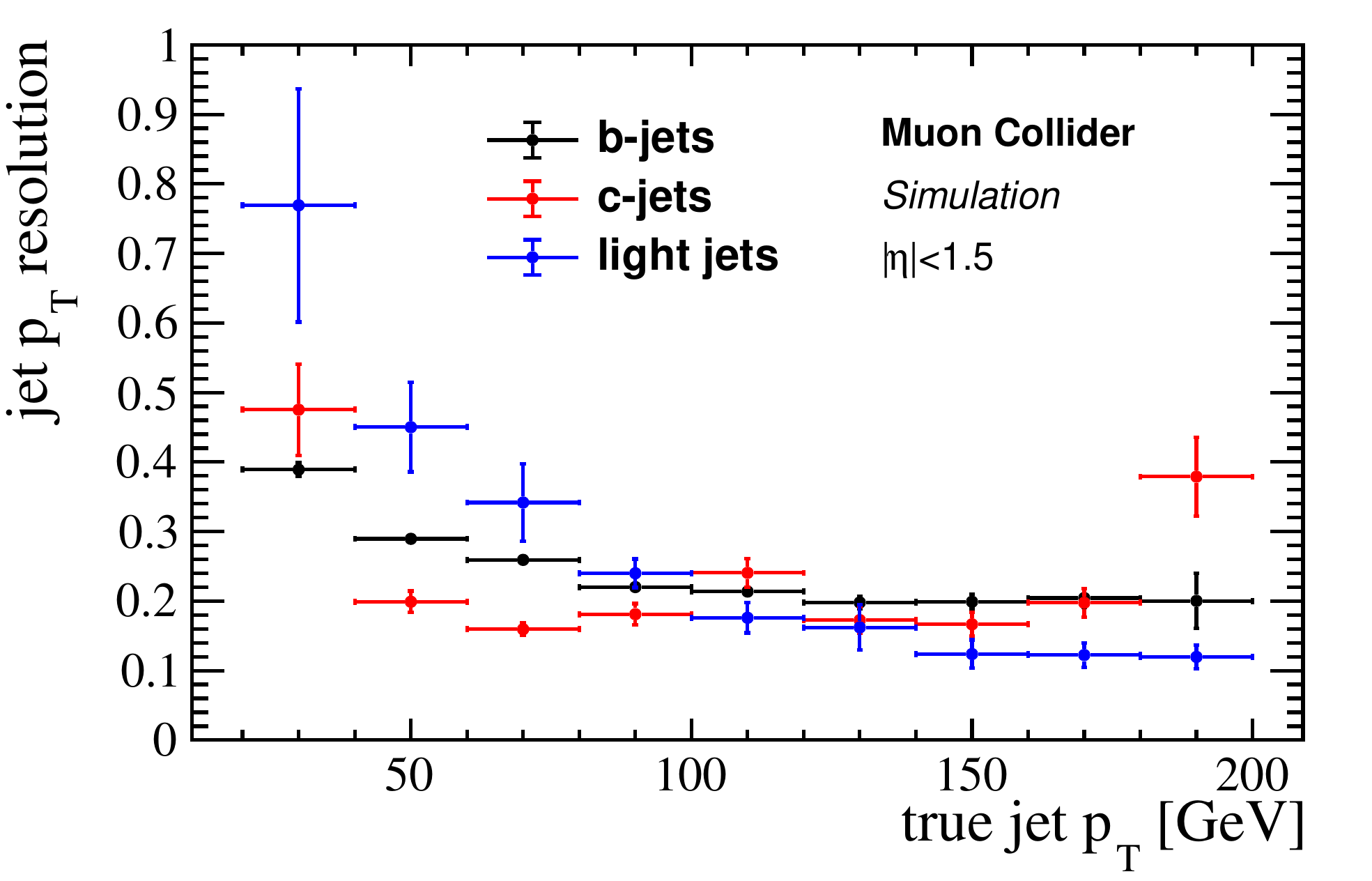}
    \caption{Distribution of the secondary vertex proper lifetime for b, c and light tagged jets (top). Distributions are normalized to the unit area. Jet transverse momentum ($p_T$) resolution as a function of the $p_T$ of the jet for $b-$, $c-$, and light jets (bottom). Differences between flavors are mainly due to different angular distributions of jets.}
    \label{fig:btagandjets}
\end{figure}

Figure~\ref{fig:btagandjets}~(left) shows the output of a simple secondary-vertex reconstruction algorithm, proving that separation between b, charm and light quarks is achievable. The performance of such a simple secondary-vertex tagger is comparable to the one of analogous algorithms in use at the LHC in terms of efficiency and rejection for each flavor. 
%TODO: would be great to have a comparison

Jet reconstruction is also affected by unique challenges due to the diffuse BIB component. While on a calorimeter-level this is analogous to a very high level of pile-up, 
%TODO: comparison of energy density with HL-LHC and FCC-hh?
the additional tracking information is useful to disentangle the main collision products, together with an average subtraction of the expected BIB energy deposition. Figure~\ref{fig:btagandjets}~(right) shows the expected jet energy resolution from a relatively simple particle-flow based jet reconstruction algorithm with minimal optimization on BIB energy subtraction. The resolution is comparable to the ones achieved by LHC when using analogous simple calibration algorithms. 
%TODO: it would be very very nice to make a comparison plot indeed in this case.

The reconstruction of muons, electrons and photons have been implemented and assessed as well in Ref.~\cite{MuonCollider2022ded}, and shown to be viable. While more advanced studies will surely be needed to establish a full set of expected performance, the current studies present, for the first time, a complete picture that builds confidence in the ability to extract satisfactory performance in the muon collider environment for the basic high-$p_T$ objects needed for the physics program outlined earlier in this report.

The aim of the these reconstruction algorithms is, at this stage, to show what minimal performance is surely achievable, while the ultimate performance is almost certain to be vastly better than the projections shown.

%% file: fullsim-Hbb.tex
In this study the full simulation of the Muon Collider experiment is used to determine the statistical sensitivity on the measurement of the $H \rightarrow b \bar{b}$ cross section at $\sqrt{s} =$3 TeV, assuming an integrated luminosity of 1 ab$^{-1}$. 

The signal $\mu^+ \mu^- \rightarrow X H(\rightarrow b \bar{b})$ has been generated with WHIZARD, where $X$ are two muons or neutrinos. The background $\mu^+ \mu^- \rightarrow X  q \bar{q}$ has been generated with WHIZARD as well, considering diagrams not mediated by the Higgs bosons, and with $q=b,c$. Contributions from light jets are considered negligible, since a heavy flavour tagging technique is applied as explained later. A number of 10k signal events and 10k background events are generated in this way.

The detector configuration and simulation is described in Sec. \ref{sec:det:current-configuration}. The beam-induced background is included. Jets are reconstructed using a Particle Flow algorithm for selecting tracks and calorimeter clusters, and the $k_t$ algorithm with radius $R=0.5$ is used for the clustering. In order to select jets in the region with the best performance in term of reconstruction efficiency and jet energy resolutions, the requirements $p_T>40$ GeV and $|\eta|<2.5|$  are applied.

The heavy flavour identification efficiencies and misidentification rate have been determined with independent samples of $b\bar{b}$, $c\bar{c}$ and light jets. Secondary vertices (SVs) are reconstructed using tracks in the jet cones, and a jet is tagged as heavy flavour if at least one SV is found. The light jet misidentification rate is found negligible, and for this reason light jets have not been included in the background. The tagging efficiencies have been determined as a function of the jet $p_T$ and polar angle ($\theta$). These efficiencies are then applied to the reconstructed $H \rightarrow b \bar{b}$ and background samples with a reweighting technique.

The dijet invariant mass distributions for signal and background are then fitted with double-gaussian pdfs, in order to obtain the signal and background models. The number of expected signal and background events is determined by considering the WHIZARD cross sections, the total selection efficiency and the integrated luminosity. In particular, about 59.5k $H \rightarrow b \bar{b}$ events and 65.4k background events are expected to be collected with 1 ab$^{-1}$.

\begin{figure}[!h]
\center
\includegraphics[width=0.8\textwidth]{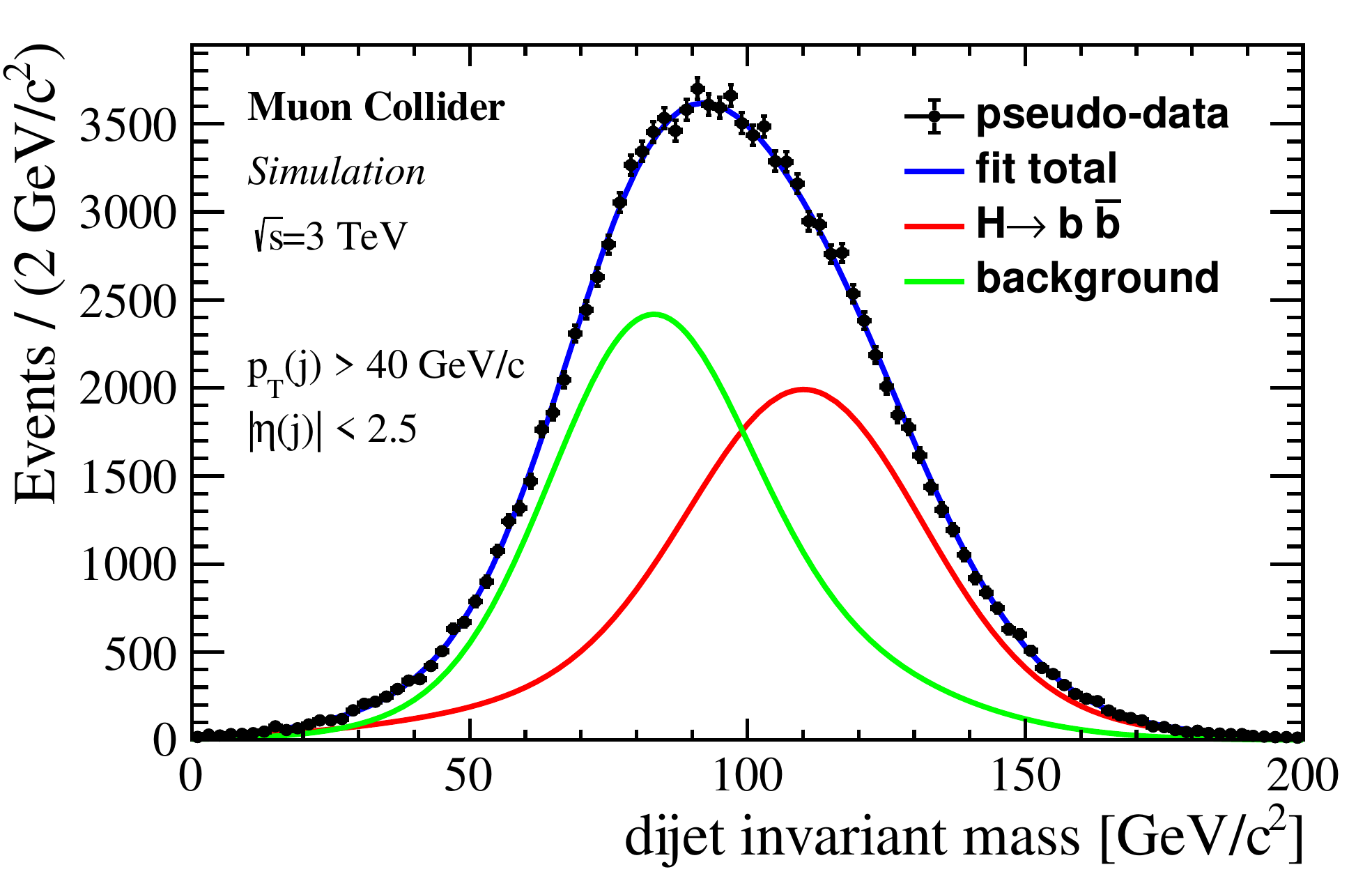}
\caption{Result of the dijet invariant mass fit used to extract the $H \rightarrow b \bar{b}$ yield and uncertainty. Pseudo-data are obtained by exploiting the Muon Collider experiment simulation at $\sqrt{s} =$3 TeV,  and assuming an integrated luminosity of 1 ab$^{-1}$.}
\label{fig:hbb_fit}
\end{figure}

The signal and background invariant mass models and the expected number of events are used to generate pseudo-data. The pseudo-data are then fitted with the invariant mass models, by using an unbinned maximum likelihood fit, and by letting the signal and background yields float. In this way the measured $H \rightarrow b \bar{b}$ yield is extracted. A result of one of this fits is shown in Fig.~\ref{fig:hbb_fit}. The uncertainty on the signal yield obtained from the fit is $0.75\%$. Several pseudo-experiments (1k) have been performed to check the stability of this result, and to rule out possible biases. This uncertainty can be taken as the statistical uncertainty on the measurement of the $H \rightarrow b \bar{b}$ cross-section.

The sensitivity in this channel, among others, has been studied in \cite{Forslund:2022xjq} including physics backgrounds using fast simulation at both 3 TeV and 10 TeV. Here we briefly summarise the analysis strategy used there at 3 TeV for comparison with the above full simulation results. Events were generated using {\sc MadGraph5} \cite{Alwall:2014} and showered with {\sc Pythia8} \cite{Sjostrand:2014zea}, with detector reconstruction and performance approximated by using the muon collider detector card included in the latest {\sc Delphes} \cite{deFavereau:2013fsa} releases. Jets were clustered using the exclusive Valencia jet clustering algorithm with $R=0.5$ and $N=2$. After applying a $p_T$ correction factor, preselection cuts of $p_T>40$ GeV and $|\eta|<2.5$ were applied.

A flat flavour tagging efficiency of 50\% was taken for $b$-jets, with a $c$-jet misidentification rate of roughly 1-3\%, and negligible light-jet mistagging rate. Events with two $b$-tagged jets passing the preselection cuts were then subject to a cut on the reconstructed dijet invariant mass of $100<m_{jj}<150$ GeV. The sensitivity to the channel was then estimated using $\Delta \sigma / \sigma = \sqrt{S+B}/S$, where $S$ and $B$ are the number of surviving signal and background events, respectively. The obtained precision is 0.76\% for the total VBF $\mu^+ \mu^- \rightarrow X H(\rightarrow b \bar{b})$ signal.

The results in \cite{Forslund:2022xjq} include some diboson backgrounds such as $\mu^+\mu^-\rightarrow \mu^\pm \nu_\mu W^\mp H$, so the precision presented there is not directly comparable to the full simulation result above. In order to obtain a closer comparison, we consider a modification of the results such that we only include the $\mu^+\mu^-\rightarrow Xq\bar{q}$ background while still considering the total $\mu^+\mu^-\rightarrow XH(\rightarrow b\bar{b})$ as the signal. Doing this yields a precision of 0.73\% in fast simulation, to be compared to the 0.75\% obtained in the full simulation fit above.

It is worth mentioning that the largest impact of the BIB on this channel's analyses is the worsened jet energy resolution, which results in a larger overlap of the $H$ and $Z$ peaks, as can be seen in Fig.~5 of Ref.~\cite{Forslund:2022xjq}, where using a worse jet energy resolution than the {\sc Delphes} default in fast simulation reproduced a very similar distribution to Fig.~\ref{fig:hbb_fit}. Of course, the fast simulation result uses less sophisticated flavour tagging and a simpler analysis strategy, so the comparison is not quite one to one. Nevertheless, we find the sensitivity for this channel in fast simulation and full simulation to be very similar, regardless of the presence of the BIB.

%% file: fullsim-DT.tex
Long-lived particles (LLPs) appear in a variety of models and yield a large range of signatures at colliders~\cite{Curtin:2018mvb, Alimena:2019zri}. Depending on the LLP quantum numbers and lifetime, these can span from LLP decay products appearing in the detector volume, even outside of the beam crossings, to metastable particles with anomalous ionisation disappearing after a short distance.

The higgsino is among the most compelling dark matter candidates, with tight connections to the naturalness of the weak scale, which could lead to charged LLPs ( $\tilde \chi^{\pm}$) being produced in particle collisions and then decaying in the volume of the tracking detectors (e.g. decay lengths between 1~mm and 500~mm).

Searches at the LHC are actively targeting this scenario~\cite{ATLAS:2022rme,Aaboud:2017mpt,Aad:2013yna,Sirunyan:2020pjd,CMS:2014gxa}, but are not expected to cover the relic favoured mass of 1.1~TeV~\cite{Hisano:2006nn, ATL-PHYS-PUB-2018-031, EuropeanStrategyforParticlePhysicsPreparatoryGroup:2019qin}. Different studies have assessed the reach of a muon collider in the search for these particles with fast~\cite{Han:2020uak,Han:2022ubw} and full~\cite{Capdevilla:2021fmj} simulation.
This document reviews these two results and provides comparison of the expected reach from each approach as a function of the main parameters of the model.

The production of pairs of electroweakinos at a muon collider operating at multi-TeV centre-of-mass energies proceeds mainly via an s-channel photon or off-shell Z-boson, with other processes, such as vector boson fusion, being subdominant~\cite{Han:2020uak}. The decay products of the $\tilde \chi^\pm$ are assumed to be undetectable.

In the work described in Ref.~\cite{Han:2020uak,Han:2022ubw}, events are selected with an ISR photon with minimal $p_T$ of 25~GeV, and one or two disappearing tracks. We assume the reconstruction probability of a signal event with one disappearing track is
\begin{eqnarray}
\epsilon_\chi (\cos\theta,\gamma, d_T^{\rm min}, d_T^{\rm max})=\exp\left(\frac {-d_T^{\rm min}} {\beta_T \gamma c\tau}\right)-\exp\left(\frac {-d_T^{\rm max}} {\beta_T \gamma c\tau}\right),\\ \nonumber
\text{with}~d^{\rm min}_T=5\ {\rm cm}~,d^{\rm max}_T=17\ {\rm cm}~{\rm and~} |\eta_\chi|<1.5
\label{eq:dtcut}
\end{eqnarray}
where $\gamma=E_\chi/m_{\chi}$
and 
$\beta_T=\sqrt{1-1/\gamma^2} \sin\theta$, which is the transverse velocity in the lab frame. 
The minimal transverse displacement of $5$~cm represents the minimal track reconstruction requirement (of two hits) for a typical muon collider detector design with pixel layers. A maximal transverse displacement of 17~cm to capture the ``disappearing'' signature and comparable to the detector simulation-based study. In principle, longer tracks are easier to be reconstructed and separated from background. Future detector design studies could further optimize the layout for Higgsino-like short tracks, e.g., moving pixel layers closer to the beam spot, that could greatly improve the muon collider sensitivities.

The study described in Ref.~\cite{Capdevilla2021fmj} investigates the prospects for such a search exploiting a detector simulation based on 
\GEANT 4~\cite{Agostinelli2002hh} for the modelling of the response of the tracking detectors, which are crucial in the estimation of the backgrounds. 
The simulated events were overlaid with beam-induced background events simulated with the MARS15 software~\cite{Mokhov2017klc}.

The analysis strategy relies on requiring one (\SRotp) or two (\SRttp) disappearing tracks in each event in addition to a 25~GeV ISR photon. Additional requirements are imposed on the transverse momentum and angular direction of the reconstructed tracklet and on the distance between the two tracklets along the beam axis in the case of events with two candidates. The disappearing track candidates are required to be reconstructed from at least four silicon detector hits (corresponding to a radius of 5.1 cm), and to have no associated hits beyond the first layer of the inner tracker (corresponding to a radius of 12.7 cm). This veto condition is relaxed to the middle layer of the outer tracker (corresponding to a radius of 115.3 cm) for one of the two candidates in \SRttp. 

\begin{figure}[t]
\centering
\includegraphics[width=0.9\textwidth]{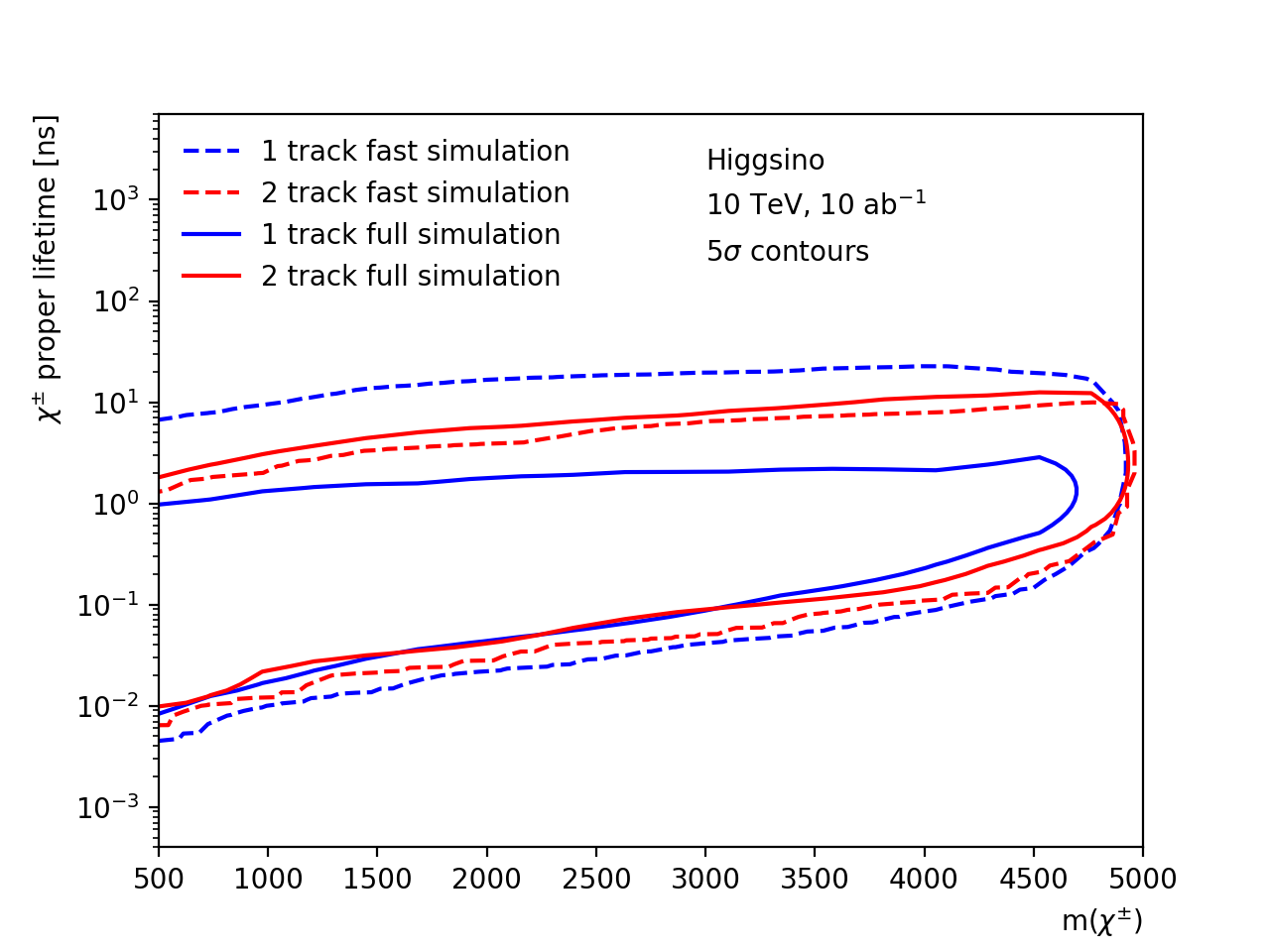}
\caption{Expected sensitivity using 10~ab$^{-1}$ of 10~TeV $\mu^{+}\mu^{-}$ collision data as a function of the $\tilde \chi^\pm$ mass and lifetime. The contours represent the $5~\sigma$ discovery expectation. The solid lines show the predictions from full simulation, while the dashed lines the predictions from the fast simulation.}
\label{fig:comparison}
\end{figure}

The results of the fast and full simulation approaches are compared in  Figure~\ref{fig:comparison} in terms of expected discovery sensitivity. The expected backgrounds are taken from the full detector simulation from Ref.~\cite{Capdevilla:2021fmj}. The five sigma discovery lines correspond to the regions of the phase space that predict more than 60 signal events in \SRotp, or more than $7.5$ signal events in \SRttp.

The results are found to be in good agreement over a wide range of higgsino masses and lifetimes, and well above the current and expected collider limits. In the most favourable scenarios, the analysis of 10~ab$^{-1}$ of 10~TeV muon collisions is expected to allow the discovery of $\tilde \chi^\pm$ masses up to a value close to the kinematic limit of $\sqrt{s}/2$. 
The results obtained with the theory analysis generally predict a wider coverage than the full simulation. This difference is attributed to the simplified description of the disappearing track reconstruction efficiency, which could be accounted for by applying an averaged signal efficiency to the reconstruction, and the larger acceptance given by the veto condition being imposed at $d_T^{\rm max}=17$~cm.  
For lifetimes above 1~ns this difference is inverted in the case of the two track selection, with the full simulation results predicting a marginally better sensitivity. This difference is explained by a difference in the event selection, where the increased event acceptance due to the looser disappearing condition (applied on the middle layer of the outer tracker) counterbalances the lower reconstruction efficiency predicted by the full simulation.

%% file: det-improv.tex
Optimization of the detector for muon collider environment is far from complete. Significant improvements are possible in all aspects of the detector. Some of the items that need early study are:
    \begin{itemize}
        \item \textbf{MDI}: Beam induced backgrounds are significant at the muon collider. Current BIB simulations are done with $E_{cm}=1.5$ TeV, with some limited extrapolation to $E_{cm}=3.0$ TeV. Proper simulation of target energy of $E_{cm}=10$ TeV is amongst the most important items for detector optimization. Options for the absorber material, shape and size determination need to be made with both background mitigation for physics object reconstruction point of view and the lifetime of the sensitive material of the planned detectors. The work requires close collaboration with accelerator lattice designers.
        \item \textbf{Magnet}: What is the size of the magnet and the field? The answer depends on the $E_{cm}$. Additional dependence is on the size and shape of the nozzle used to mitigate the beam induced backgrounds.
        \item \textbf{Tracker}: The location of the inner layers, number of tracking layers needed, segmentation size of the pixels and strips, etc. Of particular concern is the mitigation of single low energy BIB particle hits, which can result in very large data size and excessively large combinatoric problem for track reconstruction. One possible way to mitigate the BIB is to discriminate single-hits against multi-layer correlated hits characteristic of real higher momentum tracks. The optimal way to reduce this background needs to be identified. The study requires detailed simulation of the expected detector conditions, with overlapping simulated events of interest. Cost, radiation tolerance, power and cooling optimization will place constraints on number of electronics channels and sophistication of electronics needed.
        
        The ability to have precision timing information available as hit level information would improve the ability to separate hits from particles from the hard scattering interaction from BIB particles and allow filtering at the readout and trigger level. Silicon sensor with ~20-30 picosecond resolution are being studied and developed currently ~\cite{PELLEGRINI201412}. Continued development of these technologies to be able to construct future detector with picosecond level timing resolution would be expected to vastly decrease the readout requirements and reduce combinatoric confusion in tracking. 
        
        \item \textbf{Calorimeter}: Photons, electrons, taus and jets need to be reconstructed with good resolution and efficiency from few 10s of GeV to TeV scale in order to do physics at the muon collider. In addition to the usual trade-off between the electromagnetic resolution and the hadronic resolution, mitigation of fake jet background, from the BIB background deposits, is essential. Segmentation of the calorimeter in all three dimensions and signal arrival time is needed both for particle-flow particle reconstruction to improve the energy resolutions using tracking of charged-hadrons and BIB mitigation. Cost drivers are both calorimeter materials quality, from signal yield, radiation damage, calibration stability, etc. points of view, and the electronics needed for processing the signals.
        
        Different types of high granularity calorimeters with good energy resolution and precise timing information are being developed and constructed. Continued development of novel calorimeters with picosecond level timing resolution would improve rejection of BIB particles and improve photon, electron, and jet reconstruction by rejecting energy deposits from non-collison sources. 
        
        \item \textbf{Muon system}: The challenge of measuring momentum of TeV-muons is to be studied. The bend needed for TeV muons has significant cost implications due to the magnet size and field. Secondly, bremstrahlung from high momentum muons begins to degrade the ability to measure the momentum well. Perhaps, a secondary lower momentum but larger sized magnet may be appropriate but will have implications on the size of the experimental cavern and the size of the muon detectors that are needed. Deep underground location for a 10-TeV machine is another cost consideration.
        
        Micropatern gas detectors with tens of picosecond timing resolution and integrated readout electronics are currently an item of high priority in the gas detector community ~\cite{Black:MPGD}. The continued development of detectors with excellent spacial and timing resolution which do not use greenhouse gases are critical to the performance of such a detector. 
        
        \item \textbf{Data acquisition system}: The true event rate at the muon collider is low enough, and the time between collisions is long enough, to be able to operate triggerlessly provided the data volumes are under control. However, much investigation is needed to tame the data volumes by dropping the BIB hits, while not compromising on the signal hit efficiencies. The power and cooling needed for on-detector electronics is also of concern. As well the continued development of integrated electronics, fast high-bandwidth readout to handle large data volumes should also be pursued.

    \end{itemize}
    Studies are needed in all of the areas outlined. While generic detector R\&D can provide innovative solutions, dedicated studies of the muon collider environment in simulation is the key to make improvements in the design of the detector, which is currently adopted from the CLIC detector. The nature of the problem requires detailed layout of the detector and simulation in \GEANT 4, requiring talented and knowledgeable experts providing core framework support, as well as a group of students and postdocs. 
    
    While technologies needed for a muon collider detector appear feasible, dedicated R\&D efforts are necessary to make them mature and cost-effective by the time the accelerator technology is ready. The detector R\&D thrusts of timing, on-detector filtering, and radiation hardness are aligned with work needed for FCC and other future collider detectors. There is clearly a general need for R\&D on the simultaneous optimization of mass, speed, power and cooling, and mechanics for the muon collider and other future detector systems. It is therefore imperative to incorporate these needs into the existing and future detector R\&D programs.

%% file: neutrinos.tex
A muon collider offers a unique opportunity to deepen our understanding of the neutrino sector.
A neutrino beam sourced by muon decay could have significant impact in several aspects of neutrinos physics: standard and beyond standard oscillation physics; searches for novel states; BSM effects in the neutrino sector; electroweak precision physics; neutrino-nucleus interaction modeling; among others.
Since historically a neutrino experiment in which the beam comes from muon decays has been refered to as the \emph{Neutrino Factory}, we will also use the term hereafter, even when we will not be focusing on oscillation physics.
In what follows, we will briefly describe some of these opportunities and explain what would be the role of a neutrino factory in leveraging those.
Further details on the physics case of a neutrino factory and its synergy with a muon collider can be found in Ref.~\cite{Bogacz:2022xsj}.

First, let us consider uncertainties related to neutrino production and detection.
In beam neutrino experiments, neutrinos are produced primarily from the decay of mesons, which are in turn produced by a beam of protons impinging on a target.
The uncertainties related to meson production will therefore propagate to the neutrino fluxes.
On top of that, neutrinos are detected by interacting with nuclei, due to the larger cross section compared to neutrino-electron scattering.
Currently, flux and cross section uncertainties are among the dominant limiting factors in the determination of neutrino oscillation parameters~\cite{NOvA:2021nfi, T2K:2021xwb}.
Moreover, neutrino event generators do not describe well exclusive final states and differential cross sections, which are both crucial to the reconstruction of the incoming neutrino energy~\cite{MINERvA:2020zzv, CLAS:2021neh}.
Neutrino experiments mitigate these issues by leveraging a two-detector configuration: a near detector which is sensitive to the unoscillated neutrino spectrum, and thus measures the initial flux and cross section; and a far detector which measures the oscillated neutrino spectrum.
Nevertheless, the near detector is not sufficient to determine both neutrino flux and cross section, and the uncertainties associated to those will be the dominant uncertainties in future measurements of neutrino experiments such as DUNE and Hyper-Kamiokande (HK)~\cite{DUNE:2020lwj, Hyper-Kamiokande:2018ofw}.
 
Having a well known neutrino flux is the key strength of a neutrino factory.
This would allow precise determination of neutrino-nucleus interaction cross sections, which could significantly improve measurements at DUNE and HK $CP$ violation, precision determination of mixing angles and mass splittings~\cite{DUNE:2020lwj, Hyper-Kamiokande:2018ofw}, nonstandard neutrino interactions~\cite{Proceedings:2019qno}, sterile neutrinos~\cite{DUNE:2020fgq}, Earth tomography with atmospheric neutrinos~\cite{Kelly:2021jfs}, etc.
To achieve this, one would need a neutrino spectrum that overlaps with DUNE and/or HK spectra (around 1-5 GeV and 0.2-1 GeV, respectively), and near detectors with the same chemical composition (argon and water, respectively).
Note that a better determination of neutrino-nucleus interaction cross sections could be used to improve oscillation measurements even after the experiments took data, as long as data preservation plans are put in place.
On top of that, searches for new states in neutrino experiments, such as heavy neutral leptons, light axions, dark matter, and dark neutrinos~\cite{DeRomeri:2019kic, Berryman:2019dme, Brdar:2020dpr, Kelly:2021jfs, Bertuzzo:2018itn, Bertuzzo:2018ftf, Ballett:2018ynz, Ballett:2019pyw, Abdullahi:2020nyr, Datta:2020auq, Dutta:2020scq, Abdallah:2020biq, Abdallah:2020vgg, Hammad:2021mpl, Dutta:2021cip}, typically need to deal with backgrounds originating in neutrino-nucleus interactions.
A better determination of these cross sections would translate into smaller background uncertainties, improving past and current BSM searches.

At higher neutrino energies, above 10 GeV or even 100 GeV, neutrino events would be dominantly observed by deep inelastic scattering cross section (DIS).
Measuring DIS in neutrino-nucleus interactions could be relevant to the understanding of high energy atmospheric neutrinos and perhaps even to high energy cosmogenic neutrinos, as observed by the IceCube experiment~\cite{Stettner:2019tok, IceCube:2020wum}.
In particular, nuclear parton distribution functions still have large uncertainties~\cite{Hirai:2007sx,AbdulKhalek:2019mzd, AbdulKhalek:2021gbh}, and a measurement of neutrino-nucleus interactions with well understood neutrino fluxes could improve our knowledge of these nuclear pdfs.
DIS events would also allow to revisit the NuTeV determination of the weak mixing angle with improved theoretical descriptions of QCD and nuclear effects, possibly clarifying the NuTeV anomaly~\cite{NuTeV:2001whx}.
Besides, a high energy neutrino factory would be an excellent environment to probe generaic new physics in neutrino interactions via an effective field theory approach.
This would only require a near detector, and as much statistics as one can obtain.
It has been shown that experiments like FASER$\nu$ can be competitive to LHC and meson decay observables when probing certain dimension-6 EFT operators~\cite{Falkowski:2019xoe, Falkowski:2021bkq}.
In fact, a high energy muon collider could host a neutrino experiment similar to FASER$\nu$~\cite{FASER:2018eoc, Feng:2022inv},  with much smaller flux uncertainties compared to a forward detector at a hadron collider.

%% file: beamdump.tex
A muon collider facility would also provide opportunities to search for weakly coupled new particles below the weak scale. Given a high energy muon beam, an economical extension to the collider facility is a beam dump experiment, where the muon beam is directed onto a dense target with a detector placed at the end of a long decay volume. Similar extensions have been proposed for the LHC and other future colliders~\cite{Alekhin:2015byh, Feng:2022inv, Kanemura:2015cxa}. With a dense target, the cross section for forward production of new states with masses $m \ll E_{\textrm{beam}}$ can be very large, compensating the weak couplings to SM states.

The precise reach for new, weakly-coupled particles depends on the details of the experimental setup, such as the target material, shielding, the length of the decay volume, and the detector, as well as the number of muons on target. These considerations must be optimized in concert with the design of the collider facility to maximize the reach while accounting for the cost and collider facility constraints. Given the sensitivity to these choices, and the immense opportunity, it is necessary to consider this optimization in the design of a muon collider facility itself.

\begin{figure}[t]
\centering
\includegraphics[width=0.45\linewidth]{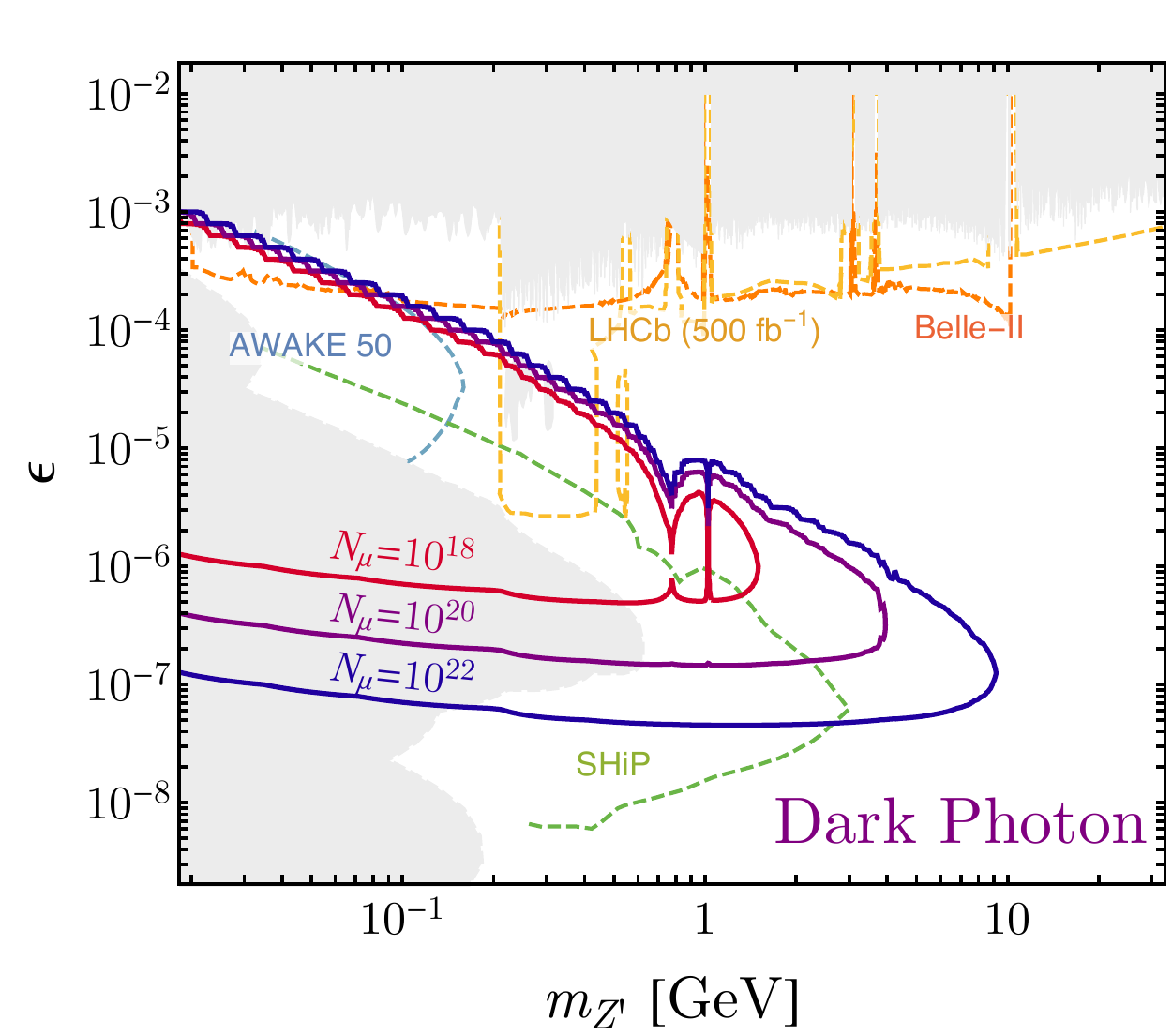}
\quad
\includegraphics[width=0.45\linewidth]{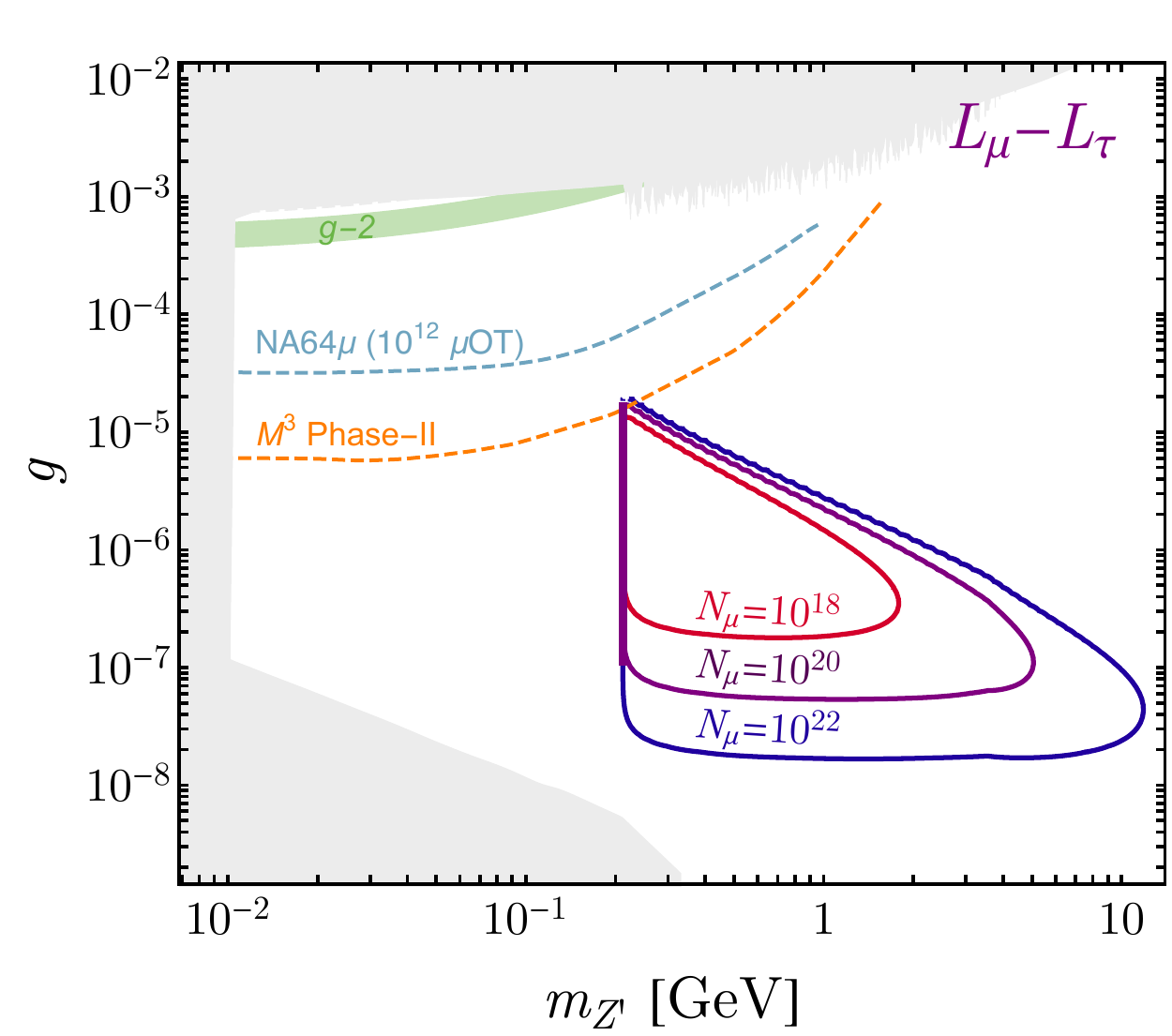}
\caption{The reach of a 1.5 TeV muon beam dump experiment for a dark photon (left) or $L_{\mu} - L_{\tau}$ gauge boson, adapted from Ref.~\cite{Cesarotti:2022ttv}.}
\label{fig:beamdump_reach}
\end{figure}

A first estimate of the potential reach, however, was provided in Ref.~\cite{Cesarotti:2022ttv}, assuming a 1.5 TeV muon beam, with variable choices of the number of muons on target and reasonable choices for the other experimental details. The results of this study are reproduced in Fig.~\ref{fig:beamdump_reach}, considering two benchmark models: a kinetically-mixed dark photon, and an $L_{\mu} - L_{\tau}$ gauge boson. In the former case, we see that the muon collider reach extends to higher masses and larger mixing angles than any other proposed experiment. This is due to the larger boost that these particles are produced with from a high-energy lepton beam, which compensates their shorter lifetimes at these masses and mixing angles so that they can survive to the detector. An even higher energy beam in e.g., a 10 TeV collider facility, could extend the reach even further. In the $L_{\mu} - L_{\tau}$ gauge boson case, a muon beam dump at these energies would probe entirely disparate parameter space from any other imagined experiment. This is both due to the boost, and from the unique setup involving a muon beam as opposed to an electron or proton. Exploration of the reach for other dark sectors, particularly those benefiting from the muon-specific opportunities, is deserving of further study.

Finally, while we have emphasized the unique opportunities of a high energy muon beam above, other opportunities for probing light, weakly coupled dark sectors exist at intermediate stages of a muon collider facility. The reach of muon beam experiments with lower energy muon beams has been studied in other contexts~\cite{Chen:2017awl, Kahn:2018cqs, Sieber:2021fue}, demonstrating their capability for probing new physics with muon-specific couplings such as the $L_{\mu} - L_{\tau}$ gauge boson. Staged approaches to the development of a muon collider facility may present the perfect opportunity for an economical, low-energy muon beam dump experiment to be performed in parallel, and potentially in combination with neutrino beam experiments discussed in the preceding Section. Ultimately, it is important that the full scope of possibilities for exploring new physics at a muon collider facility be explored, so that no stone is left unturned. 

%% file: muIC.tex
%\subsubsection{Mu-Ion Collider Concept}

\begin{figure}[t!]
\centering
\includegraphics[width=0.9\linewidth]{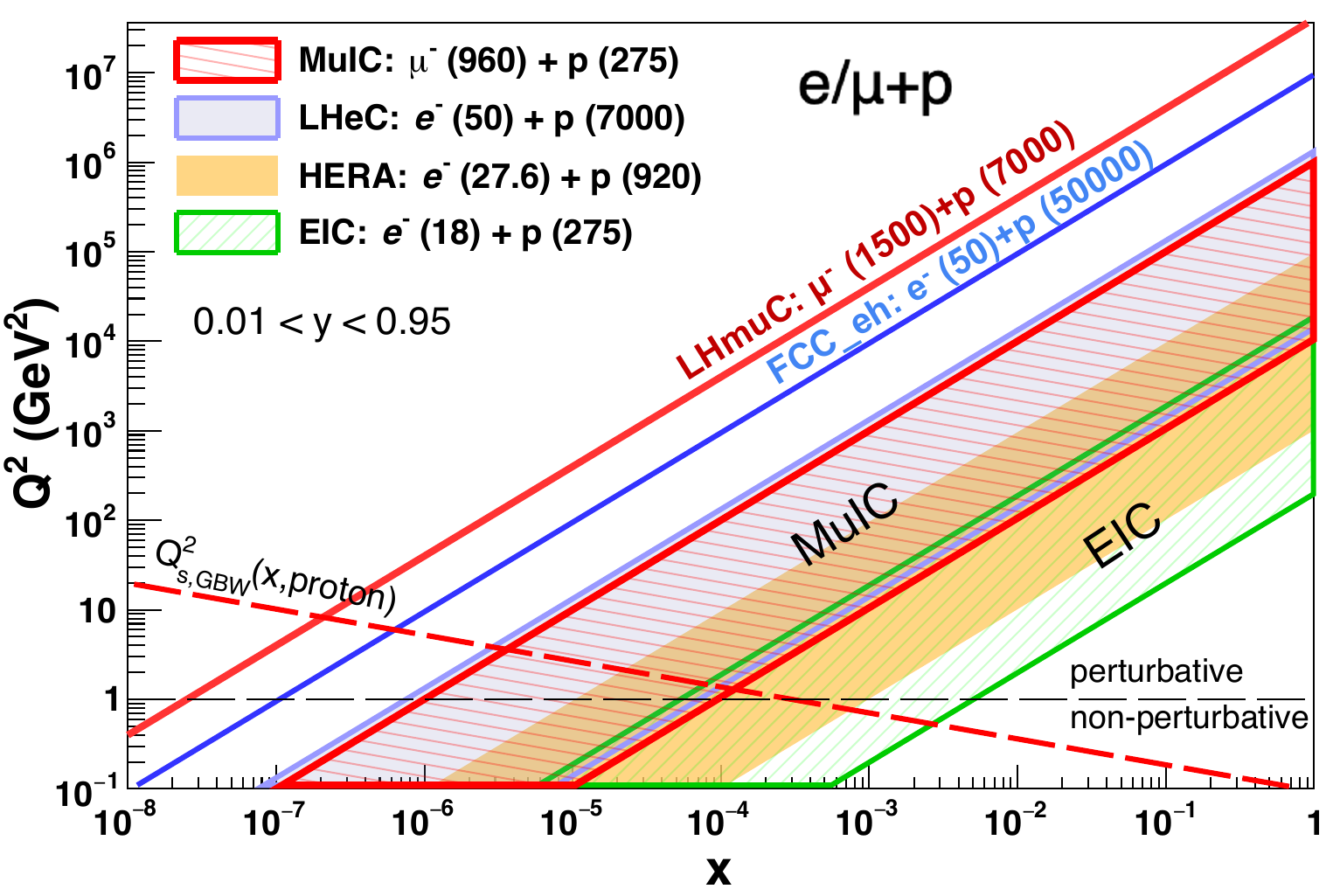}
\caption{Kinematic coverage of $Q^{2}$ and $x$ in deep 
inelastic lepton-proton scattering 
for two muon-ion collider design options and for the 
EIC at BNL, HERA at DESY, and the LHeC and FCC-eh options at CERN, 
each at their  maximum beam energies. The inelasticity ($y$) range 
is assumed to be $0.01<y<0.95$ (hatched areas). The long dashed lines
indicate the saturation 
scale as a function of $x$ in the proton from the GBW model~\cite{GolecBiernat:1998js}.}
\label{Q2x-all}
\end{figure}

A possible scientific target for an intermediate step toward the ultimate development of a multi-TeV muon collider is that of a muon-proton and muon-nucleus collider facility, referred to as a mu-ion collider, as discussed in Refs.~\cite{Acosta:2021qpx,Acosta:2022ejc}. Such a facility could utilize the existing hadron accelerator infrastructure at laboratories such as BNL (as an upgrade to the planned Electron-Ion Collider), CERN (using the LHC), or Fermilab while seeding, or leveraging, the development of a high energy and high intensity muon storage ring at the same site. A muon-proton center-of-mass energy of up to 1~TeV at BNL (6.5~TeV at CERN) can be achieved when a 1~TeV (1.5~TeV) muon beam is brought into collision with a 0.275 TeV (7 TeV) proton beam at that facility. Such a mu-ion collider would enable deep inelastic scattering measurements in completely new regimes at low parton momentum fraction $x$ and high squared four-momentum transfer $Q^2$, as illustrated in Fig.~\ref{Q2x-all}, which will further elucidate the structure of the proton and of nuclei as well as provide precision QCD and electroweak measurements complementary to those done at lepton and hadron colliders. The coverage of a mu-ion collider at BNL is nearly identical with that of the proposed Large Hadron electron Collider (LHeC) at CERN, which would use the LHC and a 50~GeV electron beam, while the coverage of a muon-LHC collider at CERN (LHmuC)~\cite{Acosta:2022ejc} would significantly exceed that of the FCC-eh option of a 50~TeV proton beam colliding with a 50~GeV electron beam.

%% Figure

The maximum muon beam energy at BNL is taken to be 1~TeV assuming that 11~T dipoles are used in the arcs of the RHIC~\cite{Harrison:2002es} ring (290~m radius). In terms of staging scenarios for the muon beam energy, we note that any energy above 93~GeV leads to a center-of-mass energy above that of the HERA $ep$ collider. In addition, beam polarization is a unique capability at the BNL facility. A muon beam energy of 1.5~TeV at CERN comes simply from assuming that one beam of a 3~TeV $\mu^+\mu^-$ collider constructed there is brought into collision with the LHC. 

The estimate on the maximum achievable luminosity at the mu-ion collider is $4.7 \times 10^{33}$~cm$^{-2}$s$^{-1}$ ($2.8\times 10^{33}$~cm$^{-2}$s$^{-1}$) for a collider at BNL (CERN). For the muon beam, the proposed parameters of the proton driver scheme from Ref.~\cite{Palmer:2014nza,Delahaye:2019omf} are taken. The muon bunch repetition frequency is taken to be 12--15~Hz. The BNL proton beam parameters are assumed to be those achieved at RHIC~\cite{Aschenauer:2014cki} or foreseen to be achieved at EIC~\cite{eic_cdr}, and the CERN beam parameters correspond to those of the  LHC~\cite{Kaya:2019ecf}. 

%\subsubsection{Mu-Ion Collider Physics Potential}

The scientific potential of the mu-ion collider is similar to the proposed LHeC~\cite{LHeCStudyGroup:2012zhm}, but with complementary scattering kinematics and complementary sensitivity to BSM processes with a muon beam as opposed to an electron beam. For deep inelastic structure function measurements, the mu-ion collider can probe parton momentum fractions $x$  in the proton as small as  ${\approx }10^{-6}$ (${\approx }10^{-8}$) for collisions at BNL (CERN), and thus should be sensitive to gluon saturation effects \cite{GolecBiernat:1998js}. The corresponding maximum reach in $Q^2$ is $10^6$~GeV$^2$ ($4\times 10^7$~GeV$^2$), which is well above the electroweak scale where neutral-current and charged-current scattering cross sections become similar. The maximum reach in $Q^2$ of the BNL mu-ion collider can be achieved with 10~fb$^{-1}$ of integrated luminosity, and is well within the machine estimation. However, to probe the highest $Q^2$ reach of the LHmuC would require ${\approx }1000$~fb$^{-1}$, which is somewhat disfavored in the beam assumptions. However, the cross sections at lower $Q^2$ and correspondingly lower $x$ are many orders of magnitude larger, and thus enable a science program in that regime with significantly less luminosity by the same order. Finally, the mu-ion collider also could provide polarization of both beams (when utilizing the BNL facility) for spin structure measurements, and provide lepton-proton and antilepton-proton collisions with similar luminosity (although switching between the two would require flipping the polarity of magnets along the muon beam line). 

Higgs boson production also opens up at the mu-ion collider through vector boson fusion, with a cross section of $80$~fb for $\mu^- p$ collisions at the BNL mu-ion collider and 1700~fb  at the LHmuC.
The Higgs boson decay products tend to be more central in the detector for collisions at the mu-ion collider than for those at the LHeC, given its more symmetric collision. For a  $H\to b\bar{b}$ measurement at the mu-ion collider, a statistical uncertainty on the signal of about 3\% is possible from a 10-year data set (400~fb$^{-1}$). Another decay channel with significant yield that is potentially measurable is $H\to \tau^-\tau^+$. We note that $H\to gg$ would be an interesting target as well. The $H\to c\bar{c}$ channel would open up with the higher cross sections available at higher energies, such as with the LHmuC. 

The production of other Standard Model particles at a mu-ion collider also are of interest. In total, the inclusive $W^\pm$ production cross section in $\mu^{-}p$ collisions for the mu-ion collider at BNL is about 20~pb, yielding $2.1\times 10^4$ leptonic $W\to \ell \nu$ decays into each lepton flavor for 10~fb$^{-1}$ of integrated luminosity. This increases by an order of magnitude for the LHmuC configuration. Both single and pair production of top quarks is also possible, the former of which directly targets the CKM matrix element $V_{tb}$. Precision mass measurements of the $W$ boson and $t$ quark also may be viable, depending on the measurement capabilities of a mu-ion collider experiment.

The mu-ion collider also offers interesting sensitivity to BSM processes. For example, the $s$-channel and $t$-channel exchange of a new particle such as a leptoquark or $Z^\prime$ boson is possible, and $\mu$-$p$ scattering measurements would be particularly relevant to test BSM models that propose to explain the potential anomaly in the muon $g-2$ measurement and in the deviations from lepton flavor universality in $B$ meson decays. While certain leptoquark cross sections at the mu-ion collider have been reported in \cite{Acosta:2022ejc}, an analysis of the constraint sensitivity still remains to be done.

%\subsubsection{Synergies}

The mu-ion collider concept expands the science opportunities at a facility featuring a high-energy, high-intensity muon accelerator and collider. When the muon beam is brought into collision with a hadron beam, a unique frontier in both particle and nuclear physics is opened up at high $Q^2$ and low $x$.  The communities involved in the funding and construction of such a facility overall could therefore increase as well, sharing the overall cost burden. Additionally, much of the science extraction can be done with modest goals for the initial muon beam energy and luminosity, making it an ideal scientific target for a muon collider demonstrator that is on the path toward the development of an ultimate ${\cal O}(10+$ TeV) $\mu^+\mu^-$ collider. The mu-ion collider is very amenable to a staged approach to the maximum energy and intensity of the muon beam. Costs for the mu-ion collider component of the facility can be further minimized by reusing existing hadron beam and collider infrastructure at an appropriately chosen site. 

The kinematics of the collisions and the experimental needs of a mu-ion collider are similar to those  for a muon collider experiment. The asymmetric nature of the collisions, however, leads to an emphasis of instrumentation along the downstream muon beam direction at small scattering angles, although the rest of the final state tends to occupy the central region of the experiment as would $\mu^+\mu^-$ annihilations. One shielding cone may be sufficient to reduce the beam induced background for a mu-ion collider experiment (on the incoming muon side, to be studied), which would potentially open up the instrumentation room at small angles on the downstream side. As the scattered muon peaks strongly in the muon beam direction for low-$Q^2$ and for vector boson fusion processes, a dedicated muon spectrometer along the beamline would be needed. But such a spectrometer design also may prove useful for experiments at a $\mu^+\mu^-$ collider to tag the nature of VBF processes.

%% file: pathforward.tex
The most fruitful path forward towards the development of conceptual design of the Muon Collider and associated detector in the USA, will involve coordination with the IMCC efforts. A US MCC needs to be formed with a goal towards preparing a proposal for the US funding agencies, assuming identification of the Muon Collider as an important path forward for the US energy frontier community. The US collaborations of the LHC experiments and the accelerator form a good example. Given the necessity of a strong coupling of the Muon Collider accelerator and detector communities, it seems most suitable to follow the IMCC model with a unified US MCC group, rather than work individually at various institutes. We anticipate that individual US institutions joining the IMCC will also be members of a new US MCC, elect its leadership and represent the combined interest of the US muon collider community to the US funding agencies.

\subsection{Engagement in IMCC} 
To foster the Muon Collider concept an International Muon Collider Collaboration (IMCC) has been initiated after the recommendation of the update of the European Strategy for Particle Physics, initially hosted at CERN. The collaboration will address the muon collider challenges and develop the concept and technologies in the coming years in order to be able to gauge if the investment into a full conceptual design and demonstration program is scientifically justified. This will allow future Strategy Processes in the different regions to make informed decisions. The collaboration welcomes active participation from the United States and encourages interested institutions to sign a Memorandum of Cooperation. Many US institutions are interested in joining the collaboration but have been waiting for the outcome of Snowmass/P5. An explicit endorsement by P5 and a collaborative agreement between CERN and DOE would allow for direct participation of US institutions in IMCC and for direct contributions from US to the program. 
    
\subsection{Contributions to Physics Studies}
The Muon Smashers Guide and other studies from the  theoretical particle physics community have made a strong impact in the burgeoning interest in the 10-TeV scale physics. The case for developing this physics case, with more realism and rigour in event generation and simulation, and addition of other physics cases of interest is very much of interest. A strong collaboration of phenomenologists and experimenters is necessary for this effort to be fruitful. We anticipate that there will be contributions which will establish the prospects for the measurements of both standard model higgs parameters and exploration of new physics sectors. Furthermore, studies of various staging scenarios will be carried out together by the accelerator, experimental, and theoretical communities in order to develop the most comprehensive integrated program. 
    
\subsection{Contributions to Detector R\&D}
The US groups should participate in determining the detector parameters for achieving the best physics results. The choice of detector technologies, size, overall geometry and segmentation are yet to be optimized. The initial studies from the IMCC, based on adaptation of the CLIC detector, is a good start, and appears suitable. Perhaps, a fresh ground up start with a second detector concept may also be useful in understanding the details of what is important and make a comparative evaluation to form a viable detector proposal. Of particular importance is mitigation of beam induced background. Presently, BIB simulation program is tricky and time consuming and restricted to lower energy colliders. A close collaboration of Fermilab based beam dynamics simulation experts and postdocs/students from US universities could yield more realistic studies for 10 TeV collider.
    
\subsection{Contributions to Accelerator R\&D}
The US based MAP collaboration has made very significant contributions to the Muon Collider concept. The IMCC has essentially taken over the direction of Muon Collider R\&D lately because of the US divestment from the MAP program. However, some of the expertise at DOE national labs remains and can quickly be reconstituted to bootstrap a substantial R\&D team. The rebuilding of the team will enable us to participate in the CERN based demonstrator efforts which are recently funded. Perhaps, some of the pieces of equipment used for MAP R\&D could be resurrected to contribute towards the IMCC demonstrator. The reconstituted accelerator R\&D team will also give us the opportunity to prepare the US demonstrator proposals, once the P5 process identifies the project priority and the DOE the scale of project budgets. 
    
\subsection{Explore US options}
The reconstituted US Muon Collaboration with the MAP participants and its expansion to include the expanding group of scientists with a new-found interest in the Muon Collider, provides an ideal opportunity to make designs of the collider for potential US siting. Fermilab could be an ideal site for a Muon Collider with a center-of-mass energy reach at the desirable 10-TeV scale. The synergy with the existing/planned accelerator complex and neutrino physics program at Fermilab is an additional stimulus for such investment of effort. A set of Muon Collider design options, with potential siting at Fermilab, could be made for discussions at the IMCC and the international committees to eventually form a global consensus decision on siting and selection of the Muon Collider. Having a pre-CDR document summarizing design for the Fermilab-sited Muon Collider in time for the next Snowmass is a good goal. The preparation of such a document will require substantial, yet affordable, investment. Such an investment will reinvigorate the US high-energy collider community and enable much needed global progress towards the next energy frontier.